\documentclass[journal]{IEEEtran}

\usepackage[utf8]{inputenc}
\usepackage{cite}
\usepackage{comment}
\usepackage{amsfonts}
\usepackage{amssymb}
\usepackage{amsmath}
\usepackage{xcolor}

\usepackage{graphicx}
\usepackage{bigdelim}
\usepackage{caption,setspace}
\usepackage[ruled,vlined]{algorithm2e}
\usepackage{booktabs}
\usepackage{subfig}
\usepackage{soul}
\usepackage{mathtools} 
\usepackage{wrapfig}
\usepackage{hyperref}
\usepackage{kbordermatrix}
\usepackage{float}
\allowdisplaybreaks
\usepackage[edges]{forest}
\usepackage{longtable}
\usepackage{supertabular}
\usepackage{bm}
\usepackage{balance}

\usepackage[acronym, nohypertypes={acronym}]{glossaries}

\usepackage{pifont}
\newcommand{\cmark}{\ding{51}}%
\newcommand{\xmark}{\ding{55}}%

\usepackage{tikz}
\tikzset{
	basic/.style  = {draw, text width=2cm, drop shadow, font=\sffamily, rectangle},
	root/.style   = {basic, rounded corners, thin, align=center, fill=red!30},
	level-2/.style = {basic, rounded corners=6pt, thin,align=center, fill=blue!20, text width=3cm},
	level-3/.style = {basic, thin, align=center, fill=orange!20, text width=2.4cm}
}

\usetikzlibrary{trees,positioning,shapes,shadows,arrows}

\usepackage{lipsum,adjustbox}
\usetikzlibrary{mindmap,calc,positioning}

\newtheorem{rem}{Remark}

\definecolor{Lightgray}{RGB}{235,235,235}
\usepackage{colortbl}
\usepackage[para]{threeparttable}

\usetikzlibrary{arrows}
\tikzset{decision/.style = {diamond, draw, fill=black!20, text width=4.5em, text badly centered, node distance=2.8cm, inner sep=0pt},
	block/.style    = {rectangle, draw, fill=blue!5, text width=5em, text centered, rounded corners, minimum height=2em},	block2/.style    = {rectangle, draw, dashed,  fill=blue!20, text width=5em, text centered, rounded corners, minimum height=2em},
	line/.style     = {draw, -latex', very thick},
	cloud/.style    = {draw, ellipse,fill=red!20, node distance=2.8cm, minimum height=2em}
}

\makeglossaries
\newacronym{3gpp}{3GPP}{3rd Generation Partnership Project}
\newacronym{5g}{5G}{fifth generation}
\newacronym{6g}{6G}{sixth generation}
\newacronym{ae}{AE}{autoencoder}
\newacronym{ai}{AI}{artificial intelligence}
\newacronym{anoma}{A-NOMA}{asynchronous NOMA}
\newacronym{atsc}{ATSC}{advanced television systems committee}
\newacronym{awgn}{AWGN}{additive white Gaussian noise}
\newacronym{bc}{BC}{broadcast channel}
\newacronym{bch}{BCH}{Bose-Chaudhuri-Hocquenghem}
\newacronym{ber}{BER}{bit error rate}
\newacronym{bicm}{BICM}{bit-interleaved coded modulation}
\newacronym{bler}{BLER}{block error rate}
\newacronym{bpsk}{BPSK}{binary phase-shift keying}
\newacronym{bs}{BS}{base station}
\newacronym{cdma}{CDMA}{code division multiple access}
\newacronym{cnoma}{C-NOMA}{code-domain NOMA}
\newacronym{csi}{CSI}{channel state information}
\newacronym{csit}{CSIT}{channel state information at the transmitter}
\newacronym{dpc}{DPC}{dirty-paper coding}
\newacronym{dvbt}{DVB-T}{digital video broadcasting terrestrial}
\newacronym{e2e}{E2E}{end-to-end}
\newacronym{fcnn}{FCNN}{fully-connected neural network}
\newacronym{fdd}{FDD}{frequency division duplexing}
\newacronym{harq}{HARQ}{hybrid automatic repeat request}
\newacronym{ici}{ICI}{inter-cell interference}
\newacronym{idma}{IDMA}{interleave division multiple access}
\newacronym{igma}{IGMA}{interleave-grid multiple access}
\newacronym{iot}{IoT}{Internet of things}
\newacronym{isac}{ISAC}{integrated sensing and communications}
\newacronym{isi}{ISI}{inter-symbol interference}
\newacronym{iui}{IUI}{inter-user interference}
\newacronym{lcrs}{LCRS}{low-code-rate
spreading}
\newacronym{ldpc}{LDPC}{low-density parity-check}
\newacronym{lds}{LDS}{low-density signature/spreading}
\newacronym{lpma}{LPMA}{lattice partition multiple access}
\newacronym{lte}{LTE}{Long Term Evolution}
\newacronym{mac}{MAC}{multiple access channel}
\newacronym{mcs}{MCS}{modulation and coding scheme}
\newacronym{mmse}{MMSE}{minimum mean squared error}

\newacronym{mimo}{MIMO}{multiple-input multiple-output}
\newacronym{ml}{ML}{machine learning}
\newacronym{mpa}{MPA}{message passing algorithm}
\newacronym{musa}{MUSA}{multi-user shared access}
\newacronym{ngma}{NGMA}{next generation multiple access}
\newacronym{noca}{NOCA}{non-orthogonal coded access}
\newacronym{nr}{NR}{New Radio}
\newacronym{ofdm}{OFDM}{orthogonal frequency division multiplexing}
\newacronym{ofdma}{OFDMA}{orthogonal frequency division multiple access}
\newacronym{oma}{OMA}{orthogonal multiple access}
\newacronym{noma}{NOMA}{non-orthogonal multiple access}
\newacronym{pam}{PAM}{pulse amplitude modulation}
\newacronym{pdma}{PDMA}{pattern division multiple access}
\newacronym{pnoma}{P-NOMA}{power-domain NOMA}
\newacronym{psk}{PSK}{phase-shift keying}
\newacronym{pdcch}{PDCCH}{Physical downlink control channel}
\newacronym{pucch}{PUCCH}{Physical uplink control channel}
\newacronym{pbch}{PBCH}{Physical broadband channel}
\newacronym{prach}{PRACH}{Physical random access channel}
\newacronym{pdsch}{PDSCH}{Physical downlink shared channel}
\newacronym{pusch}{PUSCH}{Physical uplink shared channel}
\newacronym{qam}{QAM}{quadrature amplitude modulation}
\newacronym{qos}{QoS}{quality of service}
\newacronym{qpsk}{QPSK}{quadrature phase-shift keying}
\newacronym{ran}{RAN}{radio access network}
\newacronym{rb}{RB}{resource block}
\newacronym{re}{RE}{resource element}
\newacronym{ris}{RIS}{reconfigurable intelligent surface}
\newacronym{rsma}{RSMA}{rate-splitting multiple access}
\newacronym{resma}{ReSMA}{resource spread multiple access}
\newacronym{rdma}{RDMA}{repetition division multiple access}
\newacronym{scma}{SCMA}{sparse code multiple access}
\newacronym{scsic}{SC-SIC}{superposition coding with successive interference cancellation}
\newacronym{sdma}{SDMA}{space division multiple access}
\newacronym{ser}{SER}{symbol error rate}
\newacronym{sic}{SIC}{successive interference cancellation}
\newacronym{siso}{SISO}{single-input single-output}
\newacronym{snr}{SNR}{signal-to-noise ratio}
\newacronym{svd}{SVD}{singular value decomposition}
\newacronym{tcm}{TCM}{trellis-coded modulation}
\newacronym{tcma}{TCMA}{trellis-coded multiple access}
\newacronym{tcnoma}{TC-NOMA}{trellis-coded NOMA}
\newacronym{thz}{THz}{terahertz}
\newacronym{tnoma}{AT-NOMA}{asynchronous time-domain NOMA}
\newacronym{uav}{UAV}{uncrewed aerial vehicle}
\newacronym{ucnoma}{UC-NOMA}{uncoded NOMA}
\newacronym{ue}{UE}{user equipment}
\newacronym{urllc}{URLLC}{ultra-reliable low-latency communications}
\newacronym{adc}{ADC}{analog-to-digital converter}
\newacronym{dac}{DAC}{digital-to-analog converter}
\newacronym{mmwave}{mmWave}{millimeter-wave}
\newacronym{vamos}{VAMOS}{voice services over adaptive multi-user channels on one slot}
\newacronym{aqpsk}{AQPSK}{adaptive QPSK}
\newacronym{apnoma}{AP-NOMA}{asynchronous P-NOMA}

\begin{document}

	\title{Modulation and Coding for NOMA and RSMA}
	
	\author{Hamid Jafarkhani,~\IEEEmembership{Fellow,~IEEE}, Hossein Maleki,~\IEEEmembership{Student Member,~IEEE}, and \\ Mojtaba Vaezi,~\IEEEmembership{Senior Member,~IEEE} \IEEEspecialpapernotice{(Invited Paper)}
		\thanks{Hamid Jafarkhani and Hossein Maleki are with Center for Pervasive Communications and Computing, University of California at Irvine, Irvine, CA 92697 USA (e-mails: hamidj and malekih@uci.edu). Their work was supported in part by the NSF Awards CNS-2229467 and CCF-2328075. 
  Mojtaba Vaezi is with the Department of Electrical and Computer Engineering, Villanova University, Villanova, PA 19085, USA (e-mail: mvaezi@villanova.edu). His work was supported by the U.S. National Science Foundation under Grant ECCS-2301778.
  Corresponding authors: Hamid Jafarkhani and Mojtaba Vaezi.
  }}

	\maketitle
		
 \begin{abstract}
		\Gls{ngma} serves as an umbrella term encompassing transmission schemes distinct from conventional orthogonal methods.  As a prominent candidate of \gls{ngma}, \gls{noma} emerges as a promising solution, enhancing connectivity  by allowing multiple users to concurrently share time, frequency, and space. However, \gls{noma} faces challenges in practical implementation, particularly in canceling inter-user interference. In this paper, first, we discuss the principles behind \gls{noma} and review the conventional \gls{noma} methods and results. Then, to address the above challenges, we present asynchronous transmission and interference-aware modulation techniques, leading to decoding free from successive interference cancellation. The goal is to design constellations that dynamically adapt to interference, minimizing \glspl{ber} and enhancing user throughput in the presence of inter-user, inter-carrier,  and inter-cell interference. The traditional linkage between minimizing \gls{ber} and increasing spectral efficiency is addressed, with the exploration of deep autoencoders for end-to-end communication as a new concept with significant potential for improving \glspl{ber}. 
  Interference-aware  modulation techniques  can revolutionize constellation design and communication over  non-orthogonal channels. \Gls{rsma} is another promising interference management technique in multi-user systems. Beyond addressing existing challenges and misconceptions in finite-alphabet \gls{noma}, this paper offers fresh insights to the field and provides an overview of code-domain NOMA schemes,  trellis-coded \gls{noma}, and  \gls{rsma} as other potential candidates for \gls{ngma}.  {
  Additionally, we discuss the evolution of channel coding towards low-latency 
  communication and examine the modulation and coding schemes in fifth-generation cellular networks.}  Finally, we examine future research avenues and challenges, highlighting the importance of addressing them for the practical realization of \gls{noma} from a theoretical concept to a functional technology. 
 \end{abstract}



\glsresetall
\section{Introduction and Historical Notes}\label{sec:intro}
The next generation of communication systems is expected to deliver improved end-user experience by
offering new applications and services such as industry automation, smart cities, virtual and augmented
reality, remote medical surgery, self-driving cars, and \glspl{uav}. These envisioned services pose many challenging
requirements, such as low latency, high data rates, massive connectivity, high reliability, and support
of diverse quality of service. 
The need for massive connectivity in fifth-generation (5G) wireless  networks and beyond is mainly pushed by the explosion of the \gls{iot} devices, as projected by leading industry including Cisco and Ericsson \cite{vaezi2022cellular}. 
Particularly, 6G wireless networks  require a connection density of  $10^7$ devices/$\mathrm{km}^2$, which is 1000 times higher than that of 4G and 10 times higher than that of 5G networks \cite{viswanathan2020communications,wang20236g,vaezi2022cellular}. The requirements in terms of  improving reliability, spectral efficiency, and energy efficiency are also stringent. In this context, the roles of multiple access in general and modulation and coding in particular are crucial toward achieving these goals.


{
\Gls{noma}  \cite{saito2013non}} is perhaps the most prominent candidate for \gls{ngma}. \gls{noma} increases the number of connected devices and  enhances the spectral efficiency of communication by enabling multiple users to share time, frequency, and space, thus accommodating a larger number of users compared to conventional \gls{oma} schemes.
\gls{noma} facilitates massive connectivity by allowing the concurrent service of multiple users within the same resource block, such as a time slot, sub-carrier, or spreading
code. It  has actively been considered by academia \cite{saito2013non,vaezi2018book,dai2015non,liu2017non,ding2017survey,vaezi2019interplay}, standardization  bodies,  and industry \cite{NOMA3GPP,yuan20205g,saito2013system,benjebbour2015noma}. 
 An intriguing aspect of \gls{noma} is its flexibility in integration with various technologies, including \gls{ofdm} which is the multiple access method in 4G and 5G. That means a \gls{noma} user has the capability to share a single \textit{resource block} of \gls{ofdm} with one or more additional users.

Despite its significant promise and immense academic work in this field, \gls{noma} has not been incorporated into any standards yet.  Several factors contribute to this.
A primary reason is that the theoretical gains of \gls{noma} were not achieved  in practical implementations \cite{qi2021over}. 
{There are primary challenges that make it difficult to achieve \gls{noma}'s theoretical gains in practice. They include} the difficulty of canceling  inter-user interference introduced by \gls{noma} (such as the complexity of successive interference cancellation), sensitivity of \gls{noma} to \gls{csi}, and non-synchronous nature of multi-user communication. 
Another reason for this shortfall is the absence of novel modulation  schemes addressing inter-user interference introduced by \gls{noma}. Overall, there has been limited research and innovation on finite-alphabet \gls{noma}.
  A third contributing factor is the rise of competitive solutions, such as massive  \gls{mimo}, \gls{mmwave}, and narrow-band \gls{iot}, which have effectively tackled some of the requirements for massive connectivity and spectrum efficiency through innovative solutions 
  {and the fact that many theoretical \gls{noma} results are limited to single antenna cases}.

Nonetheless, inheriting the rich theoretical background of the \gls{bc} \cite{cover1972broadcast}, \gls{noma} still holds great promise as a future multi-user transmission technique, referred to as \gls{ngma}. Besides, emerging methods like \gls{rsma} have presented themselves as new potential candidates for \gls{ngma}. \gls{rsma} utilizes \gls{sic} to decode a portion of the interference and treats the remaining interference as noise. Thus, \gls{rsma} is in between \gls{sdma}, which treats interference as noise, and \gls{noma}, which decodes the interference 
{of the users with weaker channels and removes it from the received signal}.

With a specific focus on \gls{noma} and \gls{rsma}, this article delves into current and future modulation and coding schemes for \gls{ngma}.  Modulation and coding techniques play a critical role in achieving the ultimate goal of digital communication, which is transmitting a maximum number of bits reliably, i.e., with a small \gls{ber}. Moreover, the emphasis here is on finite-alphabet and asynchronous \gls{noma}, representing a crucial step in evaluating \gls{noma}'s gains in more practical settings and advancing research toward the integration of \gls{noma} into wireless standards.

\subsection{Motivation and Objectives}

Modulation techniques, like \gls{qam}, are employed to increase the bit rate (spectral efficiency)  while resulting in an acceptable \gls{ber}. 
Current modulation techniques are, however, designed several decades ago with point-to-point communication in mind \cite{foschini1974optimization,forney1984efficient,goldsmith1997variable,barsoum2007constellation}. They have predefined, inflexible symbols and their constellation shaping is oblivious to interference, whereas  modern communication systems are limited by interference more than by any other
single effect \cite{andrews2005interference}. Interference appears in these networks in different forms such as \gls{iui},  \gls{ici}, and \gls{isi}. 
These all distort the received constellations in one way or another and thus reduce the reliability of communications by increasing the \gls{ber}. The typical practical solution is then to sacrifice the spectral efficiency and limit the number of users by allocating orthogonalized resources to each user or by using low-rate and high-energy  constellations.

The described interference scenarios share a common issue: they can displace a constellation symbol from its predefined decoding region (boundary), leading to decoding errors and, consequently, symbol and  bit errors. This challenge arises because current modulation techniques, such as \gls{qam}, were originally designed for point-to-point systems without interference. These modulation techniques feature predefined constellation symbols and their shape is insensitive to interference. The rigidity of these constellations poses a significant hurdle to improving the \gls{ber} and spectral efficiency in today's interference-limited communication systems. Therefore, there is a need for innovative interference-aware modulation techniques to meet key performance indicators in future wireless communication networks, including spectral efficiency, the number of supported devices, and high reliability.


A key goal of this paper is to present \gls{ngma} modulation and coding methods that reduce decoding \glspl{ber} and increase the number of users and their throughput for a given number of resource blocks in the presence of inter-user and inter-cell interference. Minimizing  \gls{ber} and increasing spectral efficiency are linked together \cite[Fig. 1]{forney1998modulation} and are traditionally optimized by designing modulation and coding schemes. We present a comprehensive survey of modulation schemes, both with and without coding, for both \gls{oma} and \gls{noma} scenarios. In the context of \gls{noma}, the survey critically evaluates the implications of superimposed signals on symbol overlapping and \gls{ber}. Then, we propose using deep autoencoders for \gls{e2e} \gls{noma}. Deep learning-based \gls{e2e} communication is a novel concept with significant potential.
As a concrete example, this approach outperforms the \gls{mimo} precoder in terms of \gls{ber}  with/without the channel's knowledge \cite{o2017deep,song2020benchmarking,o2017introduction,zhang2021svd}.

In line with the fundamental objective of digital communication, which revolves around the reliable transmission of a maximal number of bits, our focus in this work is on exploring various avenues that pave the way for reliable  and feasible non-orthogonal transmission. Instead of integrating \gls{noma} with emerging and existing communication technologies, our work centers around understanding the essential steps required to transform non-orthogonal transmission, specifically \gls{noma}, from a theoretical research topic into a practical and feasible technology. 
Toward this goal, our discussion encompasses practical considerations, including asynchronous transmission, \gls{sic}-free decoding, trellis-coded \gls{noma}, interference-aware constellation design, end-to-end communications, and numerous other aspects detailed in the following section.

\subsection{Contributions and Insights}

This paper  comprehensively explores existing and emerging
solutions related to synchronous and asynchronous \gls{noma}, code-domain \gls{noma}, trellis-coded \gls{noma}, uniform, non-uniform, and interference-aware constellation design, bit-interleaved coded modulation,  \gls{sic}-free decoding,  and end-to-end deep learning-based \gls{noma}. Additionally, the paper provides a state-of-the-art overview of \gls{rsma} and discusses open problems and future directions in \gls{ngma}. 
Particularly, we contribute to 
\begin{itemize}
	\item Exploring the utilization of asynchronous transmission to tackle synchronization challenges inherent in multi-user and distributed systems, such as those found in uplink and downlink \gls{noma}. 
 \item Reviewing the structure of transmitters and receivers in code-domain \gls{noma} and categorizing various code-domain \gls{noma} schemes accordingly.
	\item Investigating the effectiveness of non-uniform modulation, trellis-coded \gls{noma}, and  bit-interleaved coded modulation  with iterative decoding to approach {
 the capacity region of downlink \gls{noma}. \footnote{
 Unless otherwise stated, downlink \gls{noma} refers to the Gaussian broadcast channel.}}
 \item 	Debunking misconceptions surrounding power allocation in \gls{noma}, which may have arisen from assumptions tied to a specific type of modulation and decoding.
  \item {
  Reviewing modulation and coding schemes utilized in 5G and discussing the road to reduce latency in 6G.}

 \item Introducing the innovative concept of interference-aware constellation design and end-to-end \gls{noma} and demonstrating its  merit.
 \item Conducting a review of uplink and downlink  \gls{rsma} and exploring its interconnections with \gls{noma}.
 \item Highlighting research directions and  open problems related to the above topics, with a specific emphasis on advancing \gls{noma} as a feasible and practical technique.

\end{itemize}

In light of the above contributions, we gain diverse insights that propel \gls{noma} from a theoretical concept towards a practical technology. We hope that these insights, as listed below, may trigger advancements to pave the way for \gls{noma}'s inclusion in wireless standards in the near future.

\begin{enumerate}
\item 
Contrary to conventional wisdom, which states that asynchronous transmission results in increased overall interference and performance degradation due to extra \gls{isi}, we show that asynchronous transmission  decreases the overall interference. This unexpected outcome occurs because the reduction in \gls{iui} outweighs the impact of the added \gls{isi}.

\item We demonstrate that drawing general conclusions about \gls{noma} power allocation solely based on a specific type of modulation and decoding  may lead to misconceptions.

\item Interference-aware constellation design is a practical approach to realize non-overlapping super-constellations in \gls{noma}. Attaining this objective becomes challenging, if not impossible, when utilizing established constellations like \gls{qam} for \gls{noma} users.

\item For \gls{mimo}-\gls{noma} systems, the joint design of constellations and precoding for all \gls{mimo} sub-channels holds great potential. 

\end{enumerate}

\begin{figure}
	\scalebox{.82}{
	\begin{forest}
		for tree={
			align=left,
			edge = {draw, semithick, -stealth},
			anchor = west,
			font = \small\sffamily\linespread{.84}\selectfont,
			forked edge,
			grow = east,
			s sep = 0mm,
			l sep = 8mm,
			fork sep = 4mm,
			tier/.option=level
		}
		[Paper's \\Structure
		[{\color{blue}Section VII:}  Open Problems and Future Directions]
		[{\color{blue}Section VI: \\ RSMA}  
			[Uplink RSMA] 
			[Downlink RSMA]
		]
		[{\color{blue}Section V: \\ ML-Based \\ Modulation \\ Design}  
			[Interference-Aware Constellation Design] 
			[Autoencoder-Based E2E Communication] 
			[SIC-free NOMA]
		]
		[{\color{blue}Section IV: \\ Channel \\ Coding}  
			[Trellis-Coded Modulation \& TC-NOMA]
			[Channel Coding in 5G \& 6G]	
			[Channel Coding Evolution]
		]
		[{\color{blue}Section III: \\ Traditional \\ Modulation \\ Design}  
			[Non-Uniform Modulation Schemes] 
			[Uniform Modulation Schemes]	
			[Modulation Design Evolution]
		]
		[{\color{blue}Section II: \\ P-NOMA \& \\ C-NOMA}  
			[Integration with OFDM]
                [C-NOMA]	
			[Asynchronous NOMA] 		
			[P-NOMA]
		]
		[{\color{blue}Section I: \\ Introduction}  
			[Structure] 
			[Contributions and Insights]
			[Introduction]
		]
		]
	\end{forest}
 }
	\caption{Structure of the paper. } \label{fig:str}
\end{figure}
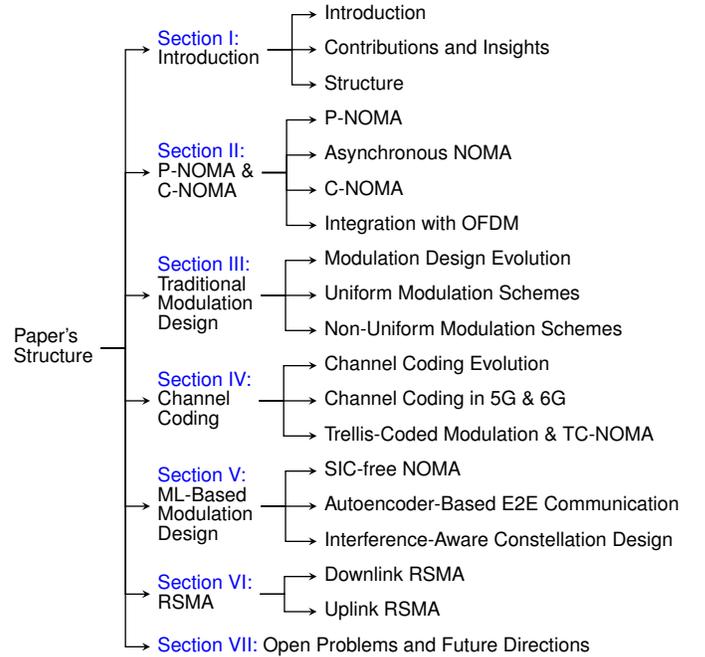

\subsection{Structure}

Figure~\ref{fig:str} illustrates the organization of the paper, providing an overview of the topics covered in each section.

In Section~\ref{sec:NOMA}, we discuss different techniques under the umbrella of \gls{noma}. These could be primarily categorized into \gls{pnoma} and \gls{cnoma} \cite{vaezi2018book}. \gls{cnoma} includes  schemes like \gls{scma} \cite{nikopour2013sparse}, \gls{pdma} \cite{chen2016pattern},  \gls{lds} \cite{mohammed2012performance}, and others. \gls{pnoma} is explored in both uplink and downlink, along with asynchronous \gls{noma}. 
{
The integration of both \gls{pnoma} and \gls{cnoma} into  \gls{ofdm}  systems is also discussed.} 
  
In Section~\ref{sec:modulation}, we explore the evolution of modulation techniques for both \gls{oma} and \gls{noma} scenarios, to achieve a desirable \gls{ber} versus \gls{snr} independent of channel coding or in conjunction with it.  This section emphasizes that simply employing existing modulation schemes for  \gls{noma}  users without modification can cause symbol overlapping and \gls{ber} issues. We then discuss non-uniform constellations and their advantages, particularly with bit-interleaved coded modulation. 
  
{
Section~\ref{sec:coding}  discusses the channel codes utilized in different generations of cellular networks, with a particular emphasis on  modulation and coding schemes in 5G networks. We explore the evolution of channel coding techniques to meet the growing demands of \gls{urllc}. Additionally, we cover trellis-coded modulation  for both \gls{oma} and \gls{noma} scenarios, emphasizing the role of joint detection methods in enhancing performance.}

  Section~\ref{sec:ML} introduces interference-aware constellation design, incorporating \gls{sic}-free \gls{noma} and autoencoder-based \gls{e2e} communication. Autoencoders are utilized to create super-constellations with distinguishable symbols, transitioning from block-by-block to \gls{e2e} communication. The section includes \gls{iui}-aware \gls{noma}, \gls{ici}-resilient \gls{noma}, and modulation strategies for \gls{mimo}-\gls{noma}.
  
  In Section~\ref{sec:RSMA}, we review \gls{rsma}, discussing its origin and advantages in both downlink and uplink scenarios. Uplink \gls{rsma} achieves the capacity region of the Gaussian \gls{mac} without time sharing, while downlink \gls{rsma} is advantageous with imperfect \gls{csi} at the transmitter. We briefly explore the use of \gls{rsma} in connection with different technologies, including integrated communication and sensing, reconfigurable intelligent surfaces, and \glspl{uav}.

  Section~\ref{sec:open} highlights research areas for \glsentrytext{ngma}, addressing non-uniform modulation in \gls{noma} with bit-interleaved coded modulation. Challenges in \gls{e2e} \gls{noma} and \gls{mimo}-\gls{noma} modulation are discussed. Limited feedback and \gls{csi} present open problems, including 
  robust \gls{mimo}-\gls{noma} system design with estimated and quantized \gls{csi}.


\begin{table}[! htbp]
\caption{List of Key Abbreviations.}
	\begin{threeparttable}
	
	\begin{tabular}{|l| l |} 
		\hline
		\textbf{Acronym} &  \textbf{Description}  \\ [0.5ex] 
		\hline \hline
		3GPP  & 3rd Generation Partnership Project \\ \hline 
		AE  & Autoencoder\\  \hline
  		AI  & Artificial intelligence\\  \hline
		A-NOMA & Asynchronous NOMA  \\  \hline
		   AWGN & Additive white Gaussian noise \\  \hline
      BC & Broadcast channel  \\  \hline
       BER &  Bit error rate \\  \hline
         BICM & Bit-interleaved coded modulation \\ \hline 
		CDMA & Code division multiple access\\  \hline
		C-NOMA & Code-domain NOMA  \\  \hline
		CSI & Channel state information  \\  \hline
		CSIT & Channel state information at the transmitter \\  \hline
		E2E & End-to-end \\  \hline
        ICI & Inter-cell interference\\  \hline
		IDMA & Interleave  division multiple access \\  \hline
		IGMA & Interleave-grid multiple access \\  \hline
		IoT &  Internet of things\\  \hline
            ISAC & Integrated sensing and communications \\ \hline
		ISI & Inter-symbol  interference\\  \hline
		IUI & Inter-user interference\\  \hline
		LDS & Low-density signature/spreading \\  \hline
            LPMA & Lattice partition multiple access \\ \hline
            LTE & Long Term Evolution \\ \hline
    MAC & Multiple access channel  \\  \hline
    MCS & Modulation and
coding scheme  \\  \hline
  MIMO &Multiple-input multiple-output \\  \hline
		ML &  Machine learning\\  \hline
		\glsentrytext{ngma} & Next generation multiple access\\  \hline
  NOMA & Non-orthogonal multiple access\\  \hline
            NR & New Radio \\ \hline
		OFDM & Orthogonal frequency division multiplexing \\  \hline
            OMA & Orthogonal multiple access\\  \hline
PDMA & Pattern division multiple access \\  \hline
		P-NOMA & Power-domain NOMA \\  \hline
		PSK & Phase-shift keying \\  \hline
  QAM & Quadrature amplitude modulation  \\  \hline
		QPSK & Quadrature phase-shift keying \\  \hline
		RB & Resource block  \\  \hline
            RIS & Reconfigurable intelligent surface \\ \hline
		RSMA &  Rate-splitting multiple access\\  \hline
		SCMA & Sparse code  multiple access  \\  \hline
		SDMA & Space division multiple access   \\ \hline 
         SER &  Symbol error rate \\  \hline
		SIC & Successive interference cancellation \\ \hline 
		SISO & Single-input single-output \\ \hline
  SNR &Signal-to-noise ratio \\ \hline
		SVD & Singular value decomposition\\  \hline
		TCM & Trellis-coded modulation \\ \hline
		TC-NOMA & Trellis-coded NOMA \\ \hline
            UAV & Uncrewed aerial vehicle \\ \hline
            URLLC & Ultra-reliable low-latency communications \\ \hline
	\end{tabular}

	\end{threeparttable}
	\label{tableABB}
\end{table}

{
\subsection{Related Works}\label{sec:related}
Numerous survey papers have appeared on \gls{noma} and \gls{rsma} in recent years. For example, see \cite{liu2017non, ding2017survey, vaezi2019interplay} for \gls{noma} and \cite{mao2022rate} for \gls{rsma}. However, many of these survey papers are not directly related to this paper as their focus is not on modulation, finite-alphabet \gls{noma}, or \gls{rsma}. The most related works are \cite{cai2017modulation, yahya2023error, shahab2020grant}.
}

		\section{\gls{pnoma} \& \gls{cnoma} }\label{sec:NOMA}

4G cellular networks have been architected around orthogonal radio resource allocation techniques not allowing for overlapping resource allocation. For instance,  a \textit{\gls{rb}} in \gls{lte}, which spans 180kHz, cannot be shared among multiple users; it must be exclusively assigned to one user. This resource allocation approach has two limitations in the context of massive \gls{iot}:
\begin{itemize}
	\item With the rapid growth of massive \gls{iot} devices, there would not be enough spectrum to allocate a dedicated \gls{rb} to each device.\footnote{{
 Interestingly, the concept of allocating more than one user to one \gls{rb} has already been successfully implemented in 2G under the name of \gls{vamos} which is used to increase the capacity of voice services by allowing multiple users to share the same time slot. In the simplest case, it is known as the \gls{aqpsk} modulation scheme. It enables scheduling two users on in-phase (I) and quadrature-phase (Q) channels, thus doubling the number of users served by a single radio resource \cite{vamos, vamosEricsson}. Additionally, allocation of different power levels for each user is possible \cite{vamos, vamosEricsson}. 

With the above definition of \gls{noma}, which involves using one \gls{rb} for multiple users' signals, \gls{vamos} is a \gls{noma} scheme. There is, however, a subtle difference. The constellation symbols of users try to be orthogonal (one is mapped to the I channel and the other to the Q channel). In other words, half of the constellation points are for the first user and the other half are for the second user. However, in the \gls{noma} modulation schemes considered in \cite{3gppTR36859} and this paper, by receiving every symbol, we can reconstruct information for both users. 
Therefore, the two schemes are different in terms of transmission rates. 
}
}
	\item Massive \gls{iot} users typically do not exhaust an entire \gls{rb}, rendering such resource allocation inefficient.
\end{itemize}

Due to the above challenges,  the communication system design has recently undergone a transformation, shifting from an orthogonal resource allocation  to a non-orthogonal one \cite{vaezi2018book}.  
This paradigm shift encompasses various aspects including waveform design, multiple access, and random access \cite{vaezi2018book}. 
Particularly, \gls{noma} has attracted significant attention as a promising multiple access technique.

It should be highlighted that the proposed \gls{noma} techniques for the uplink and downlink are distinct, rooted in the inherent differences between communication requirements for each direction. The downlink predominantly serves human-centric communications, characterized by larger packets and higher data rates. Conversely, the uplink involves an extensive array of uncoordinated devices transmitting small packets at low data rates. Consequently, addressing the demands of massive, low-rate \gls{iot} devices in the uplink mandates a unique set of techniques compared to \gls{pnoma} which is good for the downlink channels. 

 It is worth noting that, \gls{pnoma} has been studied  by the \gls{3gpp} for \gls{lte} in TR 36.859,  Release 13 \cite{3gppTR36859}. It has also been studied for \gls{nr} in TR 38.812, Release 16 \cite{3gppTR38812}. Also, there are several \gls{pnoma}-related survey papers worth mentioning, including \cite{shahab2020grant,elbayoumi2020noma,maraqa2020survey,hussain2020machine}. These works cover various topics including grant-free \gls{noma}, resource management mechanisms, and others.

	\subsection{\gls{pnoma}}\label{sec:PNOAM}

To increase the efficiency of wireless networks, the concurrent transmission of different users is unavoidable. Such a concurrent transmission in a wireless multi-user network results in interference among different signals. Therefore, one of the most distinctive features of multi-user wireless networks is the interference phenomenon. As a result, dealing with interference is very important when shifting from the single-user paradigm to a multi-user paradigm. 
To deal with interference, currently, communication standards rely on assigning separate time, frequency, and code resources to different users such that each user utilizes only one such a given resource in an orthogonal multiple access framework. This is a huge waste of resources and to curtail it cellular networks utilize frequency reuse in cells that are far from each other. Nevertheless, in each cell, and at each time, frequency, and code, only one user is served using a resource that is orthogonal to other resources. Therefore, the main source of interference will be the leak from users of adjacent resources due to the imperfection of the filters or the inter-cell interference due to the frequency reuse. 
However, \gls{noma} relies on assigning more than one user to each resource, for example by employing superposition coding at the transmitter and \gls{sic} at the receiver. {
In fact, superposition coding is not new and has been used in achieving the capacity of the degraded \gls{bc} \cite{cover1972broadcast} as well as the single-antenna Gaussian \gls{bc} which is always degraded.} In addition, combined with an appropriate transmitter, \gls{sic} at the receiver is capable of approaching the boundaries of the capacity region of both the {
degraded \gls{bc}} and \gls{mac}\cite{cover2012elements}.

	\begin{figure}
		\centering 
  \includegraphics[width=2.2in, page=1, trim=10cm 6cm 16cm 4.5cm, clip=true]{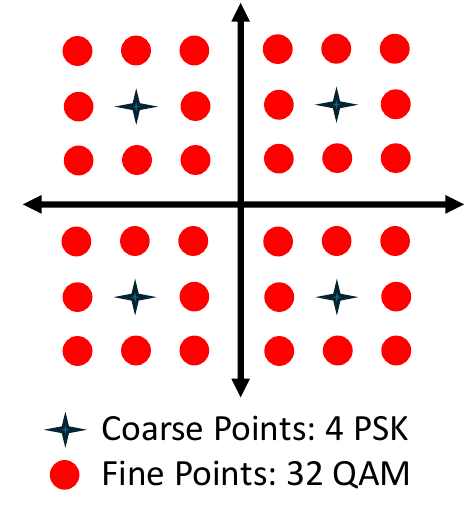}
		\caption{An example of superposition coding. 
  {The constellation for the user with
weaker channel is 4-PSK while the constellation for the user with stronger channel is 8-QAM. The overall superposed constellation is called 32-QAM.}}
		\label{constel}
	\end{figure}

To study the main principles behind \gls{pnoma}, let us take a downlink \gls{pnoma} system as an example. It consists of one \gls{bs} and $K$ users. Let us assume the transmit power of User~$k$'s signal is $P_k$ and $\sum_{k=1}^{K}P_k = P$, where $P$ is the total transmit power of the \gls{bs}. The transmitted signal $s$ is defined as
\begin{equation}
    s=\sum_{k=1}^K \sqrt{P_k} s_k, \label{eq:s}
\end{equation}
where $s_k$ is the transmitted symbol for User~$k$. 
Let us denote the channel coefficient between the \gls{bs} and User~$k$ by $h_k$. 
Then, the received signal at User~$k$ is given by
	\begin{align}
		y_k = h_k\sum_{l=1}^K \sqrt{P_l} s_l + \eta_k,\ \ \label{eq:yk}
	\end{align}
where $\eta_k$ is the additive Gaussian noise, $\eta_k\sim \mathcal{CN}(0, \sigma_k^2)$. For simplicity of the notation, we assume the same noise power for all users, i.e., $\sigma_k^2=\sigma^2$.
Obviously, different users generate interference for each other and their decoders should manage the interference. \gls{pnoma} manages the interference by decoding the signal of users 
{with weaker channels} and canceling it from the received signal. As such, in \gls{pnoma}, it is important to know the relative strength of the users' channels to understand which signals can be decoded for interference cancellation. The main principle behind \gls{pnoma} decoding is that stronger channels, i.e., channels with larger channel magnitudes $|h_k|$, will have larger capacities. Therefore, they can support higher rates. As a result, they can successfully decode the symbols of the weaker users with lower rates. 

Without loss of generality, let us assume that $|h_1|^2 \ge |h_2|^2 \ge \cdots \ge |h_K|^2$, i.e., a lower index represents a stronger user. For any $k < K$, the capacity of User~$k$ is higher than that of User~$l$, for $l = k+1,\cdots,K$. Therefore, using \gls{sic}, User~$k$ can decode the signal for Users $k+1,\cdots,K$ and remove their interference before decoding its own signal. Under the assumption of perfect \gls{sic}, the throughput of User~$k$ for a case with $K$ users can be calculated by
	\begin{align}
		R_k = \log\left(1 + \frac{P_k|h_k|^2}{|h_k|^2\sum_{l=1}^{k-1}P_l  + \sigma^2}\right),\ \ k=1,\cdots,K. \label{eq:Ri}
	\end{align}
It is clear from (\ref{eq:yk}) and (\ref{eq:Ri}) that the throughput values depend heavily on the power allocation and the channels. The co-channel interference makes the resource allocation problem in \gls{pnoma} a non-convex optimization problem. 
The power allocation for a two-user \gls{pnoma} system is studied in \cite{yang2016general}. Power allocation for multiple users sharing one channel, i.e., multi-user \gls{noma}, is investigated in \cite{wang2016power, zhu2017optimal,doan2019power}. 
The solution to the power minimization problem can be obtained using the uplink-downlink duality. 
In general, \textit{resource management} is an important aspect of the transmitter design in different \gls{noma} scenarios. 
In addition, user grouping is essential for  balancing spectral efficiency, fairness, and system throughput. While having a single group is ideal in theory \cite{vaezi2018book}, practical limitations like the complexity of SIC necessitate multiple groups \cite{ali2016dynamic}. As group size increases, the SIC complexity rises. Typically, users are grouped by channel quality, pairing those with strong conditions with those having weaker conditions to enhance performance.

While the above analysis is based on capacity formulas, i.e., error-free decoding, similar principles can be applied to a practical modulation scheme that can cause errors. In what follows, we present an example of superposition coding and \gls{sic} decoding for traditional modulation schemes.

\noindent {\bf Example:}  
An example of superposition coding using 4-\gls{psk} and 8-\gls{qam} constellations for two users is shown in Fig.~\ref{constel}. In this example, the input bits of the user with weaker channel, i.e., smaller $|h_k|$, are modulated with a 4-\gls{psk} modulation as shown by the coarse points in Fig.~\ref{constel}. The input bits for the user with stronger channel are modulated with an 8-\gls{qam} modulation. As described in \eqref{eq:s}, the transmitted symbol is the weighted sum of the two  modulated symbols. In this example, we consider equal power allocation that results in a 32-\gls{qam} constellation shown by the fine points in Fig.~\ref{constel}.
	
	\begin{figure}
		\centering
		\subfloat[Decoding at the weaker user and the first step at the stronger user.]{%
			\includegraphics[width=1.7in, page=1, trim=11.5cm 7.3cm 14.7cm 5cm, clip=true]{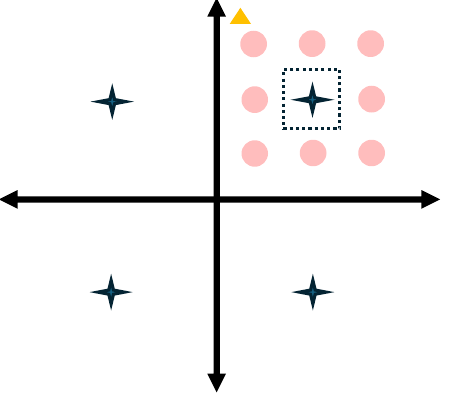}\label{decode1}}
		\hfil
		\subfloat[Second step of decoding at the stronger user.]{%
			\includegraphics[width=1.5in, page=2, trim=12.5cm 8cm 16cm 6cm, clip=true]{figures/Fig_NOMA_Constellation.pdf}\label{decode2}}
		\caption{An example of \gls{sic} decoding.}
		\label{decode}
	\end{figure}

Figure~\ref{decode} shows the \gls{sic} technique for decoding the superposed signal of the two users in Fig.~\ref{constel} at the receiver. As shown in Fig.~\ref{decode1}, the receiver of the weaker user only decodes the coarse points by mapping the received signal to the nearest point in the corresponding \gls{psk} constellation. Practically, this means that the weaker user is considering the symbol of the stronger user as noise. The stronger user is also able to decode the coarse points since its channel is stronger than that of the weak user. Therefore, the stronger user decodes the symbol of the weaker user, using the same approach, and subtracts the effects of the decoded coarse symbol from the received signal. This will cancel the weaker user's interference from the stronger user's received signal. The remaining signal is decoded using the corresponding constellation (8-\gls{qam}) as shown in Fig.~\ref{decode2}.
    {The modulation constellations in this example are chosen to make the example visually more appealing. However, the example works for any choice of constellations. A more practical example may choose a 4-\gls{qam} constellation for the stronger user, instead of 8-\gls{qam}, that results in an overall 16-\gls{qam} modulation.}
\subsection{Asynchronous \gls{noma} }\label{sec:ANOMA}
 
{
Traditional modulation and coding schemes assume perfect symbol synchronization. The performance of many existing modulation and coding schemes will severely degrade without symbol-level synchronization.}	
Because of the distributed nature of multi-user networks and the effects of multipath and propagation delays, signals from different users experience different time delays to the receiver \cite{MGHJ16}. Perfect synchronization requires feedback and coordination, which complicates the system greatly. Having multiple antennas at the receiver makes the matter more complicated. For example, assuming a multiple antenna receiver in uplink \gls{noma}, one can synchronize the received signal perfectly at one of the receive antennas but other antennas may experience imperfect timing synchronization among received signals. Even if such a complete synchronization is possible, it is not clear if it is desirable. The possible advantages along with the difficulties in achieving perfect synchronization motivate a thorough analysis and design of \gls{noma} systems under imperfect timing synchronization. In fact, symbol-level asynchrony has been advantageous in managing interference in some communication systems as discussed in \cite{SPHJ15,MAHJ15,MGHJ16}. Similar principles can be applied to \gls{noma} to design \gls{anoma} \cite{cui2017asynchronous,XZBHHJ19,MGHJwcnc19,MGXZHJ21}. In this section, we discuss the principles behind asynchronous \gls{noma} for both uplink and downlink.

	\begin{figure}
		\centering
		\subfloat[Symbol Synchronous]{%
			\includegraphics[width=2.5in, page=1, trim=8.5cm 1.5cm 9cm 1cm, clip=true]{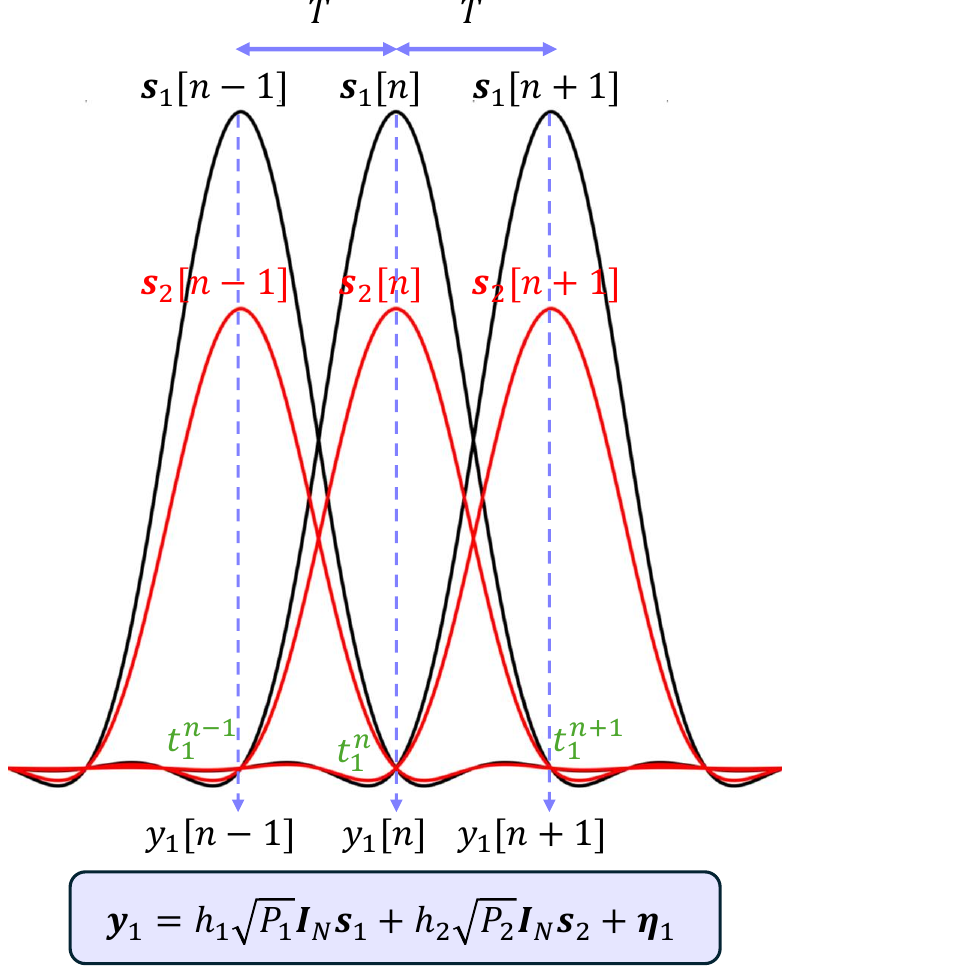}}
		\hfil
		\subfloat[Symbol Asynchronous]{%
			\includegraphics[width=2.7in, page=1, trim=10cm 0cm 8cm 0cm, clip=true]{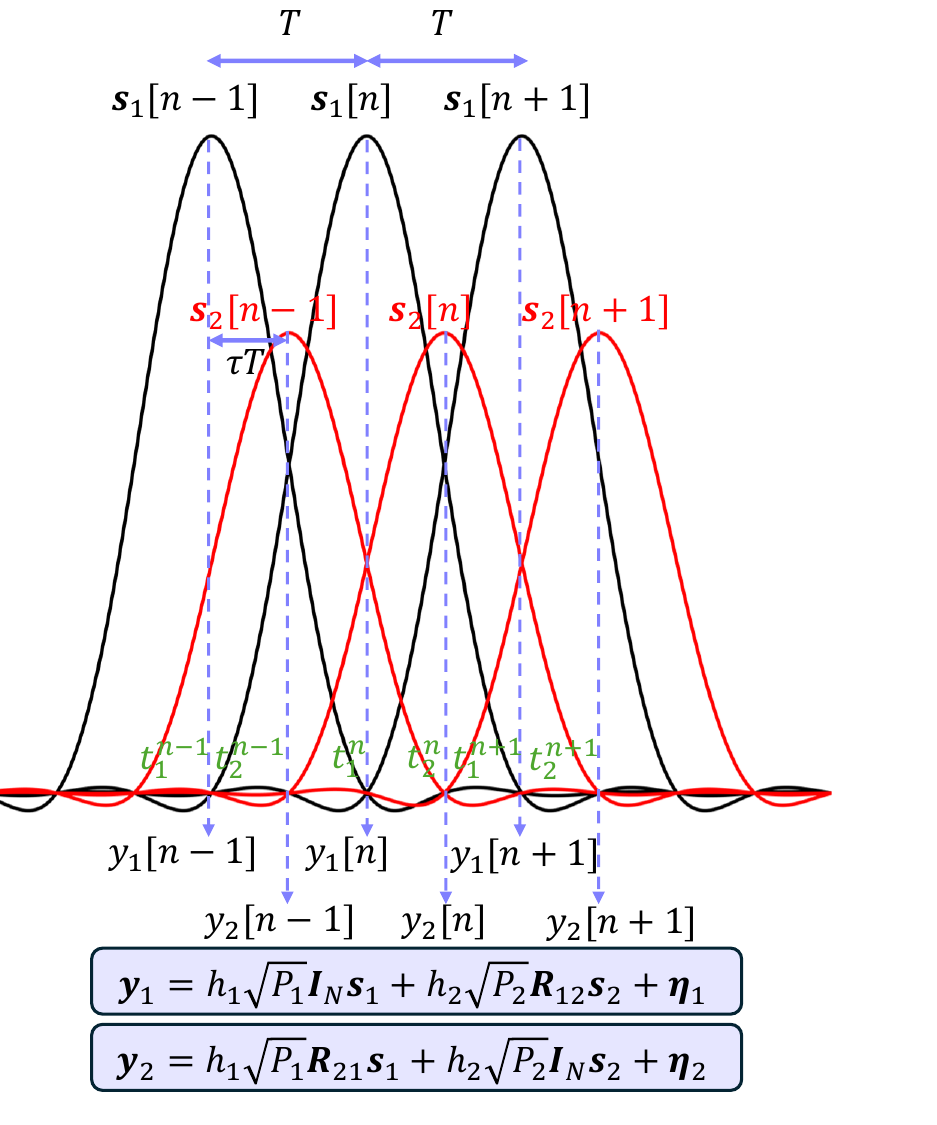}}
		\caption{Sufficient statistics for synchronous and asynchronous \gls{noma}. 
  {In asynchronous NOMA, the output of the matched filter is sampled twice, each time synched with one of the users.}}
		\label{s_vs_a}
	\end{figure}
 
\noindent {\bf Uplink \gls{anoma}:} 
Let us assume User $k$ transmits  $\sqrt{P_k}\boldsymbol{s}_k[n]$, where $\boldsymbol{s}_k[n]$ is the $n$th element of vector $\boldsymbol{s}_k$ denoting the $n$th normalized transmitted symbol and $P_k$ denotes the transmit power. 
{Assuming the pulse-shaping filter $p(.)$ and a frame length of $N$, the transmitted signal from User $k$ will be
$\sum_{n=1}^N{\sqrt{P_k}} \boldsymbol{s}_k[n]  p(t-nT)$. 
Considering the relative time delay of $\tau_k$ for User $k$, i.e., a time delay of $\tau_k T$, and the corresponding channel $h_k$, the signal from User~$k$ arrives at the receiver as $\sum_{n=1}^N{h_k \sqrt{P_k}} \boldsymbol{s}_k[n]  p(t-nT-\tau_k T)$.
}
Then, the received signal at the \gls{bs} is given by
	\begin{flalign}
		y(t)=\sum_{n=1}^N\sum_{k=1}^K{h_k \sqrt{P_k} \boldsymbol{s}_k[n]  p(t-nT-\tau_k T)+\eta(t)},
	\end{flalign}
where $\eta(t)$ denotes	the \gls{awgn} with a power spectral density of $\sigma^2$.
Unlike the case with perfect synchronization, in addition to \gls{iui}, the received signal includes \gls{isi} as well. The set of sufficient statistics is found by proper filtering at the receiver, i.e., a matched filter with the impulse response $p(t)$, and over-sampling $K$ times, each time synched with one of the users \cite{MGHJ16}. In other words, the receiver samples at $t_n=nT+\tau_k T$ to generate $K$ sets of samples, $k=1,\cdots,K$, at each discrete-time $n=1,\cdots,N$. 
\begin{figure*}
    \centering
    \includegraphics[width=.9\textwidth, trim=1cm 6cm 2cm 5cm, clip=true] {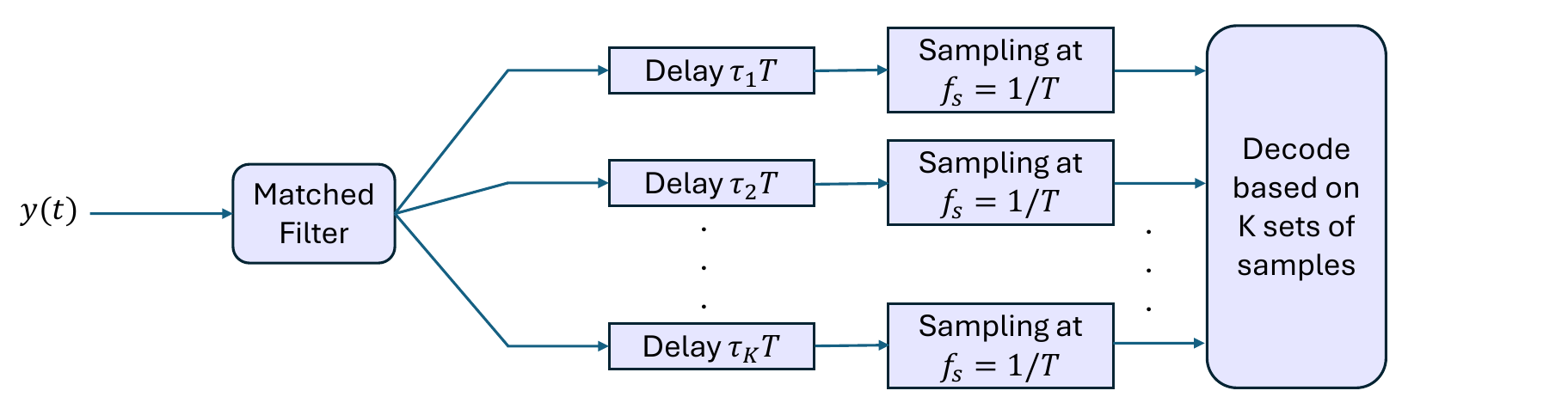}
    \caption{
    {Decoder structure for uplink A-NOMA. After matched filtering, the signal is sampled $K$ times, each time {
    synchronized}  with one of the {
    $K$} users, to generate sufficient statistics. Then, the resulting samples are used for decoding.}}
    \label{fig:ANOMA_sampling}
\end{figure*}
{Figure~\ref{fig:ANOMA_sampling} illustrates one possible structure of the decoder including oversampling. As shown in the figure, the sampling rate is still $f_s=1/T$, i.e., the same as that of the synchronous case, but there are $K$ parallel branches of sampling.}

When sampled synched with User $k$'s signal, the relative delay of User $l$'s signal is $\tau_{kl}=\tau_l-\tau_k$. The resulting samples can be collected in an $N\times 1$ vector $\boldsymbol{y}_k$ as
\begin{equation}
    \boldsymbol{y}_k=\sum_{l=1}^{K} h_l \sqrt{P_l} \boldsymbol{R}_{kl}\boldsymbol{s}_l+\boldsymbol{\eta}_k, \label{eq:ykVector}
\end{equation}
where the ($n,m$)th element of the $N\times N$ Toeplitz matrix $\boldsymbol{R}_{kl}$ is
\begin{equation}
    \left[\boldsymbol{R}_{kl}\right]_{n,m}=g(\tau_{kl} T+(m-n)T), 
    \ m,n=1,\cdots,N,
\end{equation}
in which $g(t)=p(t)*p(t)$, where $*$ denotes the convolution, and the noise vector $\boldsymbol{\eta}_k$ has the co-variance matrix  $\mathbb{E}[\boldsymbol{\eta}_k{\boldsymbol{\eta}_l}^H]=\sigma^2\boldsymbol{R}_{kl}$.
Using a square-root Nyquist pulse, $p(t)$, like the practically common root raised cosine {
(RRC)} pulse shape, $\boldsymbol{R}_{kk}=\boldsymbol{I}_N$ and  $\boldsymbol{R}^T_{kl}=\boldsymbol{R}_{lk}$.
An example of the received signals and the corresponding sufficient statistics for two users is shown in Fig.~\ref{s_vs_a}. To collect sufficient statistics in \gls{anoma}, the receiver should perform oversampling, compared with synchronous \gls{noma}. For the example of two users in Fig.~\ref{s_vs_a}, the number of samples in \gls{anoma} is twice that of the synchronous \gls{noma}. 
Note that for the case with perfect synchronization, when all delays $\tau_k$ are the same, we have $\boldsymbol{R}_{kl}=\boldsymbol{I}_N$, i.e., there is no \gls{isi}. Therefore, 
the perfect synchronization results in the conventional uplink \gls{pnoma} system model of
\begin{equation}
    \boldsymbol{y}=\sum_{l=1}^{K} h_l \sqrt{P_l} \boldsymbol{s}_l+\boldsymbol{\eta}, \label{eq:yVector}
\end{equation}
where compared to \eqref{eq:ykVector}, the index $k$ is dropped as all resulting $K$ equations are identical. 

As shown in \eqref{eq:yVector}, in a perfectly synchronized \gls{noma} system, at each time instant, only the \gls{iui}, from the same time-instant, degrades the performance. Therefore,  \gls{sic} can provide the optimal performance. However, in \gls{anoma}, not only \gls{iui} but also \gls{isi} degrades the performance. Therefore, new challenges arise with asynchronous transmission. 

The conventional wisdom suggests that because of the additional \gls{isi}, asynchronous transmission increases the overall interference and the overall performance is degraded. However, surprisingly, asynchronous transmission in fact decreases the overall interference, as the reduction in \gls{iui} outweighs the addition of \gls{isi} \cite{MGHJwcnc19}.
On the other hand, because of the \gls{isi}, the conventional  \gls{sic} is not optimal anymore and the design of efficient sequence detection methods is required \cite{MGXZHJ21}.  More specifically, to remove the stronger users' signals, all the symbols in a frame need to be decoded which adds delay and complexity to the system. 
In addition, because of the timing asynchrony, sufficient statistics results in over-sampling and the corresponding sampling diversity \cite{MGHJ16} that improves the overall performance.

To manage the above issues, 
one can collect the $K$ vector equations   \eqref{eq:ykVector} in a matrix format. 
The resulting input-output matrix equation can be represented as a virtual \gls{mimo} system
	\begin{flalign}
		\label{system_model}
		\boldsymbol{y}=\boldsymbol{R}\boldsymbol{H}\boldsymbol{s}+\boldsymbol{\eta},
	\end{flalign}
where $\boldsymbol{y}, \boldsymbol{R}, \boldsymbol{H}, \boldsymbol{s}$ and $\boldsymbol{\eta}$ represent the set of samples, the timing offsets matrix, the effective channel matrix, the transmitted symbols (including the assigned power), and the noise vector, respectively. The formulation of the problem with a virtual \gls{mimo} system enables the use of various interference mitigation methods and decoder designs developed in the literature for \gls{mimo} systems \cite{STC2005}.

	\begin{figure}
		\includegraphics[width=3.5in, page=1, trim=8cm 1.5cm 8cm 1.7cm, clip=true]{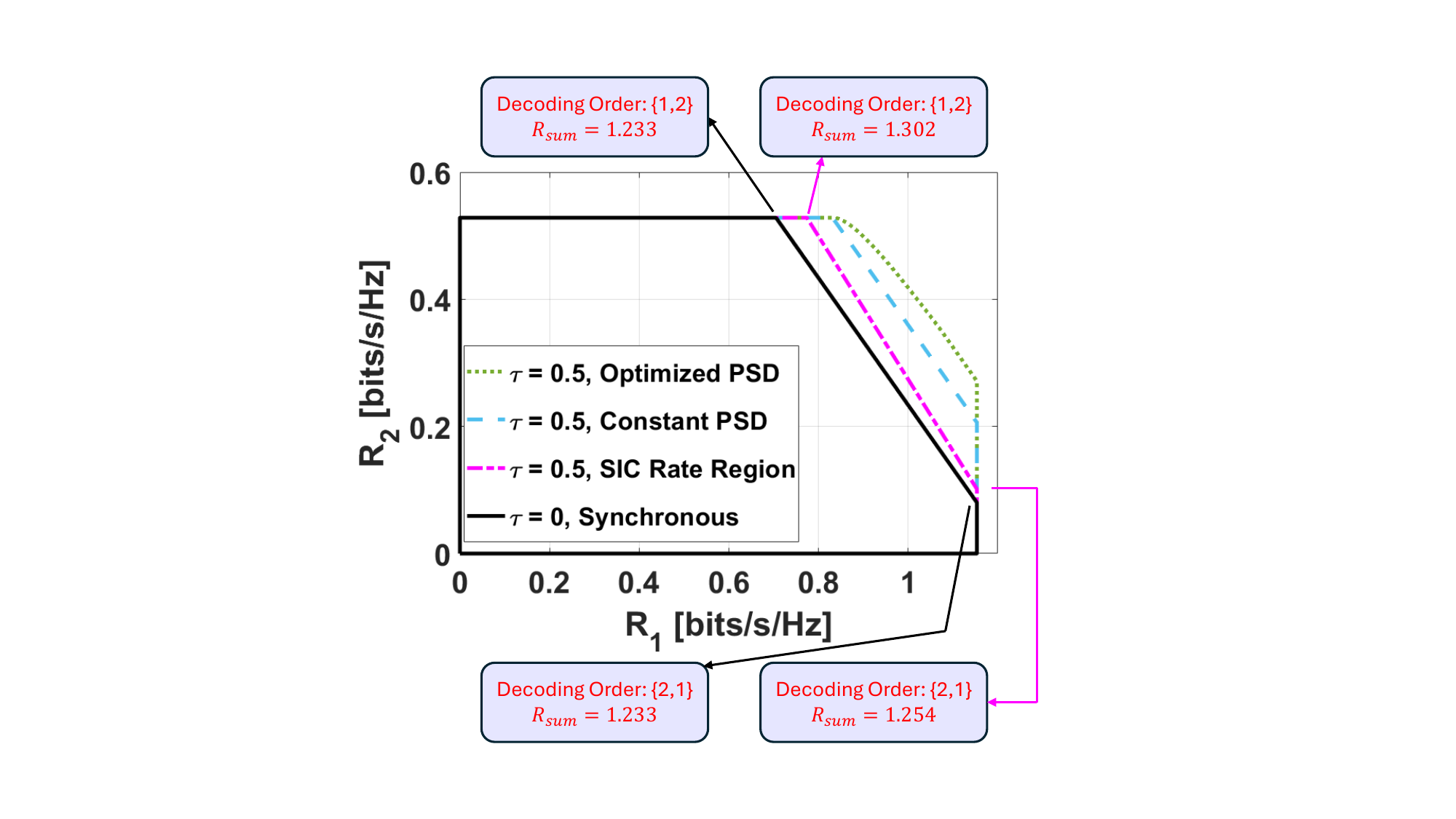}
		\caption{Different rate-regions for $P_1=P_2=10~{\rm dBm}, |h_1|^2=\sigma^2, $ {
  and} $ |h_2|^2=0.2 \sigma^2 
   $, using root raised cosine pulse shaping with $\beta=0.5$. 
   {Different decoding orders in SIC result in different sum rates in asynchronous NOMA.}}
		\label{rr-sic-uplink}
	\end{figure}
To quantify the effects of asynchronous transmission, let us focus on a two-user uplink \gls{pnoma} system. 
	Figure~\ref{s_vs_a} shows the received signals of such a system. 
	It can be modeled as an asynchronous \gls{mac} with a typical rate-region shown in Fig.~\ref{rr-sic-uplink}. As proved in \cite{MGXZHJ21} and depicted in Fig.~\ref{rr-sic-uplink}, not only does \gls{anoma} outperform synchronous \gls{noma}, but also \gls{sic} is not optimal for \gls{anoma}. There are several conclusions proved in \cite{MGXZHJ21} and shown in Fig.~\ref{rr-sic-uplink}:
 \begin{itemize}
     \item Intentionally creating a $\tau=0.5$ symbol timing mismatch between the two signals 
     {with double time oversampling} can enlarge the rate-region. 
     \item While the maximum sum-rate using \gls{sic} for uplink \gls{noma} is independent of the decoding order, the maximum sum-rate of \gls{anoma} depends on the decoding order. For example, in Fig.~\ref{rr-sic-uplink}, the \gls{sic} pentagon rate-region vertices corresponding to the decoding orders $\{1,2\}$ and $\{2,1\}$ for \gls{anoma} provide sum-rates of 1.302 and 1.254, respectively. On the other hand, the corresponding vertices for \gls{noma} provide the same sum-rate of 1.233. 
     \item When perfect synchronization is lacking, \gls{sic} is not optimal anymore. 
 \end{itemize}
	
	\begin{figure}
		\includegraphics[width=8.5in]{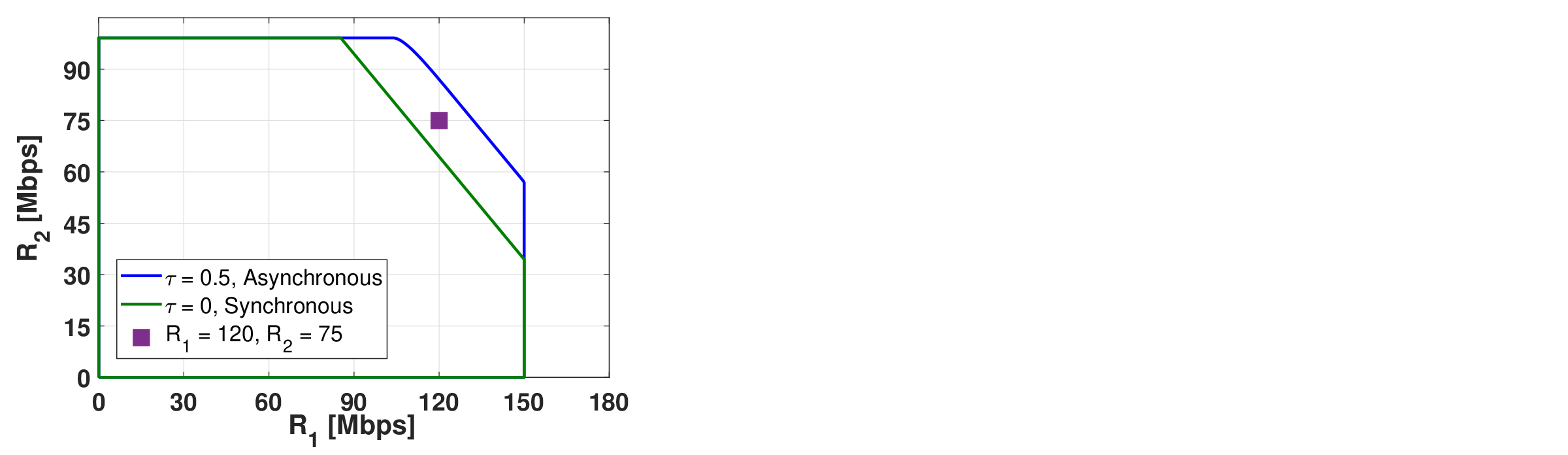}
		\caption{{
  An example of} asynchronous \gls{noma} operational point that is not achievable by synchronous \gls{noma}.}
		\label{fig:ANOMA}
	\end{figure}
	The above observations can guide the principles behind designing practical transceivers for asynchronous multiple access systems. For example, a practical \gls{anoma} transceiver that creates an intentional $\tau=0.5$ symbol asynchrony between the two users' signals has been designed in \cite{MGXZHJ21}. The operational point achieved by such an \gls{anoma} transceiver, shown in Fig.~\ref{fig:ANOMA}, is outside of the synchronous \gls{noma}'s capacity region and therefore is not achievable by any synchronous \gls{noma} transceiver. 
	
\noindent {\bf Downlink \gls{anoma}:} 
In a downlink \gls{noma} system, the superposition is performed at the transmitter and the transmitted signal is received by the intended users. Therefore, unlike uplink \gls{noma}, the transmitter has full control of the time asynchrony and can intentionally add any desired set of time delays. In fact, adding intentional time delays at the transmitter is beneficial and can improve the performance \cite{MGHJwcnc19}. Using the same notation developed for the uplink \gls{anoma}, the transmitted signal, including the added intentional time delays, can be written as 
 \begin{equation}
     s(t)=\sum_{n=1}^N\sum_{k=1}^K{\sqrt{P_{k}}s_k[n]p(t-nT-\tau_k T)},
 \end{equation} 
and the received signal by User $k$ is 
	\begin{flalign}
		y_k(t)=h_k  s(t)+\eta_k(t).
	\end{flalign}
Similar to the case of uplink \gls{anoma}, the set of sufficient statistics includes over-sampling $K$ times, each time synched with the signal of one of the users. After proper match filtering and over-sampling, the set of samples at the $k$th receiver can be represented in the following vector \cite{MGXZHJ20} 
\begin{equation}
    \boldsymbol{y}_k= h_k \sum_{l=1}^{K}  \sqrt{P_l} \boldsymbol{R}_{kl}\boldsymbol{s}_l+\boldsymbol{\eta}_k. \label{eq:ykDown}
\end{equation}
Note that the perfect synchronization results in the conventional system model of $y_k[n]=h_k\sum_{l=1}^K \sqrt{P_l}s_l[n]+\eta_k[n]$.
Using \gls{sic}, each user decodes all the signals from weaker users and removes them from the received signal. Then, it considers the remaining interference from the stronger users as noise and decodes its own symbols. This system is called \gls{apnoma} in Fig.~\ref{fig:TNOMA} and Fig.~\ref{fig:TNOMAres}.  

\tikzstyle{myblock} = [rectangle, draw, fill=blue!5, 
    text width=6em, text centered, rounded corners, minimum height=3em]
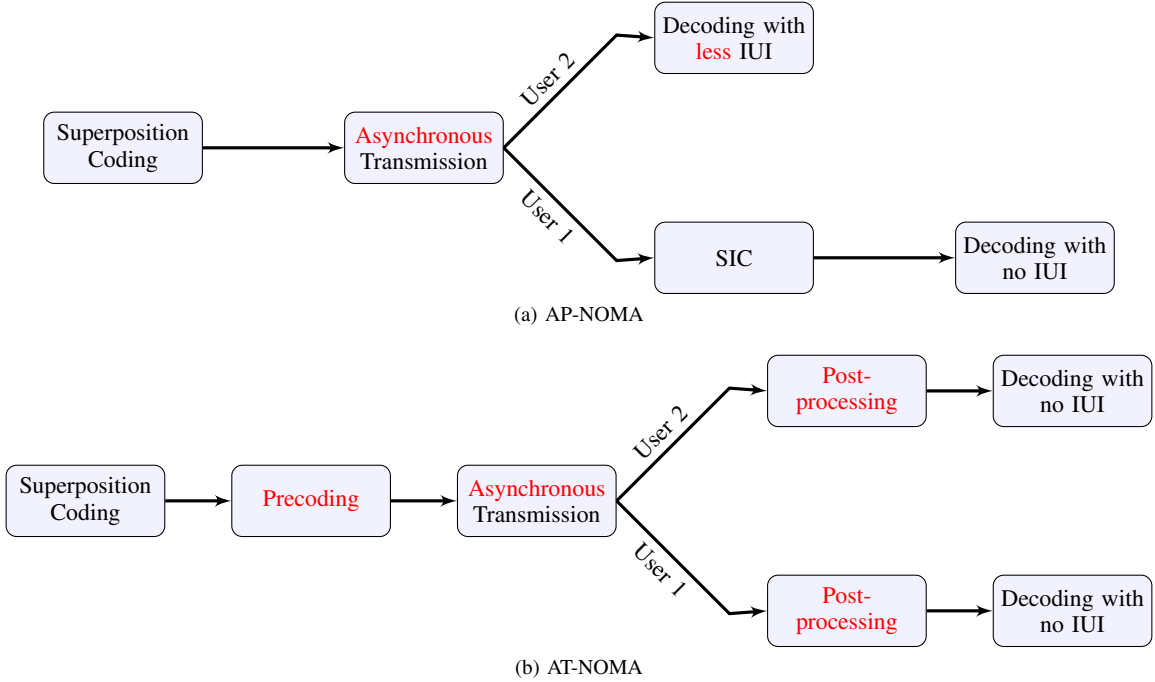
\begin{figure*}[t]
    \centering \small
    \subfloat[AP-NOMA]{%
    \begin{tikzpicture}[node distance=2cm, auto]
        \node [myblock] (st1) {Superposition Coding};
        \node [myblock, right of=st1, node distance=4cm] (st2) {\textcolor{red}{Asynchronous} \\ {Transmission}};
        \node [myblock, above right=0.5cm and 2cm of st2] (st3) {Decoding with \\ {\color{red}less} IUI};
        \node [myblock, below right=0.5cm and 2cm of st2] (st4) {SIC};
        \node [myblock, right of=st4, node distance=4cm] (st5) {Decoding with \\ no IUI};

        \path [line] (st1) -- (st2);
        \path [line] (st2.east) -- ++(1.5cm, 1.5cm) node[midway, above, sloped] {User 2} -- (st3.west);
        \path [line] (st2.east) -- ++(1.5cm, -1.5cm) node[midway, below, sloped] {User 1} -- (st4.west);
        \path [line] (st4) -- (st5);
    \end{tikzpicture}}

    \subfloat[AT-NOMA]{%
    \begin{tikzpicture}[node distance=2cm, auto]
        \node [myblock] (st1) {Superposition Coding};
        \node [myblock, right of=st1, node distance=3cm] (st2) { {\color{red}Precoding}};
        \node [myblock, right of=st2, node distance=3cm] (st3) {\textcolor{red}{Asynchronous} \\ {Transmission}};
        \node [myblock, above right=0.5cm and 2cm of st3] (st4) {{\color{red}Post-\\processing}};
        \node [myblock, below right=0.5cm and 2cm of st3] (st5) {{\color{red}Post-\\processing}};
        \node [myblock, right of=st4, node distance=3cm] (st6) {Decoding with \\ no IUI};
        \node [myblock, right of=st5, node distance=3cm] (st7) {Decoding with \\ no IUI};

        \path [line] (st1) -- (st2);
        \path [line] (st2) -- (st3);
        \path [line] (st3.east) -- ++(1.5cm, 1.5cm) node[midway, above, sloped] {User 2} -- (st4.west);
        \path [line] (st3.east) -- ++(1.5cm, -1.5cm) node[midway, below, sloped] {User 1} -- (st5.west);
        \path [line] (st4) -- (st6);
        \path [line] (st5) -- (st7);
    \end{tikzpicture}}
    \caption{ {
    {Downlink asynchronous NOMA schemes.  (a)  AP-NOMA: {
    The} user with stronger channel decodes the signal of the other user and removes its interference. (b)  AT-NOMA: {
    The} transmitter sends the signal toward the eigenvectors of the effective virtual MIMO system. {
    The} receiver uses post-processing, in the direction of the corresponding eigenvector, to recover its signal.}}}
    \label{fig:TNOMA}
\end{figure*}
In downlink \gls{noma}, 
{the virtual \gls{mimo} system in \eqref{eq:ykDown} resembles a multiuser system. As such, apart from \gls{sic}, other techniques for multiuser communications can be applied as well. For example,} 
since the transmitter has access to all signals, after superposition coding, it can perform precoding in the direction of matrix $\boldsymbol{R}$'s eigenvectors to further improve the performance. The decoder of such a system 
{is} named \gls{tnoma}
{. At User $k$, \gls{tnoma} uses post-processing, in the direction of the corresponding eigenvector, to recover its signal. Figure~\ref{fig:TNOMA} shows the block diagram of such a system that }does not 
{include a} \gls{sic} 
{block for decoding} \cite{MGHJwcnc19}. The 
corresponding rate regions are shown for two users in 
Fig.~\ref{fig:TNOMAres}. 
As shown in Fig.~\ref{fig:TNOMAres}, \gls{tnoma}'s rate-region is larger than that of \gls{apnoma} and both of them enlarge the achievable rate-region of \gls{pnoma}.

\begin{figure}
	\includegraphics[width=3.5in]{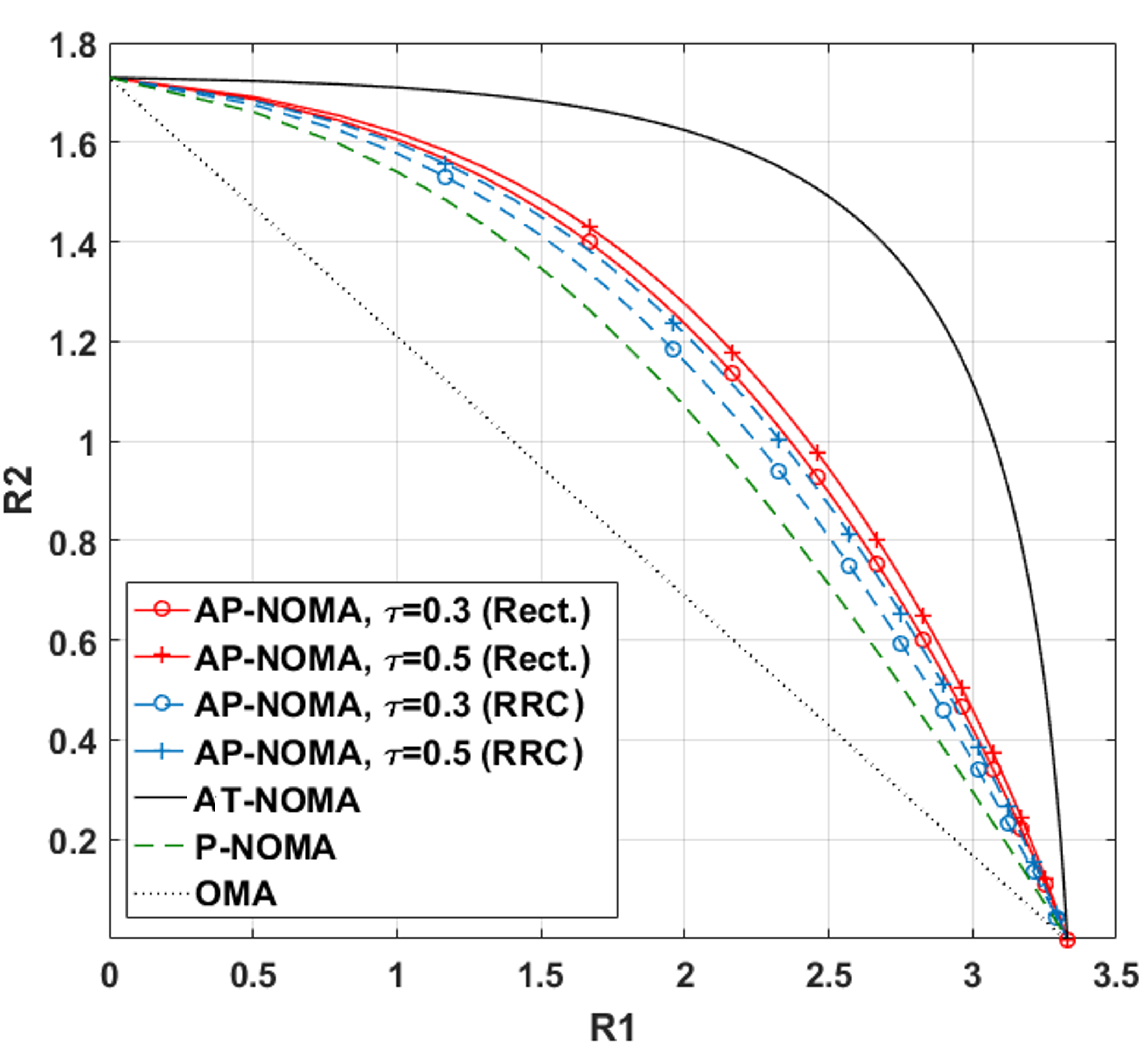}
	\caption{Rate regions of different \gls{noma} systems for $|h_1|^2= 10 \sigma^2$ and $|h_2|^2=\sigma^2$.}
	\label{fig:TNOMAres}
\end{figure}

In addition, for systems with $M$ transmit antennas, transmit beamforming can be implemented as well. Considering an $M\times 1$ beamforming vector $\mathbf{W}_k$ for User $k$, the asynchronous transmitted signal, including intentional time delays, is \cite{MGXZHJ20}
\begin{equation}
     s(t)=\sum_{k=1}^K \boldsymbol{W}_{k} \sum_{n=1}^N s_k[n]p(t-nT-\tau_k T).
\end{equation} 
Then, the received signal at User $k$ is
\begin{equation}
	y_k(t)=\boldsymbol{h}_k^H  \sum_{l=1}^K \boldsymbol{W}_{l} \sum_{n=1}^N s_l[n]p(t-nT-\tau_l T)+\eta_k(t),
\end{equation}
where $\boldsymbol{h}_k$ is User $k$'s channel. As before, the set of sufficient statistics after match filtering and over-sampling can be collected in the following vector  \cite{MGXZHJ20} 
\begin{equation}
    \boldsymbol{y}_k= \sum_{l=1}^{K}    \boldsymbol{R}_{kl} \boldsymbol{h}_k^H \boldsymbol{W}_{l} \boldsymbol{s}_l+\boldsymbol{\eta}_k. 
\end{equation}

The above input-output relationship is very similar to that of a \gls{mimo} system and the corresponding receiver designs from the \gls{mimo} literature can be used for decoding 
{\cite{STC2005,MGXZHJ20}}.

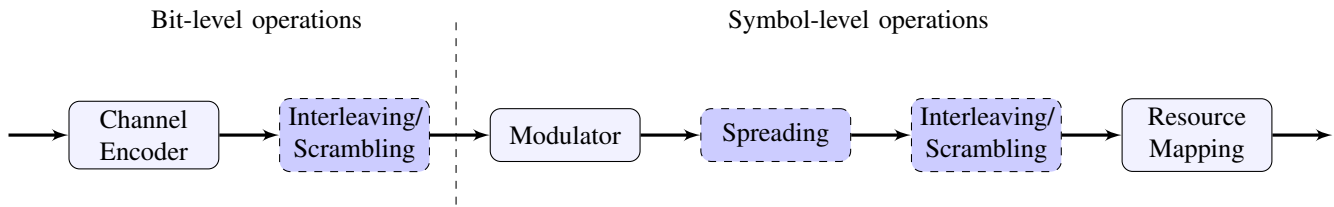
\begin{figure*}[t]
	\begin{tikzpicture}[node distance = 2cm, auto]
	\node [block] (st1) {Channel Encoder};
	\node [block2, right of=st1, node distance=2.8cm] (st2) {Interleaving/ Scrambling};
	\node [block, right of=st2, node distance=2.8cm] (st3) {Modulator};
	\node [block2, right of=st3, node distance=2.8cm] (st4) {Spreading};
	\node [block2, right of=st4, node distance=2.8cm] (st5) {Interleaving/ Scrambling};
	\node [block,  right of=st5,  node distance=2.8cm] (st6) {Resource Mapping};
	\path [line] (-1.8,0) -- (st1);
	\path [line] (st1) -- (st2);
	\path [line] (st2) -- (st3); 
	\path [line] (st3) -- (st4);
	\path [line] (st4) -- (st5);
	\path [line] (st5) -- (st6);
	\path [line] (st6) -- (15.8,0);
	\draw[dashed]  (4.15,1.5) -- (4.15,-1);
	\node[] at (1.5,1.5) {Bit-level operations};
	\node[] at (9.5,1.5) {Symbol-level operations};
	\end{tikzpicture}
	\caption{{
 The general structure of a \gls{cnoma} transmitter. It includes essential components represented by solid blocks, which are necessary for any \gls{cnoma} scheme. In addition, individual \gls{cnoma} schemes may include one or more of the dashed blocks, depending on the specific requirements and characteristics of the scheme. Therefore, the presence of the dashed boxes depends on the type of \gls{cnoma} scheme being implemented.}}
	\label{fig:CNOMAencoder} 
\end{figure*}

\subsection{\gls{cnoma}}\label{sec:C-NOMA}

In the uplink, \gls{cnoma} refers to a diverse range of non-orthogonal transmission techniques \cite{vaezi2018book,NOMA3GPPNR,YuanIndustry}. \gls{cnoma} techniques revolve around  the concept of allowing more than one user to share the same resource block, marking a distinct departure from \gls{oma} techniques. On the other hand, \gls{cnoma} distinguishes itself from \gls{pnoma} by requiring a signature, such as a spreading code, for the differentiation of users and the cancellation of inter-user interference. In contrast, \gls{pnoma} achieves this through the variation in power allocation to each user. Given the predominant emphasis on uplink communication in \gls{iot}, \gls{cnoma} emerges as a promising way to accommodate \gls{iot} users, especially those with limited resources such as simple sensors.

 The {
 range} of \gls{cnoma} techniques is extensive. A major category within \gls{cnoma} schemes  is notably influenced by \gls{cdma}.
  \gls{cdma} is a multiple access technique where data symbols are spread over a set of user-specific, mutually-orthogonal codes. An adaptation of \gls{cdma} is \gls{lds}-\gls{cdma}, where spreading codes exhibit low density, meaning only a small fraction of code elements are non-zero \cite{hoshyar2008novel}. 
 Resource \textit{overloading}  is the common theme of many non-orthogonal access methods which allows  the  number of supportable users  to be more than the number of available resources, i.e.,
$K$ users share $N$ resources in a non-orthogonal fashion ($K>N$). 
 With this, each scheme needs a multiple access \textit{signature}  to differentiate users and cancel inter-user interference.

We next explore the   transmitter and receiver  structures and  classification of \gls{cnoma} techniques.

\subsubsection{Transmitter Structure} 

A simplified, generic structure of a \gls{cnoma} transmitter is depicted in Fig.~\ref{fig:CNOMAencoder}  \cite{chen2018toward,YuanIndustry,NOMA3GPPNR}. Different \gls{noma} schemes apply their signatures in one or more of the dashed blocks in Fig.~\ref{fig:CNOMAencoder}.   
For example, the \gls{lds}-\gls{cdma} signature is a short spreading sequence applied in the `spreading' block followed by a unique interleaver for each user, resulting in a sparse signature matrix. Thus, \gls{lds}-\gls{cdma} is a symbol-based \gls{cnoma} applying both spreading and interleaving after modulation. Sparse \gls{re}  mapping has emerged as a category of multiple access signatures within several \gls{cnoma} schemes, such as \gls{scma}, \gls{pdma}, and \gls{igma}. In these schemes, the intentional transmission of zeros occurs in specific \glspl{re} to make resource mapping sparse. An illustration of such a sparse resource mapping is provided in the matrix below:
$$\small \kbordermatrix{
 & {\rm UE1} & {\rm UE2} & {\rm UE3} & {\rm UE4}  & {\rm UE5} & {\rm UE6} & {\rm UE7}  & {\rm UE8} & {\rm UE9} \\
{\rm RE1} & 0&    1&    0&    0&    1&    0&    1&    0 &   0 \\
{\rm RE2} & 1 &   0 &   0 &   0 &   0 &   1 &   0 &   1 &   0 \\
{\rm RE3} & 0 &   1&    0&    1&    0&    0&    0&    0&    1 \\
{\rm RE4} & 0&    0&    1&    0&    1&    0&    1&    0 &   0 \\
{\rm RE5} & 1&    0&    0&    1&    0&    0&    0&    0 &   1 \\
{\rm RE6} & 0&    0&    1&    0&    0&    1&    0&    1 &   0 \\
}.$$

According to the above resource allocation matrix, a total of 9 \glspl{ue} are mapped to 6 resource elements. Each \gls{ue} is assigned access to two \glspl{re}, and each \gls{igma} is accessed by three different UEs. For instance, RE1 accommodates UE2, UE5, and UE7 while UE4 utilizes RE3 and RE5. Further, it is noteworthy that  \gls{ue}-specific signatures exhibit, at most, a one-position overlap, because a pair of columns do not overlap in more than one position.

\subsubsection{Receiver Structure} 

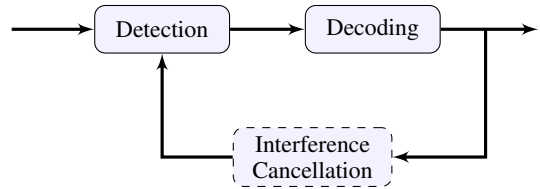
\begin{figure}
	\centering \small
	\begin{tikzpicture}[ node distance=1cm,>=latex']
	block/.style    = {rectangle, draw, fill=blue!5,  text centered, rounded corners, minimum height=2em},	
	\node [block] (st1) {Detection};	
	\node [block, right of=st1, node distance=2.8cm] (st2) {Decoding};	
	\draw (2,0) node[] (empty){} ;
	\node [block, text width=6em, dashed, below  of=empty, node distance=1.7cm] (st3) {Interference Cancellation};
	\path [line] (-2,0) -- (st1);	
	\path [line] (st1) -- (st2);
	\path [line] (st3) -- (-.0,-1.7) -- (st1);
	\path [line] (st2) -- (5,0);	
	\path [line] (4.3,0) -- (4.3,-1.7)  -- (st3);		
	\end{tikzpicture}
	 \caption{A high-level structure of a \gls{cnoma} receiver.}
		\label{fig:CNOMAdecoder} 
\end{figure}

The fundamental components of a high-level \gls{cnoma} receiver consist of three key building blocks: a detector, a channel decoder, and interference cancellation \cite{NOMA3GPPNR}, as depicted in Fig.~\ref{fig:CNOMAdecoder}.
In the following, we briefly elaborate on the functionality of each block.

The detection block, more precisely, the multi-user detection block, addresses challenges arising from multiple users sharing the same channel. {
It could employ various techniques \cite{NOMA3GPPNR}, including adaptive filtering and optimization algorithms like \gls{mmse} \cite{zaidel2018sparse}, matched filter, maximum a posteriori \cite{hu2018nonorthogonal}, and message passing algorithm \cite{bayesteh2015low}, to separate signals from different users. }

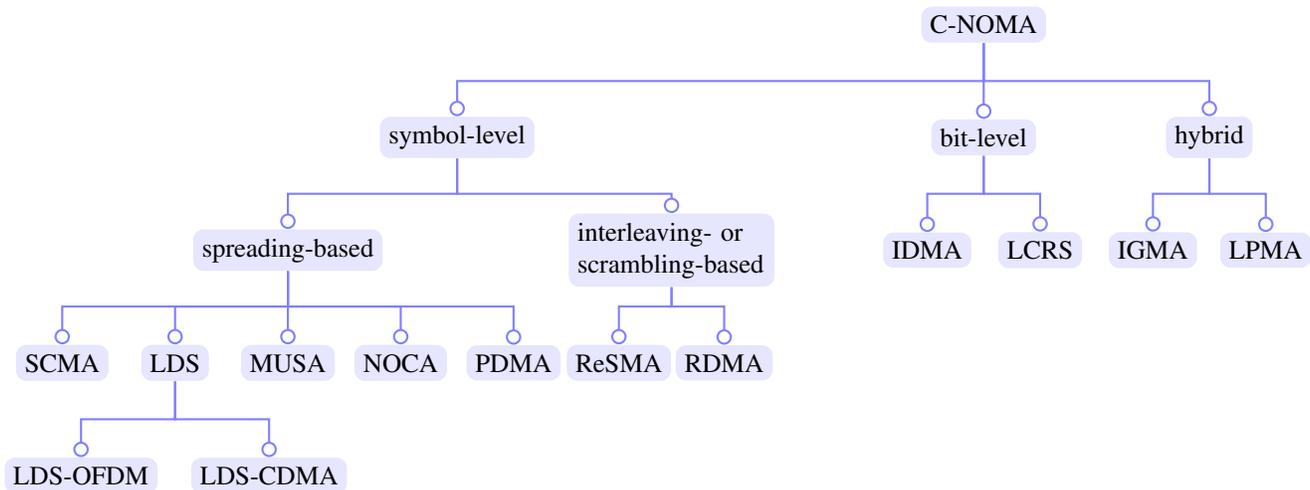
\begin{figure*}[th]
	\centering
	\begin{tikzpicture}[edge from parent fork down]
	\tikzstyle{every node}=[fill=blue!10,rounded corners]
	\tikzstyle{edge from parent}=[blue!50,-o,thick,draw]
	\tikzstyle{every sibling}=[fill=blue!90,rounded corners]
	\tikzstyle{level 1}=[sibling distance=70mm]
	\tikzstyle{level 2}=[sibling distance=30mm]
	\tikzstyle{level 3}=[sibling distance=14mm]
	\tikzstyle{level 4}=[sibling distance=10mm]
	\node {C-NOMA}
	child {node[sibling distance=40mm]{symbol-level}
		child [sibling distance=45mm] {node {spreading-based}
			child  [sibling distance=15mm] {node {SCMA}}
			child [sibling distance=15mm] {node {LDS}
				child [sibling distance=25mm] {node {LDS-OFDM}}
				child [sibling distance=25mm] {node {LDS-CDMA}}
			}		
			child [sibling distance=15mm] {node {MUSA}}		
			child [sibling distance=15mm] {node {NOCA}}	 	
			child [sibling distance=15mm] {node {PDMA}}	
		} 
		child [sibling distance=57mm] {node[text width=2.5cm] {interleaving- or scrambling-based}
			child {node {{
   ReSMA}}}
			child {node {RDMA}}
		} 
	}
	child [sibling distance=10mm] {node {bit-level}
		child [sibling distance=15mm] {node {IDMA}}
		child [sibling distance=15mm] {node {LCRS}}
	}
	child [sibling distance=30mm] {node {hybrid}
		child [sibling distance=15mm] {node {IGMA}}
		child [sibling distance=15mm] {node {LPMA}}
	};
	\end{tikzpicture}
	\caption{Various \gls{cnoma} schemes proposed in \gls{3gpp} 5G adio access network meetings. \gls{cnoma} techniques encompass operations such as interleaving, scrambling, and spreading, which can be executed at the symbol-level, bit-level, or a combination of both.}

	\label{fig:CNOMAtree}
\end{figure*}

While the detection block handles scenarios where multiple users transmit simultaneously, and is commonly used in multiple access systems, the \textit{decoding block} serves as a fundamental process in communication systems. Its applicability extends to both single-user and multi-user environments, encompassing error correction and the reconstruction of the original message or data. The decoding block ensures the accurate recovery of information from received signals.

Interference cancellation block may or may not exist. As an example, to decode a \gls{ue}'s data packet, only \gls{mmse} detection and channel decoding could be executed.
Nonetheless, interference cancellation is commonly used. This cancellation process can occur successively, in parallel, or through a hybrid approach. \gls{mmse}-\gls{sic} is a well-known scheme in this context  where interference cancellation is performed successively.
This technique is commonly utilized in symbol-level spreading schemes, where interference cancellation can take either a `hard' or `soft' form. In the hard interference cancellation approach, interference is subtracted once a user's signal is successfully decoded. On the other hand, in the soft interference cancellation method, the output of the decoder includes soft information. This soft information is then used to reconstruct symbols, involving the consideration and processing of probabilistic or continuous-valued information, rather than relying on discrete, hard decisions.

\subsubsection{Categories} \gls{cnoma} schemes can be classified in different ways. As can be seen in  
 Fig.~\ref{fig:CNOMAtree}, \gls{cnoma} related operations can be done in the \textit{bit-level} (before modulator), \textit{symbol-level} (after modulator), and \textit{hybrid}.   
For this reason, as also shown in Fig.~\ref{fig:CNOMAencoder},  \gls{cnoma} schemes may be categorized  as follows:

\begin{itemize}

\item \textit{Symbol-level spreading-based:} Most \gls{noma} schemes fall into this category where a spreading sequence serves as a signature to distinguish users. Earlier, we discussed examples of such signatures in the context of a resource allocation matrix. Notable instances of this category include \gls{scma} \cite{nikopour2014scma} proposed by Huawei, \gls{noca} \cite{NOCA} proposed by Nokia, \gls{musa} \cite{yuan2016multi} proposed by ZTE, \gls{pdma} \cite{PDMA} suggested by CATT, and \gls{lds}-\gls{cdma}, as well as \gls{lds}-\gls{ofdm} \cite{hoshyar2008novel,hoshyar2010lds}.

	\item \textit{Symbol-level interleaving/scrambling-based:} In this category, a symbol-level interleaver/scrambler is used to distinguish the users. 	
{
\Gls{resma} \cite{RSMA}} and \gls{rdma} \cite{RDMA} are examples of this category. 

	\item \textit{Bit-level interleaving/scrambling-based:} This category involves bit-level operations, specifically utilizing an interleaver or a scrambler at the bit-level to distinguish users \cite{ping2006interleave}.  A bit-level scrambler is advantageous in terms of lower processing delay and memory requirements compared to a bit-level interleaver \cite{3gppTR38812}. Examples of this category include \gls{idma} \cite{ping2006interleave} suggested by InterDigital and \gls{lcrs} \cite{LCRS} proposed by Intel. Release 15 \gls{nr} already supports a bit-level scrambler for randomization, which can be leveraged for \gls{cnoma} as well.
					
\item \textit{Hybrid:} 	Certain \gls{noma} schemes may combine multiple methods. For instance, \gls{igma} \cite{hu2018nonorthogonal}, proposed by Samsung, incorporates both bit-level interleaving and sparse mapping in its design. \Gls{lpma} is another example in this category \cite{fang2016lattice}.
\end{itemize}

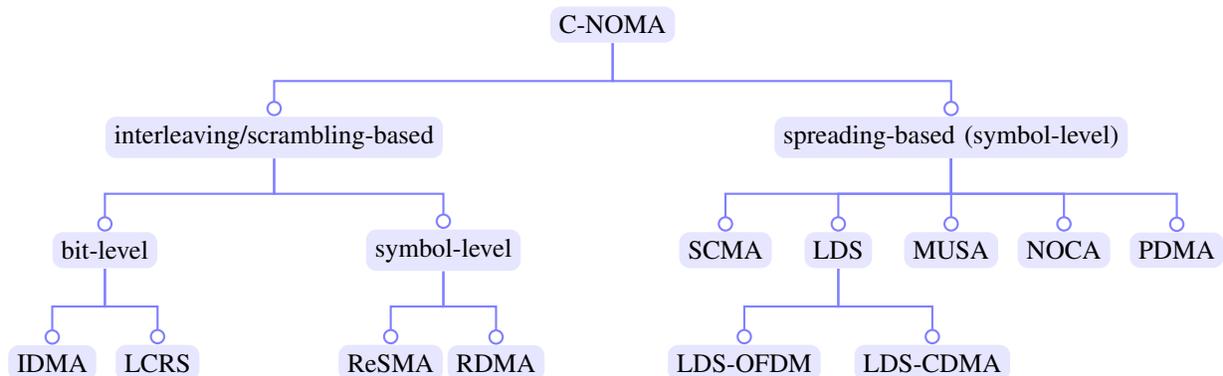
\begin{figure*}[!tbp]
\centering
\begin{tikzpicture}[edge from parent fork down]
\tikzstyle{every node}=[fill=blue!10,rounded corners]
\tikzstyle{edge from parent}=[blue!50,-o,thick,draw]
\tikzstyle{every sibling}=[fill=blue!90,rounded corners]
\tikzstyle{level 1}=[sibling distance=90mm]
\tikzstyle{level 2}=[sibling distance=30mm]
\tikzstyle{level 3}=[sibling distance=14mm]
\tikzstyle{level 4}=[sibling distance=10mm]
\node {C-NOMA}
child {node{interleaving/scrambling-based}
	child [sibling distance=45mm] {node {bit-level}
			child {node {IDMA}}
			child {node {LCRS}}
} 
	child [sibling distance=45mm] {node {symbol-level}
		child {node {{
  ReSMA}}}
		child {node {RDMA}}
} 
}
child {node {spreading-based (symbol-level)}
	child  [sibling distance=15mm] {node {SCMA}}
	child [sibling distance=15mm] {node {LDS}
	child [sibling distance=25mm] {node {LDS-OFDM}}
	child [sibling distance=25mm] {node {LDS-CDMA}}
	}		
	child [sibling distance=150mm] {node {MUSA}}		
	child [sibling distance=15mm] {node {NOCA}}	 	
	child [sibling distance=15mm] {node {PDMA}}	
};
\end{tikzpicture}
\caption{\gls{cnoma} schemes classified based on scrambling/interleaving and spreading operations.}
\label{fig:CNOMAtree2}
\end{figure*}

\gls{cnoma} schemes can be categorized differently, as illustrated in Fig.~\ref{fig:CNOMAtree2}. In this representation, the classification is based on the utilization of scrambling/interleaving and spreading operations. {
Thus, methods that use spreading are in one group, while methods that use interleaving--whether at the bit or symbol level--are in another group.}

\subsubsection{Prominent Schemes} Here, we explore several prominent \gls{cnoma} schemes in more details.
\begin{itemize}
	\item \texttt{\gls{lds}-\gls{cdma}:} This \gls{cnoma} scheme represents a non-orthogonal variant of \gls{cdma}. While in \gls{cdma}, data symbols are spread over user-specific, mutually-orthogonal codes, \gls{lds}-\gls{cdma} deviates from this norm by utilizing non-orthogonal codes. 	
	\gls{lds} underscores that spreading codes exhibit low density, i.e., only a small fraction of code elements are non-zero \cite{hoshyar2008novel}. This low-density feature allows for the utilization of near-optimal message passing algorithms with practical complexity. However, despite its moderate detection complexity, \gls{lds}-\gls{cdma} may encounter performance degradation for constellation sizes larger than \gls{qpsk}.

\item \texttt{\gls{scma}:} As one of the most renowned \gls{cnoma} schemes, \gls{scma} uses a multi-dimensional codebook where incoming data bits are directly mapped  to codewords selected from a layer-specific codebook \cite{taherzadeh2014scma}. Each codeword represents a spread transmission layer. In contrast to \gls{lds}-\gls{cdma} (and \gls{cdma}), where spreading and bit-to-symbol mapping are conducted separately, \gls{scma} integrates these two steps by directly mapping incoming bits to a spread codeword within the \gls{scma} codebook sets.

Similar to \gls{lds}, the sparsity of codewords in \gls{scma}  allows for the implementation of low-complexity reception techniques \cite{sun2019ldpc}. \gls{scma} also leverages additional degrees of freedom in the design of multi-dimensional constellations, outperforming \gls{lds} \cite{taherzadeh2014scma}. Importantly, the sparser \gls{scma} codewords can tolerate more overloading, facilitating massive \gls{iot} connectivity in the uplink. However, it is worth noting that a sparser code results in lower coding gain.

While \gls{scma} can be used for the downlink as well \cite{nikopour2014scma,tang2016low,Ma2019}, the decoding complexity remains high for low-cost devices, limiting \gls{scma}'s ability to efficiently support massive \gls{iot} connectivity in the downlink \cite{Ma2019}.
\end{itemize}

\subsubsection{\gls{sdma}} 
{
\gls{sdma} is a technique used to exploit the spatial dimension for improving system capacity and performance.} Multi-user transmit beamforming or \gls{sdma} allows simultaneous  communication with different users by directing radio frequency signals towards specific users. {
In \gls{sdma}, multiple users are served simultaneously on the same frequency channel but are separated spatially using \gls{mimo} techniques.} A  careful design of  beamforming vectors, utilizing  antenna arrays, results in minimizing the average transmit power while maintaining a desired quality of service for each user \cite{roy1997spatial}. {
Each user has a distinct spatial signature, which is leveraged to create separate beams or spatial channels. By employing beamforming techniques, the base station can direct the signal energy to specific users, thereby reducing inter-user interference and enhancing signal quality.}
\gls{sdma} systems are usually designed to beam directly towards a user while reducing the interference experienced by other users. For example, \gls{sdma} uses linear precoding to separate users in the spatial domain and any interference from the other users will be treated as noise. {
Suppose \( \mathbf{H} \) represents the channel matrix between the base station with \( M \) antennas and \( K \) users, where \( \mathbf{H} \in \mathbb{C}^{K \times M} \). The received signal \( \mathbf{y} \) can be expressed as
\[
\mathbf{y} = \mathbf{H} \mathbf{W} \mathbf{x} + \boldsymbol{\eta},
\]
where \( \mathbf{W} \) is the beamforming matrix, \( \mathbf{x} \) is the transmitted signal vector, and \( \boldsymbol{\eta} \) is the noise vector. The beamforming matrix \( \mathbf{W} \) is designed to maximize the \gls{snr} for each user while minimizing interference. Typically, the beamforming vector \( \mathbf{w}_k \) for User \( k \) is chosen to be the dominant eigenvector of the user's channel covariance matrix. This approach ensures that the signal intended for User \( k \) is maximized at the user's location while minimizing the leakage to other users.} In a broadcast channel, multi-user linear precoding  is often useful when users experience similar channel strengths and the channels are semi-orthogonal or orthogonal. \gls{sdma} based on MU-LP is a well-established multiple access technique that builds up the core principle behind many spatial techniques in 4G and 5G such as multi-user \gls{mimo} CoMP, coordinated beamforming, network \gls{mimo}, millimeter-wave \gls{mimo}, and massive \gls{mimo} \cite{mao2018rate}. 
{
\gls{sdma}'s effectiveness heavily depends on accurate \gls{csi} at the transmitter. With perfect \gls{csi}, the base station can precisely steer beams, achieving significant spatial multiplexing gains. However, in practical scenarios, obtaining perfect \gls{csi} is challenging due to factors like channel estimation errors, limited feedback, and feedback delays. To mitigate these issues, robust beamforming techniques and adaptive algorithms are employed, allowing the system to dynamically adjust to varying channel conditions and maintain high performance. Therefore, \gls{sdma}, represents a powerful method to enhance the capacity and efficiency of modern wireless communication systems, particularly in high-density user environments.}

{

\subsection{Integration with \gls{ofdm}}

NOMA schemes can be seamlessly integrated into \gls{ofdm}-based systems to enhance spectral efficiency and support more users \cite{tusha2020hybrid,parida2014power,xie2021deep}. OFDM’s fine-grained frequency division and ability to handle frequency-selective fading and multipath interference make it a suitable platform for NOMA, as subcarriers can be dynamically allocated based on channel conditions. However, the complexity increases due to the need for SIC at the receiver. In what follows, we briefly explain how different NOMA schemes fit into the OFDM structure:

\subsubsection{PD-NOMA in OFDM} In PD-NOMA, each OFDM subcarrier carries a superimposed signal from multiple users, instead of a symbol from an individual constellation \cite{tusha2020hybrid,parida2014power,xie2021deep}. This is achieved by using different power levels for each user.

\subsubsection{CD-NOMA in OFDM} Instead of using power levels to differentiate users, CD-NOMA assigns  specific code sequences to different users \cite{liu2021sparse,rajasekaran2019resource,yang2018clipping}, allowing them to share the same OFDM subcarrier.  At the receiver, the design of efficient multi-user detection schemes is crucial to separate and decode each user’s signal based on the codes.

}

\section{Traditional Modulation Design 
}\label{sec:modulation}

      This section investigates the design and evolution of modulation techniques for \gls{oma} communication and their extension to \gls{noma} communication. 
       Modulation design, or constellation design,  deals with the ways symbols can be arranged in constellations, with a particular focus on achieving a desirable \gls{ber}. This process can be approached independently of channel coding or in conjunction with it. We will discuss both approaches in this section.

\subsection{Modulation Design  Evolution}

Consider a complex-input, complex-output point-to-point \gls{awgn} channel. The channel input $s$ and output $y$ are complex random variables related by
\begin{align}
	y = s + \eta,
\end{align}
where $\eta$ is a complex random variable whose real and imaginary parts are independently and identically distributed (\textit{i.i.d.}) Gaussian random variables. The capacity of this channel is given by the well-known Shannon formula  $ \texttt C = \log (1 + \texttt{SNR})$ where $\texttt{SNR}$ denotes the signal power over the noise
power and $ \texttt C$ is the capacity measured in nats per channel use.  

The proof for achieving the capacity of this channel utilizes a random codeword formed with  \textit{i.i.d.} Gaussian components. 
However, employing a Gaussian codeword in practice is impractical, as the decoding would require an exhaustive search throughout all codewords in the codebook to determine the most probable candidate. 
This has resulted in the adoption of signaling constellations, or digital modulation techniques, such as \gls{psk} and \gls{qam}, comprised of a finite number of points in the complex plane.
Extensive literature exists on the problem of choosing a set of  $M$ symbols with in-phase and quadrature components for transmission. 
Foschini \textit{et al.} were among the first researchers who studied the signal constellation design that minimizes the probability of error on the \gls{awgn} channel under an \textit{average power constraint} \cite{foschini1974optimization}.
Other notable contributions to constellation design can be found in works such as  \cite{forney1984efficient,goldsmith1997variable,barsoum2007constellation,essiambre2010capacity}.

A constellation with $M$ symbols can carry a maximum of $\log_2 M$ information bits per symbol.
The average power of a constellation is defined by the mean of the squares of all symbol amplitudes, i.e.,  $\mathbb{E}{\{ |s|^2}\}$.  Consequently, constellations with larger $M$ must position their points in closer proximity to each other \cite{essiambre2010capacity}. As such, distinct constellations exhibit different \gls{ber} versus \gls{snr} performance for the same channel. The primary objective of constellation design is to identify configurations with the smallest \gls{ber}. Another highly relevant parameter in this context is the \gls{ser}.

Constellation design for modulation schemes has a long, rich history. Initial works primarily concentrated on developing constellations that demonstrated both spectral and power efficiency independent of channel coding  \cite{essiambre2010capacity,davarian1989multiple}.
While the pursuit of spectral and power-efficient modulation schemes remains  as relevant today as it was then, there has been notable improvement and evolution.  The  notion of treating coding and modulation as a unified entity \cite{massey1974coding}, called coded modulation, led to the development of \textit{\gls{tcm}} \cite{ungerboeck1982channel,viterbi1989pragmatic,costello2007channel} and \gls{bicm}\cite{zehavi19928,caire1998bit,li1999bit,i2008bit} as two prominent examples. Due to its efficiency, \gls{bicm} is now a standard in various modern communication systems, including WiMax and 4G/5G cellular networks. 


\subsection{Uniform Modulation Schemes}
\label{subsec:unif}

\subsubsection{\gls{oma}}

By adopting \gls{oma} as the multiple access scheme, modulation techniques originally designed for point-to-point communication can be used without necessitating major modifications. These modulation methods, such as \gls{psk} and  \gls{qam}, are designed to enhance spectral efficiency (bit per symbol) while maintaining an acceptable \gls{ber} for a specified power constraint \cite{foschini1974optimization,forney1984efficient,goldsmith1997variable,barsoum2007constellation}.
Various ways that symbols can be arranged in constellations, with a particular focus on achieving a desirable \gls{ber}, in the absence of channel coding  are discussed in \cite{essiambre2010capacity}. The paper also studies capacity of those constellations on the \gls{awgn}
channel assuming optimum coding.

\subsubsection{\gls{noma}}

{Non-orthogonal transmission is known to be optimal for Gaussian \gls{bc} \cite{cover2012elements,ElGamal2011network,vaezi2019nomachap}, which is also referred to as  \gls{noma} in this paper.}
We consider the two-user \gls{noma} for illustration. Assume   $s_1$ and  $s_2$ are the signals for User~1  and User~2 and $h_1$ and  $h_2$ are their corresponding complex channel gains, respectively. The \gls{bs} broadcasts superimposed signal $\sqrt{\alpha P}s_1 + \sqrt{\bar \alpha P}s_2$, where $P$ is the \gls{bs} power and  $\alpha $, $0\le \alpha \le 1$,
and $\bar \alpha \triangleq 1- \alpha $ are the fractions of total power allocated to the signals of User~1 and User~2, respectively. This is a special case of \eqref{eq:s}, with $K=2$ users where $P_1=\alpha P$ and $P_2=\bar \alpha P$. The received signal at User~$k$, $k\in\{1,2\}$, is given by
\begin{align}\label{eq:main}
y_k = h_k (\sqrt{\alpha P}s_1 + \sqrt{\bar \alpha P}s_2) +\eta_k,
\end{align}
where $\eta_k$ is the complex noise at User~$k$. For decoding, assuming $|h_1| \ge |h_2|$, User~1 (the user with
a stronger channel gain) first decodes the other user's message and then uses \gls{sic}
to decode its message free of interference, whereas User~2 (which has a  weaker channel gain) treats the signal of the stronger user as noise.

\begin{figure} 
	\centering	\includegraphics[width=0.47\textwidth]{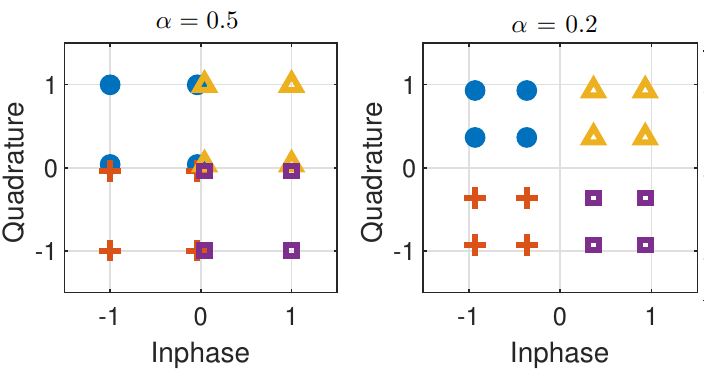}
	\caption{Superimposed constellations for two different values of $\alpha$ where both \gls{noma} users employ a \gls{qpsk} constellation. It is observed that $\alpha = 0.5$ leads to a super-constellation with overlapping symbols, i.e., a non-bijective mapping. On the other hand, when $\alpha = 0.2$, a bijective mapping is achieved. }
\label{fig:IQnoma}
\end{figure}

Similar to \gls{oma}, achieving the capacity region of \gls{noma} involves the use of random Gaussian codewords, whose decoding is impractical as it requires an exhaustive search over the entire codebook to identify the most probable candidate.
In practice, $s_1$ and  $s_2$ are chosen from a discrete and finite-alphabet set like \gls{qpsk} modulation. This simplification comes with its costs. For example, for certain values of power allocation coefficient $\alpha$, the constellations of the two users may overlap. In such cases, the mapping is non-bijective and  decoding (\gls{sic} or maximum-likelihood decoding) with zero error may not be possible.
Figure~\ref{fig:IQnoma} represents the noiseless superimposed signal when both users use a \gls{qpsk} constellation. While for $\alpha=0.2$, we can draw distinctive detection boundaries (bijective mapping), this is not possible for $\alpha=0.5$.  The issue stems from predefined constellations (in this case, \gls{qpsk}) being individually designed for each user, rather than for the transmitted signal, which is a superposition of the signals from both users. Consequently, the overlap of the superimposed constellation has not been taken into account in the constellation design.

Existing literature extensively relies on the above adoption. Especially, \gls{ber} and \gls{ser} analysis for two-user \gls{noma} over \gls{awgn} channel with regular constellations is studied by various researchers. To denote some,  exact \gls{ser} expressions for pulse amplitude modulation and \gls{qam} are derived in \cite{he2019closed}.  Analytical bit error probability expressions  are derived  in \cite{cejudo2017power} when both users employ a \gls{qpsk} constellation and  \cite{assaf2020exact} for \gls{qam}. Similarly, \gls{ber} expressions  
for \gls{bpsk} modulation are studied in \cite{aldababsa2020bit}. 
\gls{ser} and \gls{ber} in \gls{noma} with rotated constellations are studied in \cite{han2021study}. \gls{ber} of \gls{noma} with \gls{bpsk} and \gls{qpsk} modulations over fading channels was studied in \cite{kara2018ber}.
A list of papers evaluating \gls{noma} can be found in \cite{yahya2023error}. 

Despite their findings, these studies have also contributed to one of the most prevalent misconceptions in \gls{noma} literature. This is the myth that a user with a smaller channel gain should receive higher power. Specifically, many papers have assumed that in a two-user \gls{noma}, $\alpha$ should be smaller than 0.5, where $\alpha$ is the fraction of total power allocated to the user with a stronger channel. While this misconception has been debunked in \cite{vaezi2019non} based on Gaussian inputs, this has also been shown to be incorrect even in the finite-alphabet case by several independent works \cite{huppert2007achievable, almohamad2021novel,qi2024ciss}. 

The goal of the above approach is to prevent super-symbol overlapping when utilizing uniform constellations by adjusting the power allocation coefficient and decoding. However, it has long been known that even using uniform constellations that result in overlapping (non-bijective) or partially overlapping super-symbols, it is still possible to decode both \gls{noma} users' information with carefully designed turbo channel coding and iterative decoding  \cite{hoeher2011superposition}. Specifically, when employing \gls{bicm} with iterative decoding, non-bijective mapping may even outperform bijective ones, as elucidated in Myth~1 in \cite{hoeher2011superposition} and other works \cite{wo2011superposition,ma2004coded,fuentes2015low}.

Before proceeding to non-uniform constellation design, we would like to emphasize the above discussions in the following remarks. 

\begin{rem}\label{rem:power}
Drawing general conclusions about \gls{noma} power allocation solely based on a specific type of modulation and decoding is not appropriate and may lead to misconceptions. For instance,
 as discussed in \cite{qi2024ciss}, even within a given modulation scheme like \gls{qam}, successive interference cancellation and maximum likelihood decoding suggest different acceptable values for the power allocation coefficient $\alpha$.
\end{rem}

\begin{rem}\label{rem:BICM}
By leveraging \textit{bit-interleaved coded modulation} with iterative decoding, it is not necessary, or even favorable, to design non-overlapping super-symbols, as detailed in \cite[Myth~1]{hoeher2011superposition} and other references therein. This clarification indicates that there may not be a strict constraint on the power allocation coefficient in \gls{noma}, even within the framework of finite-alphabet inputs. This fact is well-established theoretically and has been noted independently in  \cite[Myth~1]{vaezi2019non}.   
\end{rem}

So far, we have established that within the framework of finite-alphabet \gls{noma}, it is not mandatory to allocate power in the reverse order of the users' channel gains, contrary to the common belief in many \gls{noma} papers. Specifically, in a two-user \gls{noma} scenario, $\alpha \le 0.5$ is not a necessity.
 Such misconceptions arise from limiting assumptions, such as using uniform constellations with \gls{sic} decoding, without exploring other established techniques like \gls{bicm} and maximum likelihood decoding.

With this clarification in mind, in the next subsection, we introduce a less-discussed topic in \gls{noma} literature. Namely, we shift the focus from employing predefined, \textit{uniform} constellation schemes for \gls{noma} users to exploring the possibility of designing/employing \textit{non-uniform} constellations for \gls{noma} users. The goal of designing non-uniform constellations for \gls{noma} users is to ensure that the resulting super-constellation provides a bijective mapping from the beginning.

\subsection{Non-Uniform Modulation}
\label{sec:nonuniform}

The constellations discussed in the previous subsection exhibit regular  shapes. In such a design, the emphasis is almost exclusively on constructing large set of symbols in 2D space with the objective of maximizing the minimum Euclidean distance between them.\footnote{{
It is worth noting that the concept of 2D modulation can be extended to multi-dimensional modulation in various ways \cite{mesleh2008spatial, guo2017generalized,choi2018spatial,zhang2023multi,kang2011construction}. In \gls{mimo} systems, spatial modulation utilizes both the antenna index and complex symbols to form a 3D constellation for efficient information transmission \cite{guo2017generalized}. Other forms of multi-dimensional modulation are explored in \cite{kang2011construction,choi2018spatial,zhang2023multi,4432260,5351720}.}
}  For example, the  conventional uniform  \gls{qam} employs signal points on a regular orthogonal grid. 
In contrast, \textit{non-uniform constellations}, by relaxing this constraint, provide an added shaping gain that facilitates reception, even under lower \glspl{snr} \cite{fuentes2015low,loghin2016non,fragouli2001turbo}.
Non-uniform constellations have found recent applications in digital broadcasting systems.

\subsubsection{\gls{oma}}
  Much of the work in non-uniform constellation is related to point-to-point channels trying to improve the performance at lower \glspl{snr} \cite{fuentes2015low,loghin2016non,fragouli2001turbo}.
 Non-uniform constellation has been particularly important in the context of video btoadcasing and is included in standards like  \gls{dvbt} and \gls{atsc}. 
 The utilization of non-uniform constellations in \gls{mimo} channels, along with two signal processing algorithms for \gls{mimo} precoding, can be found in \cite{gomez2016mimo}. Using a different approach,   guessing random additive noise decoding  is recently used to design  non-uniform constellations based on channel statistics \cite{ozaydin2022grand}. This approach is code-agnostic, focusing on identifying the impact of noise on the data.

 \subsubsection{\gls{noma}}
As highlighted in Remark~\ref{rem:BICM}, non-uniform constellations combined with \gls{bicm}  hold potential advantages in \gls{noma}. While it has been noted that superposition of individual uniform constellations generally leads to non-uniform constellations \cite{kara2018ber,abbas2018multi, han2021study, fuentes2018non,barrueco2021constellation},  there has not been an explicit exploration of this opportunity in the existing literature of \gls{noma}. 
The related topic of \textit{hierarchical modulation}, which involves constellations with non-uniformly spaced signal points, holds a more prominent  presence \cite{vitthaladevuni2003recursive,xiao2018joint,yang2020spatially}.
Constellation rotation \cite{zhang2016downlink,ye2017constellation,khorov2022phase} is another related topic in this area. In fact, the idea of rotating constellations goes back to the design of $\pi/4$-\gls{qpsk} for differential coherent detection in \cite{Baker}. To avoid higher dimensional constellation points collapsing on top of each other and losing diversity, the idea of constellation rotation has been used extensively in the space-time coding literature \cite{1193802,1381439,1545875}.
Although the hierarchical modulation and constellation rotation methods may help overcoming rate losses observed in \gls{noma} implementation \cite{qi2021over}, the design and implementation of non-uniform constellations with \gls{bicm} could be transformative. 

{
It should be noted that \gls{bicm} improves error performance of uncoded modulation but presents implementation challenges, especially in low-resource \gls{iot} devices. The increased receiver complexity, including de-interleaving and soft-decision decoding, along with higher computational demands and power consumption—particularly when using iterative decoding methods like turbo codes—are significant hurdles. These processes also introduce processing delays and potential latency, crucial considerations in real-time communication systems.

Despite these hurdles, \gls{bicm} implementation in \gls{iot} devices is achievable with a focus on efficient algorithms, hardware support, and  complexity-performance tread-off. Optimized algorithms for interleaving and decoding, energy-efficient error-correcting codes tailored for low-power operation, and hardware accelerators for key functions can effectively mitigate these challenges. For example, narrowband \gls{iot} demonstrates a practical application of \gls{bicm} for low-resource devices, employing simplified modulation and coding strategies to enhance reliability while efficiently managing complexity \cite{stierstorfer2010optimizing,yang2021protograph}.}
We will discuss such possible avenues in Section~\ref{sec:nonunifNOMA}.

{
\section{Channel Coding}
\label{sec:coding}

	Channel coding, also referred to as forward error correction, is an indispensable component of today's wireless communications. These coding techniques introduce redundancy into transmitted data using error correction and error detection codes. The redundancy enables receivers to detect and correct errors that may occur during transmission, thereby enhancing the robustness of communication systems and approaching the capacity limits of wireless channels \cite{costello2007channel}. 
	It improves cellular network performance metrics such as reliability, throughput, coverage, spectral efficiency, and energy efficiency \cite{rowshan2024channel}. In addition, channel coding ensures reliable data transmission by detecting and correcting errors caused by noise and interference, reducing the need for re-transmission and as a result improving the throughput and latency. It also extends coverage by enhancing data transmission over longer distances and under challenging radio conditions.

	\subsection{Channel Coding Evolution from  2G to 5G}

 Channel coding has evolved significantly since its inception. In early digital communications, simple error detection methods like parity checks were used, while Hamming codes were among the first error-correcting codes applied in computer systems and data protocols.
	With the advent of more sophisticated coding schemes like \gls{bch} and Reed-Solomon codes \cite{blahut2003algebraic} in the 1960s and convolutional codes \cite{viterbi1971convolutional} in the 1970s, error correction capabilities improved substantially. The development of turbo codes \cite{berrou1993near} and \gls{ldpc} codes \cite{gallager1962low,shokrollahi2004ldpc} in the 1990s represented a significant advancement, achieving performance close to the theoretical limits derived by Shannon. Polar codes \cite{arikan2009channel}, introduced in 2009, further advanced the field by offering provably capacity-achieving performance with low-complexity encoding and decoding algorithms.

	Channel coding has been integral to digital communication in cellular networks since the emergence of 2G technology. With each generation of cellular technology, channel coding has advanced and evolved to meet the increasing demands for reliability and efficiency in data transmission. In addition, different channel coding techniques have been adopted for \textit{control channels} and \textit{data/traffic channels} because of their distinct requirements. 
	Specifically, channel coding for control channels emphasizes robust error detection and correction to ensure reliable transmission of critical signaling information with minimal latency, while coding for data/traffic channels prioritizes maximizing throughput and spectral efficiency.

The tree diagram in Fig.~\ref{fig:channelcodes2} illustrates the channel codes employed in each generation of cellular systems for both control and traffic channels. A more detailed explanations follows.

	\begin{itemize}
		\item 
		\textbf{2G:} 
		Convolutional codes were introduced to enhance the reliability of control signaling, improve the quality of digital voice, and support text messaging.   
		
		\item \textbf{3G:} More advanced coding techniques like turbo codes were introduced for data channels, providing better error correction capabilities and supporting higher data rates for mobile internet access. Despite the above change, 3G systems  use  convolutional codes for control channels.
		
		\item \textbf{4G:} Turbo codes with enhancements like \gls{harq} and flexible rate matching were used. By combining error correction coding with re-transmissions, 4G \gls{lte} improves data reliability and efficiency. Further, coding schemes in 4G are more adaptable to varying channel conditions and user requirements, allowing for better utilization of available bandwidth and more efficient data transmission.
		
		\item \textbf{5G:}  \gls{ldpc} codes and polar codes are used for data channels and control channels, respectively, offering enhanced performance for a wide range of applications, including \gls{iot}, autonomous driving, and \gls{urllc} communications. \gls{ldpc} codes were adopted for data channels due to their superior performance in error correction and lower decoding complexity at high data rates.

	\end{itemize}

\begin{figure*}[!tbp]
	\centering
	\begin{tikzpicture}[edge from parent fork down]
	      grow'=0,
	  	\tikzstyle{every node}=[fill=blue!10,rounded corners]
	\tikzstyle{edge from parent}=[blue!50,-o,thick,draw]
	\tikzstyle{every sibling}=[fill=blue!90,rounded corners]
	\tikzstyle{level 1}=[sibling distance=90mm]
	\tikzstyle{level 2}=[sibling distance=30mm]
	\tikzstyle{level 3}=[sibling distance=14mm]
	\node {channel codes}
	child {node {data channel}
	child  [sibling distance=18mm] {node {2G} 	
		child [sibling distance=25mm] {node {convolutional }}
	}	
	child [sibling distance=15mm] {node {3G}
		child [sibling distance=25mm] {node {turbo }}
	}		
	child [sibling distance=15mm] {node {4G}
			child [sibling distance=25mm] {node {turbo }}
	}			
	child [sibling distance=15mm] {node {5G}
			child [sibling distance=25mm] {node {LDPC }}
	}		 	
}
	child {node {control channel}
		child  [sibling distance=25mm] {node {2G}
				child [sibling distance=25mm] {node {convolutional}}
		}
		child [sibling distance=25mm] {node {3G}
			child [sibling distance=25mm] {node {convolutional}}
		}		
		child [sibling distance=25mm] {node {4G}
				child [sibling distance=25mm] {node {convolutional}}
		}		
		child [sibling distance=20mm] {node {5G}
				child [sibling distance=20mm] {node {polar}}
		}	 	
	};
	\end{tikzpicture}
	\caption{Evolution of channel codes from 2G to 5G. {
 Data  and control channels typically apply different types of channel codes.} }
	\label{fig:channelcodes2}
\end{figure*}
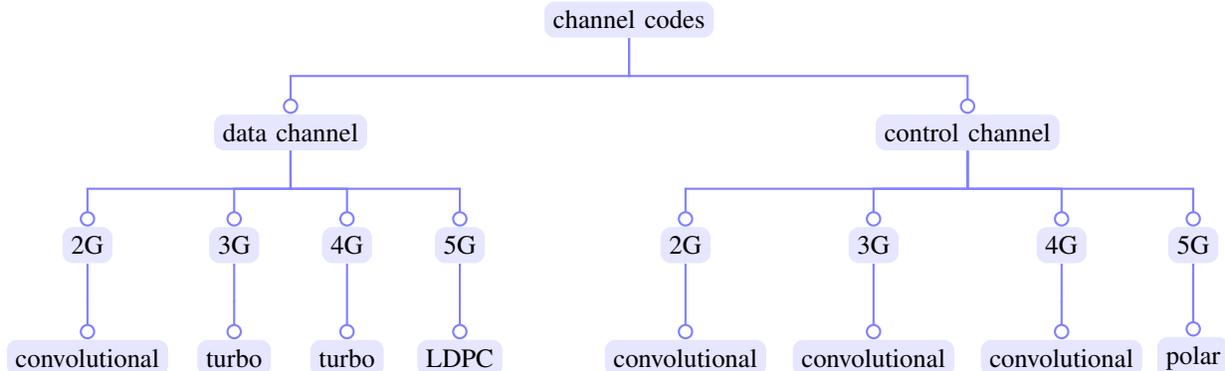

\subsection{Channel Coding and Modulation in 5G \& 6G} 
\label{sec:mod5G}

Here, we first present some details of the channel coding and modulation techniques used in 5G systems. We then discuss possible road maps for the channel coding in 6G. 
\subsubsection{Channel Coding and Modulation in 5G} 
 In wireless systems, a \textit{physical channel} refers to the actual medium via which data is transmitted between the network and the \gls{ue}.
 In total, there are six physical channels in 5G \gls{nr}, as described below: 
\begin{itemize}
	\item[] \noindent \begin{itemize}
	\item \gls{pdcch}: Transmits downlink control information.
	\item \gls{pucch}: Carries uplink control information.
	\item \gls{pbch}: Handles \gls{ue} synchronization and broadcasts essential system information.
	\item \gls{prach}: Ensures initial network access and resource allocation.
	\item \gls{pdsch}: Transports downlink data payloads.
	\item \gls{pusch}: Transmits uplink data payloads.
\end{itemize}\end{itemize}

Two of these channels,  \gls{pdcch}  and \gls{pucch}, carry most of the control information in the downlink and uplink, respectively. \gls{pbch} and  \gls{prach} are responsible for tasks such as \gls{ue} synchronization initial access to the network. As their names suggest, \gls{pbch} and \gls{prach} are downlink and uplink channels, respectively. Actual information payloads, including text messages, audio/video call data, and web streams, are transmitted over shared channels \gls{pdsch} and \gls{pusch}, for downlink and uplink, respectively.


\begin{table}[! htbp]\centering \caption{Channel Coding and Modulation in 5G Physical Channels}
	\begin{threeparttable}
		\begin{tabular}{@{}lclcc@{}}
			\toprule
			&Channel&\multicolumn{1}{c}{Coding}&\multicolumn{1}{c}{Modulation}&\multicolumn{1}{c}{Adaptive}\\
			\midrule
			\midrule
			& PBCH  & Polar&  QPSK  & \xmark \\
			Downlink   & PDCCH & Polar & QPSK & \xmark \\
			& PDSCH & LDPC &  $M$-QAM\tnote{*}   &  \cmark  \\
			\midrule
			\addlinespace[4pt]
			&  PRACH &  --- & ---  & \xmark \\
			Uplink & PUCCH & Polar & BPSK, QPSK &  \xmark \\
			& PUSCH & LDPC  &  $M$-QAM\tnote{*} &  \cmark\\ 
			\bottomrule
			\end{tabular}
		\begin{tablenotes}\footnotesize
			\item[*]  $M \in \{4, 16, 64, 256, 1024 \} $ 
		\end{tablenotes}
	\end{threeparttable}
	\label{tableMSC}
\end{table}

	Table~\ref{tableMSC} summarizes the channel coding and modulation used in each channel. 
	\Gls{prach} does not employ channel coding or modulation; instead, it utilizes Zadoff-Chu sequences for tasks such as initial access synchronization, random access, uplink control information, and channel sounding \cite{lin20215g}.
The table also indicates whether adaptive coding and modulation is used in each channel. An index, called \gls{mcs} index, determines how data is encoded and modulated before transmission and thus decides on the number of useful bits per symbol based on radio signal quality.
A higher signal quality allows  sending more data per symbol. \gls{mcs} index selection depends on radio conditions and \gls{bler}. This is determined by a quantity called \textit{channel quality indicator} and is adjusted dynamically by the \gls{bs}.  5G \gls{nr} supports $M$-\gls{qam} modulation with $M \in \{4, 16, 64, 256, 1024\}$ for the \gls{pdsch} \cite{3GPPNR}. There are 32 \gls{mcs} indices (0-31), with indices 29-31 reserved for re-transmissions. \gls{3gpp} Specification 38.214 \cite{3GPPNR} has provided four tables for \gls{pdsch} \gls{mcs} indices.
With $M$-\gls{qam} modulation and target coding rate $R$, the spectral efficiency of transmission is given by $r \log_2 M$. The highest spectral efficiency is achieved with 1024-\gls{qam} and a coding rate $r = \frac{948}{1024}$, resulting in 9.2578 bits per symbol. The lowest spectral efficiency is achieved with 4-\gls{qam} (\gls{qpsk}) and a coding rate $r = \frac{30}{1024}$, resulting in 0.0586 bits per symbol.

It also worth noting that the \gls{nr} \gls{ldpc} coding process encompasses several stages \cite{lin20215g} such as code block segmentation, cyclic redundancy check attachment, \gls{ldpc} encoding, rate matching, and  bit interleaving. These steps collectively ensure robust and efficient data transmission in 5G networks.

\subsubsection{Channel Coding Road to 6G}
As wireless systems evolve toward 6G, latency requirements become increasingly stringent. Particularly, \gls{urllc} is one of the main use cases of 5G and beyond, envisioned to bridge  the digital and physical realms. This ensures that a given data packet will be delivered within a very short time frame, such as in the order of 1~ms, and with a very high reliability, e.g., 99.999\% \cite{vaezi2022cellular}. End-to-end delay consists of three components:  1) the access delay, 2) the computation delay, and 3) the transmission delay \cite{popovski2022perspective,fantacci2021end}. These tasks involve transmitting essential data and performing computations on both ends, such as compressing data at one end and decompressing it at the other end as well as   channel encoding and decoding. These elements form a latency budget that must meet strict real-time requirements.

    This has led to the exploration of channel coding techniques with shorter block lengths and reduced complexity. Arıkan \cite{arikan2019sequential} made significant progress in this direction by introducing polarization-adjusted convolutional codes and demonstrating that a convolutional pre-transformation can effectively enhance \gls{bler} for short codes with sequential decoding\cite{moradi2020performance}. There has been several recent improvements including  deep learning-based polar codes \cite{choi2024deep}. Polar codes are not the only family of codes being studied for low-delay communication. Various other channel codes, including \gls{ldpc}, turbo, and convolutional codes, have been considered \cite{shirvanimoghaddam2018short}. Other channel codes, such as analog fountain codes, analog \gls{bch} codes, and Raptor codes could also be considered toward this goal.

Finally, we should  emphasize that channel codes are developed independently of whether multiple access techniques are orthogonal or non-orthogonal. Current research in the intersection of \gls{noma} and channel coding predominantly focuses on evaluating \gls{noma} performance using specific channel codes tailored for this purpose.
Additionally,  joint source-channel coding represents another approach for achieving low-latency communication, as explored in studies such as \cite{persson2011joint,vaezi2014distributed,yilmaz2023distributed}.

}
  \subsection{Trellis-Coded Modulations} 
  
\subsubsection{\gls{oma}}
The main idea behind \gls{tcm} is to combine coding and modulation to increase the coding gain~\cite{ungerboeck1987trellis}. To achieve this, a given constellation is expanded and then partitioned into a hierarchy of subsets with increasing minimum Euclidean distances. For encoding, a trellis, representing a finite-state
machine, decides which subset should
be used to maximize the coding gain at each time slot. \gls{tcm}'s finite-state machine and underlying modulation can be represented using convolutional  codes as well. The main design challenge is how to assign subsets to  trellis paths to maximize the coding gain. 
{A \gls{tcm} codeword includes a sequence of transmitted symbols chosen through the finite-state machine and the corresponding set partitioning. The free distance between two possible codewords is defined as the squared Euclidean distance between the two coded sequences. Analogous to how the minimum Euclidean distance determines the performance of a modulation scheme, the minimum free distance specifies the performance of a \gls{tcm}}~\cite{ungerboeck1987trellis}. 

Figures~\ref{fig_conv_encoder} and \ref{fig_4_state_trellis} show an example of a 4-state \gls{tcm} using the 8-\gls{psk} constellation.   
For maximum-likelihood decoding, the Viterbi algorithm is utilized to find the most likely valid path, a path starting at State 0 and merging to State 0, and the corresponding transmitted bits \cite{ungerboeck1987trellis}. 

\subsubsection{\gls{noma}}
In this section, we discuss \gls{tcma} \cite{765488} and \gls{tcnoma} \cite{XZMGHJwcl}.
While the main principles work for any number of users, for the sake of brevity, we focus on a downlink system with two users. 
Also, while the choice of \gls{tcm} for each user can be different, we utilize the \gls{tcm} encoder in  Fig.~\ref{fig_4_state_trellis} for both users. 
The \gls{bs} modulates the input bits using \gls{tcm} to generate two sets of symbols, $a_1(n)$ and $a_2(n)$ for Users 1 and 2, respectively. To improve the performance, $a_2(n)$'s constellation can be rotated, 
{for example} by 
{the optimal rotation} $\frac{\pi}{8}$ 
{\cite{XZMGHJwcl}}. In \gls{tcma}, the encoder superimposes the two outputs $a_1(n)$ and $a_2(n)$ by transmitting  $a_1(n)+a_2(n)$. On the other hand, similar to \gls{pnoma}, \gls{tcnoma} superimposes the symbols of different users on different power levels and transmits $\sqrt{P_1} a_1(n)+\sqrt{P_2} a_2(n)$. Optimal power allocation can be done to maximize the resulting minimum free distance \cite{ungerboeck1982channel} and maintain a constraint on the total power, i.e., $P_1+P_2\leq P$

To recover the transmitted bits, the receiver can employ \gls{sic} to separately decode the two bit streams, as discussed for \gls{pnoma} before. In other words, the stronger user, User 1, decodes the symbols of the weaker user, User 2, and cancels the corresponding interference before decoding its own bits. The weaker user decodes its signal by assuming the symbols of the other user as noise.

\begin{figure}[t b]
	\centering
	\includegraphics[width=3.1in, page=2, trim=2cm 9.5cm 14.3cm 4cm, clip=true]{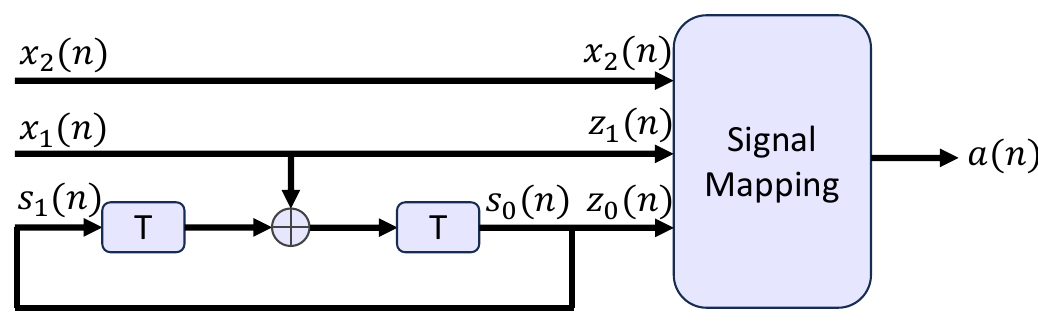}
	\caption{Illustration of an 8-\gls{psk} 4-state \gls{tcm} encoder. 
 {The coded bits go through a rate-1/2 convolutional code.}}
	\label{fig_conv_encoder}
\end{figure}

\begin{figure}[t b]
	\centering
	\subfloat[]{%
		\includegraphics[width=2in, page=2, trim=3cm 9cm 22cm 1.5cm, clip=true]{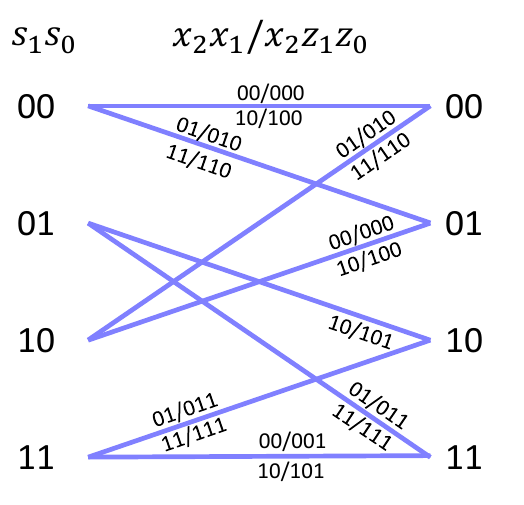}}
	\hfil
	\subfloat[]{%
		\includegraphics[width=1.3in, page=2, trim=10cm 6cm 16cm 4cm, clip=true]{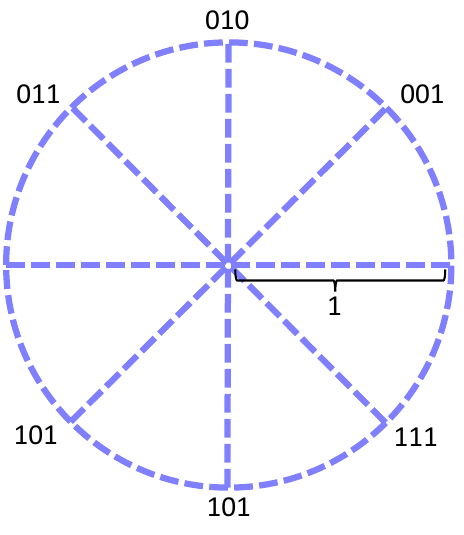}}
	\caption{
 {Encoder and bit-mapping for an 8-PSK 4-state TCM. {
 (a) Trellis representation of 8-\gls{psk} 4-state \gls{tcm}, (b) The mapping of 8-\gls{psk} constellation.}}}
	\label{fig_4_state_trellis}
\end{figure}

\begin{figure}[t b]
	\centering
	\includegraphics[width=2in, page=1, trim=2.5cm 1.3cm 22.2cm 0.9cm, clip=true]{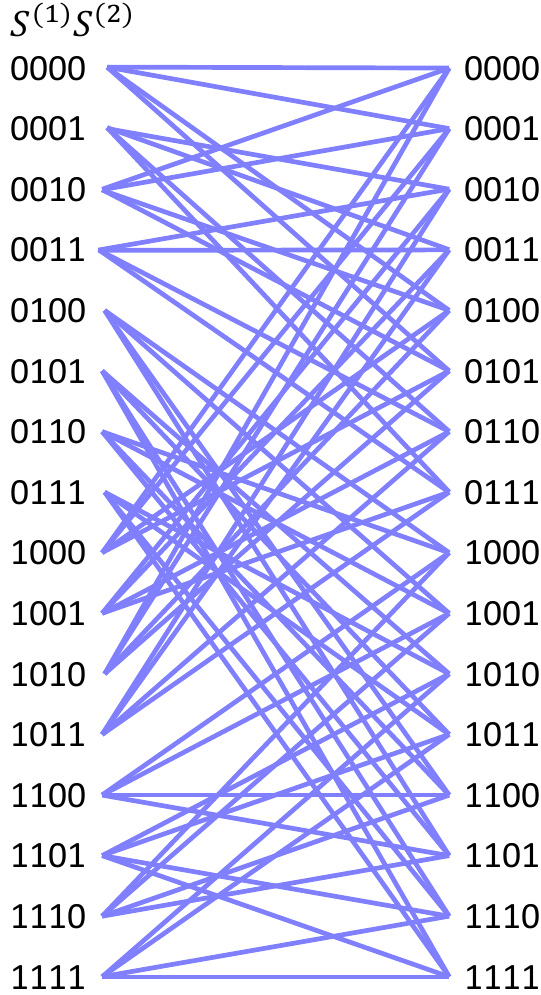}
	\caption{
 { The underlying 16-state trellis generated by the tensor product of two 4-state trellises.}}
	\label{fig_16_state_trellis}
\end{figure}

In addition, joint detection using tensor product of trellises is also possible and greatly improves the performance \cite{XZMGHJwcl}.
First, we describe the underlying tensor product of  trellises~\cite{jafarkhani1999multiple,jafarkhani1999design} which is essential for the joint detection. Let us assume User 1 in \gls{tcnoma} utilizes trellis $T_1$ with $r_1$ states $S_i^{(1)}, \ i=1,\cdots,r_1$. Similarly, $T_2$, the trellis of User 2, contains $r_2$ states $S_j^{(2)}, \ j=1,\cdots,r_2$. Then, the tensor product $T_1 \otimes T_2$ is defined as a ($r_1\times r_2$)-state trellis with states  $S_i^{(1)} S_j^{(2)}$, $i = 1, \cdots, r_1$, $j = 1, \cdots, r_2$. A state transition from $S_i^{(1)} S_j^{(2)}$ to $S_k^{(1)} S_l^{(2)}$ in $T_1 \otimes T_2$ exists if and only if transitions from $S_i^{(1)}$ to $S_k^{(1)}$ and from $S_j^{(2)}$ to $S_l^{(2)}$ exist in $T_1$ and $T_2$, respectively. A similar definition for the tensor product of more than two trellises works as well. 
Figure~\ref{fig_16_state_trellis} shows the tensor product of the trellis in Fig.~\ref{fig_4_state_trellis} by itself. 

\begin{figure}
	\centering
	\includegraphics[width=3.5in, page=1, trim=1cm 7.8cm 2cm 7cm, clip=true]{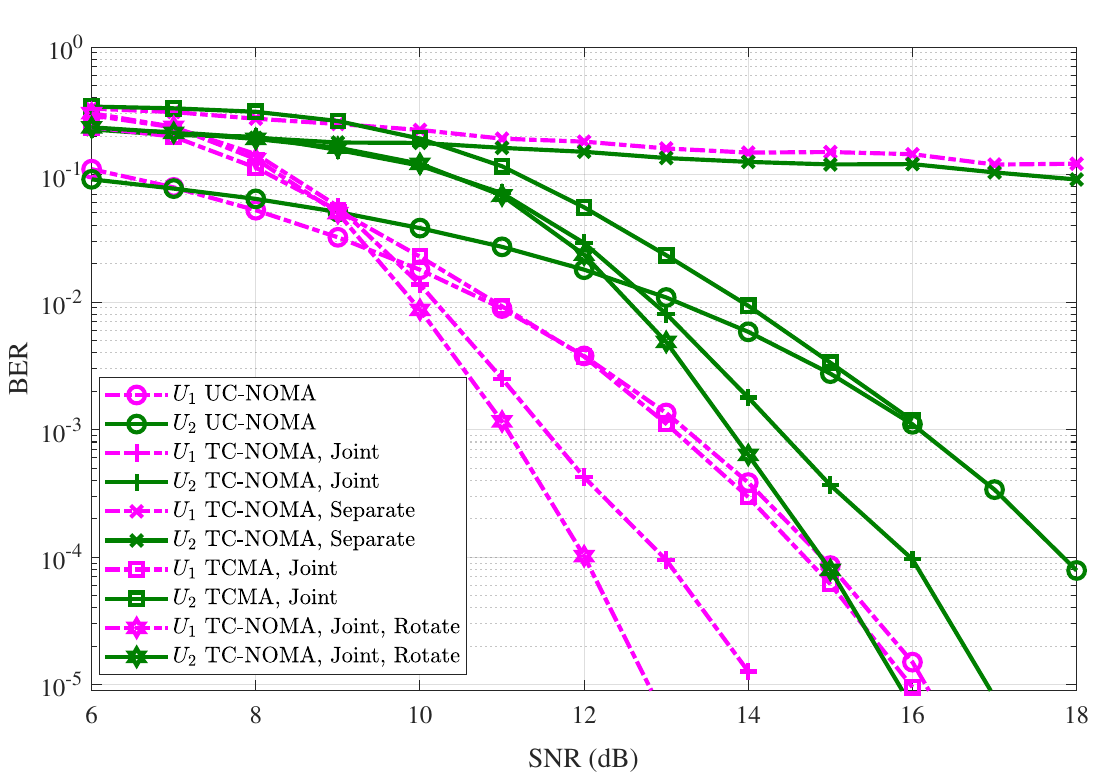}
	\caption{\gls{ber} vs \gls{snr} of \gls{tcma}, \gls{ucnoma}, and \gls{tcnoma} for $P_1 = 0.3$, $P_2 = 1$, $|h_1|^2 = 2$, and $|h_2|^2 = 1$.}
	\label{fig:TCNOMA}
\end{figure}

\gls{tcnoma} is equivalent to a \gls{tcm} using the trellis in Fig.~\ref{fig_16_state_trellis} including transitions from $S_i^{(1)} S_j^{(2)}$ to $S_k^{(1)} S_l^{(2)}$ with the superimposed symbol $\sqrt{P_1}a_1 + \sqrt{P_2}a_2$. 
Therefore, the Viterbi algorithm can be applied to the equivalent \gls{tcm} for joint detection. Overall performance of the \gls{tcnoma} can be optimized by appropriate power allocation. Additional fairness criteria can be included as well \cite{XZMGHJwcl}.
Figure~\ref{fig:TCNOMA} depicts \gls{ber} vs \gls{snr} of different \gls{noma} systems using 8-\gls{psk} and the 4-state \gls{tcm} encoder in Figs.~\ref{fig_conv_encoder} and \ref{fig_4_state_trellis} for $P_1 = 0.3$, $P_2 = 1$, $|h_1|^2 = 2$, and $|h_2|^2 = 1$.
Joint detection of \gls{tcnoma} symbols outperforms the
uncoded \gls{noma} and \gls{tcma} schemes.

\section{Machine Learning-Based Modulation Design} 
\label{sec:ML}

A common characteristic of modulation techniques, as discussed in Section~\ref{sec:modulation}, is their initial design with point-to-point communication in mind. These techniques are characterized by predefined, inflexible symbols, and their constellation shaping is oblivious to interference.
In contrast, one notable shift in today's communication systems, especially in cellular networks, is the move from point-to-point to multi-user communication with multiple transmitters. With this paradigm shift,  additional challenges are presented because  inter-user and inter-cell interference are becoming crucial factors in modern communication system design. Despite this evolution,  constellations designed decades ago are still employed and interference is often addressed through orthogonal resources or by treating interference as noise. 

Machine learning, particularly its rapidly advancing subset, deep learning \cite{goodfellow2016deep}, is  becoming a cornerstone of communication systems \cite{zhang2021multi}. In general, integrating \gls{ai} into cellular networks has already been initiated with 5G Advanced, and \gls{ai} is anticipated to play a pivotal role in shaping 6G networks \cite{lin2022overview}. Machine learning has found different applications across various settings in modulation-related challenges.
For instance, machine learning techniques have been effectively applied in automatic modulation classification \cite{o2018over, peng2021survey, tekbiyik2020robust, wang2023strategies} 
and constellation design \cite{sohrabi2018one, han2021deep, lopez2020survey, ma2021joint,madadi2023ai}, among other applications.

 In this section, we introduce a novel approach to constellation design for \gls{noma}, termed \textit{interference-aware} 
constellation design.
We aim to design \gls{noma} super-constellations  that inherently account for inter-user interference, as opposed to relying on traditional interference-oblivious constellations and attempting to eliminate inter-user interference  at the receiver. 
The objective is to achieve the most distinguishable superimposed symbols for any power allocation such that decoding can be completed without needing \gls{sic}. The main performance metric is \gls{ber} but simplicity of the decoding is as critical.

{
Designing interference-aware super-constellations using traditional methods is very challenging because the constellation shape at the transmitters needs to be adjusted depending on the interfering signal. Otherwise, symbols of superimposed constellations may overlap, as shown in Fig.~\ref{fig:IQnoma}, which is not desired.} \gls{ai}-based approaches appear as an alternative.  Our interference-aware constellation design is essentially an end-to-end \gls{ae}-based communication. 
{
This is because unlike noise, interference has a structure and autoencoders are useful in exploiting structures in data  \cite{zhang2023deep,nartasilpa2018communications}. In addition \gls{ae}-based \gls{e2e} communications simplifies block-by-block communication. }
Before delving into the details of the design, we will motivate it by explaining the significance of \gls{sic}-free decoding in \gls{noma} and provide an introduction to \gls{ae}-based \gls{e2e} communication in the following two subsections.

\subsection{\gls{sic}-Free \gls{noma}}

\textit{\Gls{scsic}} is an optimal approach for achieving the capacity region of the downlink \gls{noma}. In the proof,  Gaussian inputs are employed as optimal codewords. Inspired by this theoretical result, \gls{scsic} is then applied to scenarios where inputs have finite-length and uses finite-alphabet modulations, such as \gls{qam}  \cite{saito2013system, he2019closed, assaf2020exact, qi2021over}.
However, as discussed earlier, it is important to note that such an application may not necessarily uphold the same theoretical assertions \cite{huppert2007achievable, deshpande2009constellation}. At times, this discrepancy has contributed to misconceptions such as the belief that users with smaller channel gains should receive higher power allocations \cite{he2019closed, assaf2020exact, islam2016power}. Although  these misconceptions have been disproved in \cite{vaezi2019non}, the subtle distinctions between these theoretical premises and the practical finite-alphabet inputs are of significant importance \cite{qi2024ciss}. We use the distinctions between the two premises—theoretical and finite-alphabet—to motivate \gls{sic}-free decoding in the following.

Particularly, while theoretically feasible to achieve successful decoding for both users at any $\alpha \in [0, 1]$ \cite{vaezi2019non}, the use of a finite-alphabet input, such as a \gls{qam} constellation, requires  careful selection of the $\alpha$ value, as noted in Remark~\ref{rem:power} in Section~\ref{subsec:unif}. Thus, using \gls{sic} decoding with finite-alphabet inputs limits power allocation choices, and thereby the possibility of achieving the entire capacity region.

In scenarios involving finite-alphabet inputs, the research community has explored alternative approaches to bypass the need for \gls{sic}. Examples of these approaches are the utilization of lattice-based techniques \cite{qiu2019downlink}, index modulations \cite{almohamad2021novel}, and maximum-likelihood  decoding \cite{qi2024ciss}. These approaches may collectively be referred to as \textit{\gls{sic}-free decoding}. They may offer several advantages such as  outperforming \gls{sic} with finite-alphabet inputs and succeed in cases where \gls{sic} falls short. 
In our experimental work  \cite{qi2021over}, we conclude that addressing constellation overlapping or finding a better way to implement or bypass \gls{sic}  is necessary for advancing \gls{noma} as a practical technology. 
The end-to-end \gls{noma} introduced in this paper is an effective \gls{sic}-free \gls{noma}, empowered by  autoencoders.

\subsection{Autoencoder-Based \gls{e2e} Communication}

\subsubsection{A Primer to Autoencoders}
An \textit{\gls{ae}} is a learning technique employed to discover a low-dimensional representation of the input data.  In words, it creates a layer that has less features than the input layer.  
It first compresses (encodes) its input data into a lower dimension and then makes use of this lower dimensional representation to recreate (decode) the original  data \cite{baldi2012autoencoders}. 
While autoencoders compress the input via unsupervised learning,
autoencoders  are used to improve system performance through a training process that tries to minimize the reconstruction error---the difference between the input and reconstructed data. 

{
The loss function quantifies the difference between actual and predicted values.  \textit{Binary cross-entropy} is the most common loss function used in classification problems. It treats each element of the \gls{ae} output as a zero/one classification task. For a training sample \(\bm{y}\), the cross-entropy loss is expressed as $\sum_{i=1}^{n} \left( y_i \log p_i + (1 - y_i) \log (1 - p_i) \right),$
where \(y_i\) is the \(i\)-th element of the training sample \(\bm{y}\), \(n\) is the length of the vector \(\bm{y}\), and \(p_i\) is the predicted probability corresponding to \(y_i\). The value of \(p_i\) is obtained by passing \(\hat{y}_i\) through a sigmoid activation function \cite{goodfellow2016deep}, where \(\hat{y}_i\) is the predicted value of \(y_i\). To go from the loss for one training sample to the batch loss \(\mathcal{L}\), the average over a batch of samples is calculated. } 
Alternatively, one may use the \textit{mean squared error} loss function, which calculates the squared differences between the actual and predicted values, i.e., 
$\sum_{i=1}^{n} (y_i - \hat{y}_i)^2,
$
where \(y_i\) and \(\hat{y}_i\) are respectively the actual and predicted values for the \(i\)-th training sample. Again, we average the above value over multiple training samples to get the batch loss \(\mathcal{L}\).

\subsubsection{Related Works} 
\begin{figure}[h] \centering
	\includegraphics[width=0.41\textwidth]{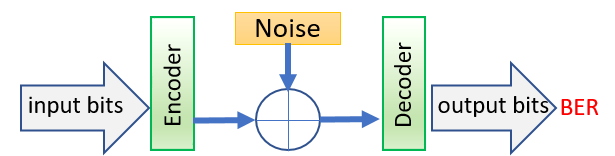}
	\caption{A simplified diagram of the autoencoder-based point-to-point communication. 
		\label{figDAE2} 
	} 
\end{figure}

Utilizing autoencoders for end-to-end communication is a novel concept with significant potential for improving \glspl{ber}. This approach provides a fresh alternative to the prevalent block-by-block design philosophy in contemporary communication systems \cite{o2017deep,song2020benchmarking,o2017introduction,zhang2023deep,alberge2018constellation,ninkovic2023weighted,Aboutaleb2014NOMA
}. As illustrated in Fig.~\ref{figDAE2}, the concept is simple,  mirroring the fundamental objective of digital communication, i.e, reliably transmitting a maximum number of bits.

Notably, this approach outperforms state-of-the-art \gls{mimo}  precoders in terms of \gls{ber} both with and without the channel's knowledge \cite{o2017deep,song2020benchmarking,o2017introduction,zhang2021svd}. 
In \cite{zhang2021svd}, an end-to-end communication is designed for the point-to-point \gls{mimo} channel when the autoencoder learns from the
\gls{csi} and the transmitted symbols to eliminate the interference at the receiver
and estimate the 
transmitted symbols with small errors.  The result is remarkable as the autoencoder system exceeds the performance of the well-known \gls{svd}-based \gls{mimo}  for practical \gls{snr}s.  The gain is attributed to two factors: 1) \gls{ae} optimizes the transmission based on finite-alphabet, finite-length inputs whereas \gls{svd} is designed for infinite-length Gaussian inputs, and 2) \gls{ae} enjoys the freedom of non-uniform constellation shapes and  is not limited to regular constellations.

\subsection{Interference-Aware Constellation Design}

Leveraging autoencoders is a promising approach for crafting  a super-constellation with distinguishable symbols, regardless of power allocation coefficients among \gls{noma} users. This contrasts with the traditional methods where each \gls{noma} user is assigned a predefined constellation (e.g., \gls{qpsk}), and the superposition of these constellations forms the super-constellation. As mentioned earlier, in the conventional approach, the effectiveness of \gls{noma} may be limited due to potential symbol overlap in the super-constellation, depending on power allocation. In contrast, \gls{ae}-based approach anticipates interference during the design phase, resulting in constellations with a robust minimum distance, regardless of power allocation. Crucially, this enables \gls{sic}-free decoding, particularly vital for resource-limited devices.

\subsubsection{\gls{iui}-Aware \gls{noma}}

Shifting from the traditional block-by-block approach to end-to-end communication allows for the simultaneous design and optimization of all components, encompassing constellation design. This innovative approach has the potential to substantially reduce \glspl{ber}. While prior studies \cite{o2017deep, song2020benchmarking, o2017introduction, zhang2021svd} establish a foundation for autoencoder-based end-to-end communication, they primarily concentrate on point-to-point transmissions.

Extending these findings to point-to-\textit{multipoint}  scenarios like \gls{noma} poses challenges.
 The involvement of multiple receivers in \gls{noma} immediately makes the problem more involved as each receiver is only interested in its own message and will have its own loss. Hence, in the training process, each receiver will only feed back  the loss of its own message to the transmitter.
The first difficulty here is that the transmitter needs to process the losses and adjust its transmission accordingly. The two \gls{noma} users  have conflicting interests as they both want to exploit the shared environment for their own benefits. {
Using autoencoders for communication over \gls{noma} channels in different settings is studied in \cite{alberge2018constellation,ninkovic2023weighted,Aboutaleb2014NOMA}.}

\begin{figure*}[th] 
	\centering
	\includegraphics[scale=.62]{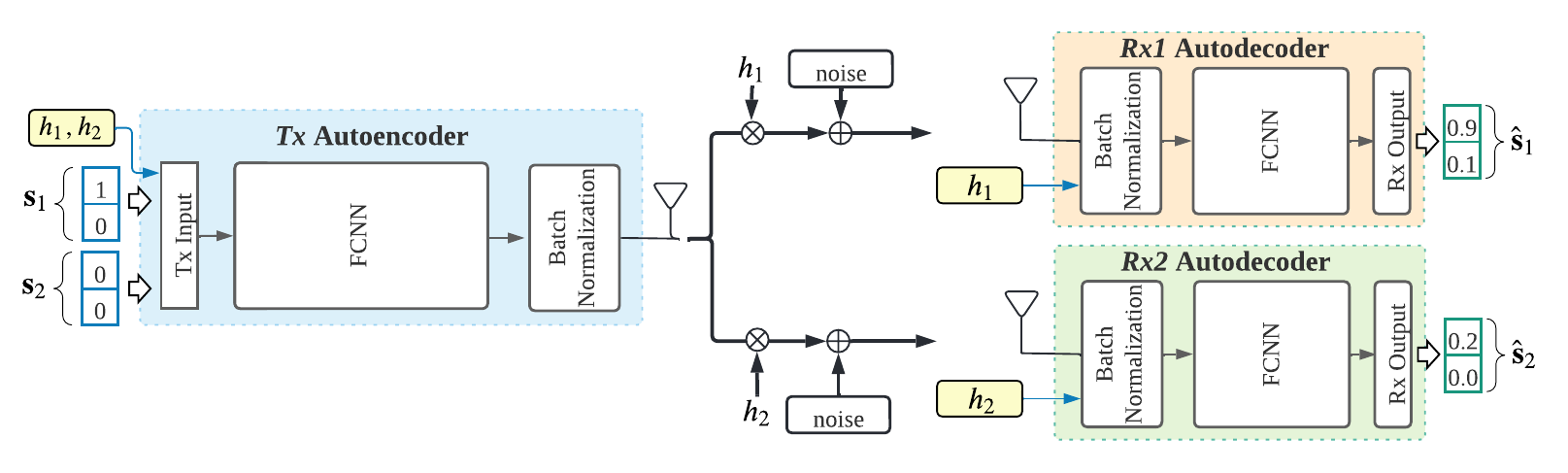}
	\caption{A conceptual architecture of the two-user  \gls{noma} network implemented 
		autoencoder. Both $h_{1}$ and $h_{2}$ are given to the encoder but they are known only in the corresponding decoders.  This end-to-end systems will be trained to form a super-constellation (containing the superposition of the desired  and interfering symbols) with separable symbols.}
	\label{fig:AE-NOMA} 
\end{figure*}

A conceptual architecture of the two-user  \gls{noma} network implemented 
autoencoder is depicted in 	Fig.~\ref{fig:AE-NOMA}. The encoder and decoders include batch normalization and \glspl{fcnn}.
The channel state information ($h_{1}$ and $h_{2}$)  are given to the encoder but they are known only in the corresponding decoders.  This end-to-end systems will be trained and tuned to minimize the difference between transmitted ($s_k$) and received symbols ($\hat s_k$), $k=1,2$. As a result,  a super-constellation (containing the superposition of the desired  and interfering symbols) will be formed. A well-designed and well-trained system will get separable super-constellation symbols.

\gls{ae} designs typically consist of stacked \glspl{fcnn}, which may not be suitable for structures resembling \gls{sic}. While constructing \gls{sic}-like \gls{ae} structures is feasible, our specific design, as illustrated in Fig.~\ref{fig:AE-NOMA}, aims to achieve \textit{\gls{sic}-free} decoding for \gls{noma}. The key objective is to devise an \gls{ae} structure capable of creating a super-constellation with distinguishable symbols for any given power allocation coefficient $\alpha$.

\begin{figure}[htbp] 
	\centering
	\includegraphics[width=0.45\textwidth]{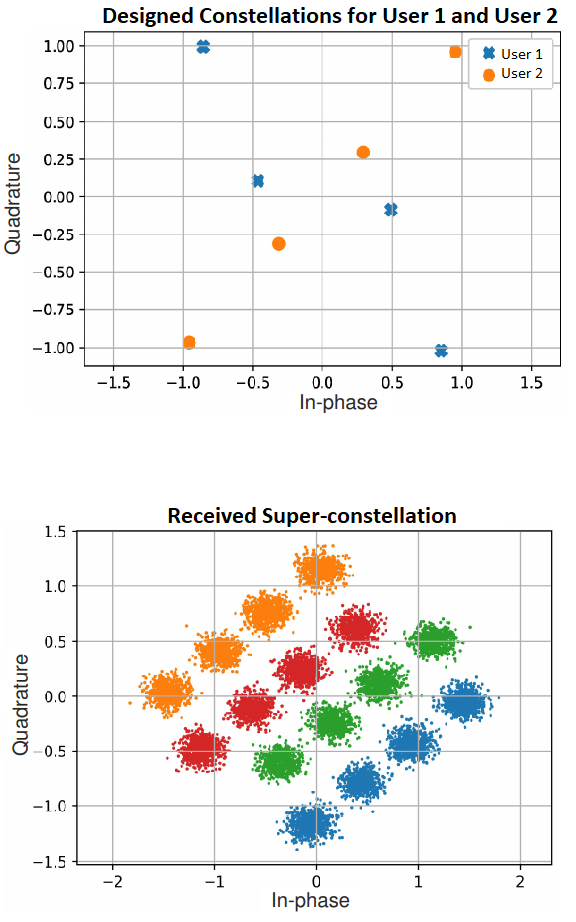}
	\caption{\label{fig:IQnomaDAE}  An example illustrating  \gls{ae}-generated constellations. (top) constellations of the two transmitters. (bottom) super-constellation at the receivers (representing a noisy version of the superimposed constellation at the transmitter).  Each color represents a combination of one symbol of user 1 with four distinct symbols from user 2. }
\end{figure}

Assume that  both users want to transmit 2 bits/symbol, i.e., $s_k$ has two bits each, as illustrated in Fig.~\ref{fig:AE-NOMA}. Then, unlike Fig.~\ref{fig:IQnoma}, where each user wants to make its own maximum separable 4-symbol constellation and superimpose them, the \gls{ae} will be trained to build a maximum separable 16-symbol super-constellation. The primary focus is on the shape of the super-constellation rather than that of the individual constellations. An illustrative instance of such \gls{ae}-generated constellations is presented in Fig.~\ref{fig:IQnomaDAE}. 
It is noteworthy that these \gls{ae}-generated constellations vary for each power allocation coefficient $\alpha$, thereby introducing a dynamic and intelligent aspect to the system.  Also, in contrast to traditional constellations such as a fixed 16-\gls{qam} constellation, which has inflexible and predefined shapes, the \gls{ae} has the capability to generate an extensive array of constellations. This brings a new \textit{shaping gain} and lowers \glspl{ber} at both users, thereby optimizing performance in diverse  scenarios.

\begin{figure}[h] 
	\centering
	\includegraphics[width=0.49\textwidth]{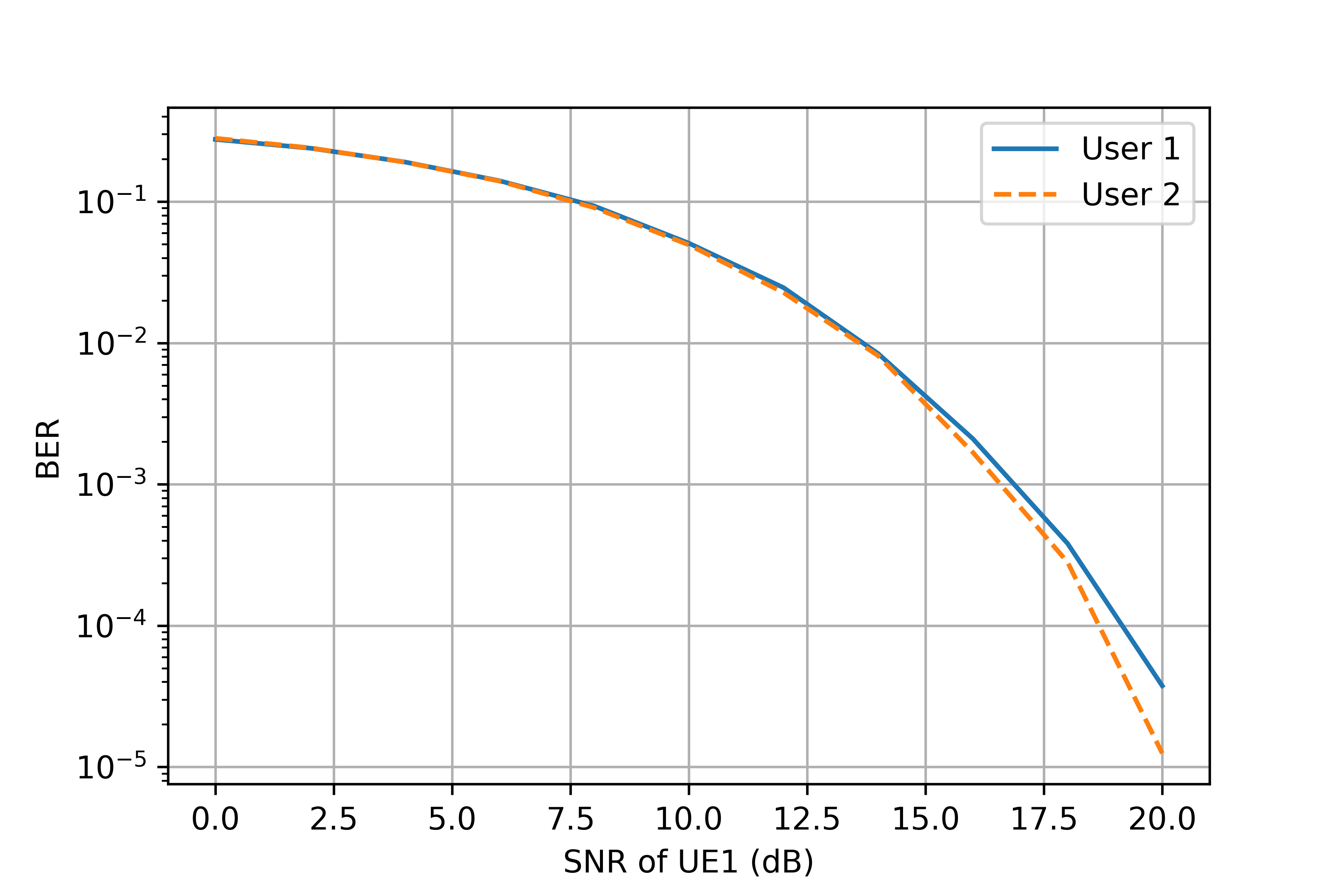}
	\caption{\gls{ber} versus \gls{snr} of UE1 for a two-user \gls{noma} with $h_1=1$ and $h_2=2$. The results are for \gls{ae}-based \gls{noma}. }
	\label{fig_BER_SNR} 
\end{figure}

The performance of the \gls{ae}-generated \gls{noma} constellation is evaluated next. The \gls{ae}-\gls{noma} network is trained with $h_1=1$, $h_2=2$, \gls{snr}$_1=10$dB, and a loss weight of 10. The noise powers at both receivers are the same. The testing results are shown in Fig.~\ref{fig_BER_SNR}.
{
Not surprisingly, the $E_b/N_0$ requirement for this scenario is much higher than that of \gls{qpsk} modulation transmitted over a point-to-point \gls{awgn} channel
where the \gls{ber} is given by $P_e = \frac{1}{2} {\rm erfc}(\sqrt{E_b/N_0})
$ \cite{haykin2008communication}. However, comparing the \glspl{ber} obtained in Fig.~\ref{fig_BER_SNR} with those obtained using \gls{qpsk} constellations for a two-user \gls{noma} channel, as seen in the literature such as in \cite[Fig. 7]{yahya2021exact}, reveals that the \gls{ae}-designed constellations demonstrate much lower \glspl{ber} at the same $E_b/N_0$. This improvement arises because the combination of two \gls{qpsk} constellations can lead to overlapping symbols, which degrades the \gls{ber}, as previously discussed. In contrast, our constellations are specifically designed to have distinct symbols, thereby reducing the likelihood of symbol overlap and improving \gls{ber} performance.
}

{
We should also emphasize that, similar to traditional constellations, the learned modulations use a finite set of symbols during each transmission. The key difference is that the symbols may vary from one transmission to another depending on the channel gains. Such a change in constellation symbols poses a significant challenge for traditional decoders, as they need to be informed about the constellation's position each time. However, in the \gls{ae}-\gls{noma}, this process is handled internally, eliminating the need to inform the decoder about the transmitters' constellation. This is because the  transmitter and receivers (\textit{Tx}, \textit{Rx1}, and \textit{Rx2} autoencoders in Fig.~\ref{fig:AE-NOMA})  are trained jointly and can handle those variations.   }

Numerous research questions revolve around the network's construction and training and the definition of the overall loss function.
 One possible loss function in Fig.~\ref{fig:AE-NOMA} is 
$\mathcal{L}=w_1 \mathcal{L}_1+ w_2\mathcal{L}_2$,
where $\mathcal{L}_1$ and $\mathcal{L}_2$ are the losses  at \textit{Rx1} and \textit{Rx2},   respectively, and $w_1$ and $w_2$ are their weights \cite{alberge2018constellation}.
In a previous work on a point-to-point channel \cite{zhang2021svd}, we trained the network for  \gls{snr}$_1 = 10dB$ and tested it across various \gls{snr}s. Given the complexity of the current problem, training multiple AEs, each tailored for distinct \gls{snr} or $\alpha$ ranges, may be necessary. Since this problem is more involved, we may need to train multiple \glspl{ae} each for a certain range of \gls{snr} or $\alpha$.  
We note that despite the time-consuming nature of training, it is performed offline and the developed model can be used for real-time over-the-fly tests, thereby minimizing computational demands during operational use.

It is worth noting that the goal of the above network is to achieve the most distinguishable super-constellation. This approach may not be the  best as emphasized in Remark~\ref{rem:BICM}, in Section~\ref{subsec:unif}. Specifically, a potentially improved \gls{ber} could be achieved by  adopting \gls{bicm} with iterative decoding which allows for overlapping super-symbols. An example of such a design can be found in \cite{pradhan2003distributed} within the context of point-to-point channels. {
It is also worth indicating that \gls{ae}-based NOMA is expected to outperform traditional designs in terms of latency, as \gls{e2e} communication is generally faster than the block-by-block approach.}

\subsubsection{\gls{ici}-Resilient \gls{noma}}

Thus far, our discussion has centered on single-cell \gls{noma} transmission where the spectrum is distributed among multiple users within a single cell. However, contemporary cellular networks operate in a multi-cell environment, aiming to reuse the same frequency band across many or all cells to enhance spectral efficiency. This shared frequency allocation leads to inter-cell interference, causing outages at cell edges and posing a substantial challenge to achieving high throughput \cite{el2013practical, sun2013interference, vaezi2023drl}. The issue is worsened by the three-dimensional expansion driven by \glspl{uav} \cite{vaezi2023drl}. The design of inter-cell interference-aware \gls{noma} becomes notably complex in light of these evolving complexities within contemporary cellular networks.

 While  multi-cell \gls{noma} has been studied in many papers \cite{shin2016coordinated,shin2017non,guo2021qos,rezvani2021optimal,sung2021distributed}, these are based on Shannon-theoretic principles and  are not directly applicable to finite-alphabet inputs.  Recent works on end-to-end communication in interference channels focus on \gls{oma} \cite{erpek2018learning, vishwakarma2018mitigation, wu2020deep}, often comparing their results with basic baselines like \gls{qpsk}. However, as discussed before, it is known that when both users employ  \gls{qpsk} constellations,  \gls{ber} performance is notably inferior to that where one user employs \textit{rotated} \gls{qpsk}. Thus, those comparisons are less competitive in terms of \gls{ber}. In contrast, the approach in \cite{zhang2022daezic,zhang2023interference} leverages the structure of co-channel interference through interference-resilient constellations adaptable to various interference regimes and topologies, resulting in minimized \gls{ber}. Indeed, the above works  also focus on addressing {
 Z-interference channel \cite{vaezi2016simplified}, also known as one-sided interference channel, in which only one of the users suffers from interference.}
Therefore, a significant gap exists in the literature regarding constellation design for multi-cell \gls{noma}.

\begin{figure*}	
	\centering 
	\includegraphics[scale=0.56]{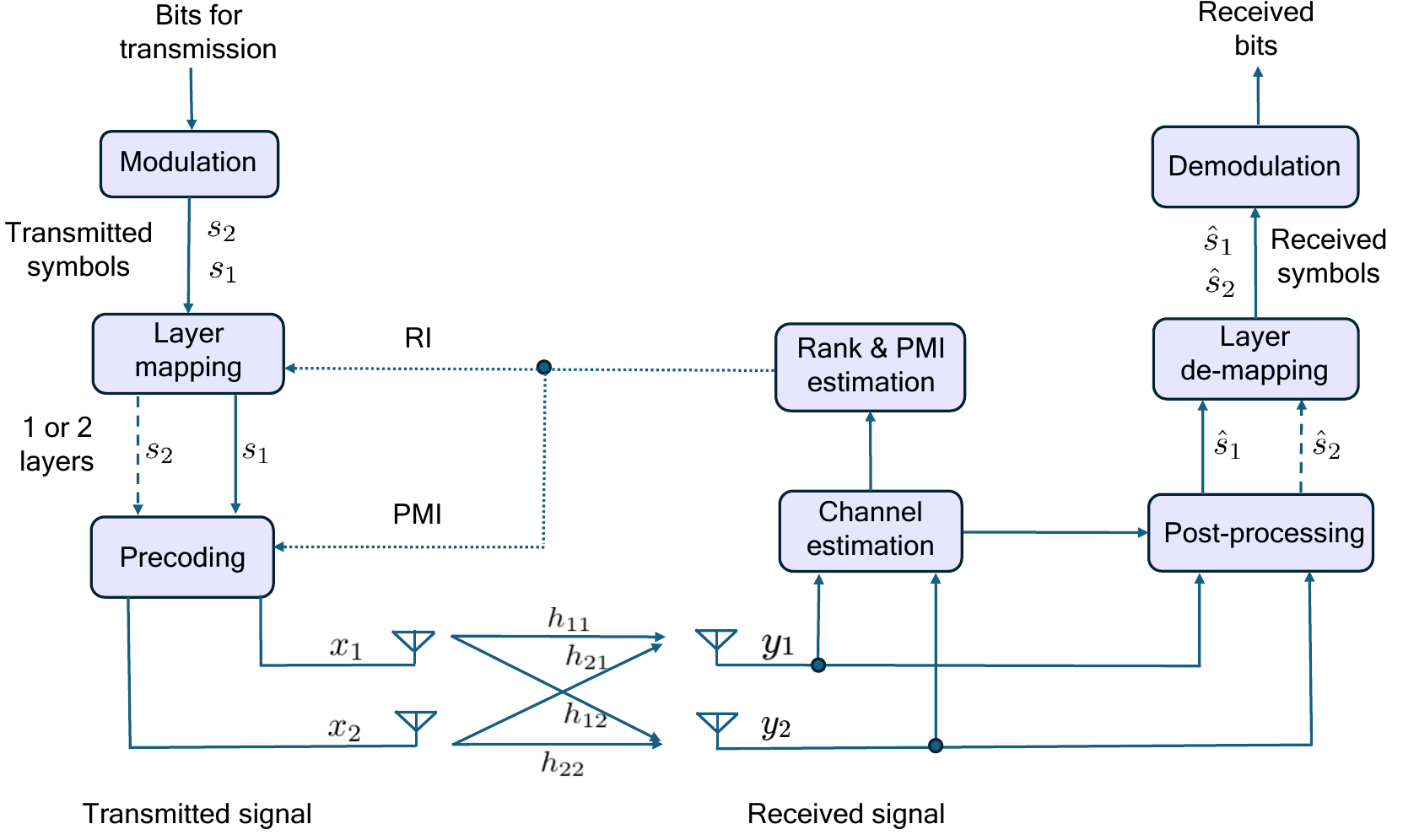}
    \caption{Block diagram of a $2 \times 2$ closed-loop MIMO system. The rank indicator (RI) and precoding matrix indicator (PMI) provide feedback from the receiver to the transmitter. If the channel rank is one, only a single symbol (layer) $s_1$ is transmitted. If the rank is two, both symbols $s_1$ and $s_2$ are transmitted, enhancing data throughput. This scheme can be generalized to an arbitrary number of antennas. Here, modulation and precoding are designed separately. }
 \label{fig:MIMOmod}
\end{figure*}

\subsubsection{Modulation for \gls{mimo} Channels}
\label{subsec:MIMO-NOMA}

{

MIMO techniques can be broadly divided into open-loop and closed-loop systems \cite{STC2005}. In an open-loop system, there is no feedback from the receiver to the transmitter regarding the channel conditions. 
Hence, modulation symbols can be directly assigned to different transmit antennas without CSI at the transmitter.  We have already covered the modulation techniques used for data channels in 5G NR in Section~\ref{sec:mod5G} and summarized them in Table~\ref{tableMSC}.
Advanced open-loop MIMO techniques, such as space-time block coding \cite{alamouti1998simple,VTHJRC1999,VTHJRC99}, have been used as modulation schemes for MIMO channels in various scenarios \cite[Chapter 4]{cox2014introduction}. These techniques are particularly useful for achieving spatial diversity and improving signal reliability.
Space-time block coding is the main block in constructing multiple-antenna differential modulation schemes \cite{857917,945280} and trellis codes for MIMO, like super-orthogonal space-time trellis codes \cite{1193802}. Original goal of space-time block coding was to achieve diversity. However, it is also a building block in closed-loop \gls{mimo} systems with limited feedback, like the beamforming/precoding codebooks in WiFi and 5G \cite{9363693,3GPPrelease15}.

Closed-loop MIMO uses feedback from the receiver to inform the transmitter about channel conditions, enabling techniques like beamforming or precoding to adapt. As shown in Fig.~\ref{fig:MIMOmod}, modulation symbols are mapped to layers before precoding. Assuming a channel rank of two, the layer mapper takes two symbols ($s_1$, $s_2$) and creates two data streams. For a rank of one, it selects one symbol ($s_1$) and sends it to the precoding block or antenna mapper. With precoding and post-processing, the channel is converted to parallel channels so that two independent data steams, without any coupling between them, are sent. This simplifies the design of the receiver.

As explained above, and shown in Fig.~\ref{fig:MIMOmod}, separating precoding and post-processing blocks, modulation schemes used in \gls{mimo} channels are the same as those developed for \gls{siso} channels. The main rational behind such a separation is the optimality of the precoding and post-processing matrices, obtained from \gls{svd}  decomposition, to convert the \gls{mimo} channel into parallel \gls{siso} channels. In such a system, either symbols from a single modulation scheme, such as \gls{qpsk}, are used across all channels, or symbols from different modulation schemes are assigned based on channel characteristics  \cite{xiao2011globally,zhou2005mimo,zhou2010adaptive,xia2020note,prabhu2010energy}.
} Such techniques separate modulation and precoding schemes, as shown in Fig.~\ref{fig:MIMOmod}.   These techniques and their representative works can be summarized as, adaptive modulation with \gls{svd} precoding, along with optimal bit and power allocation \cite{song2020benchmarking}, linear precoding for finite-alphabet \cite{xiao2011globally}, bit allocation with \gls{svd} precoding and water-filling  power 	allocation, bit allocation with \gls{svd} precoding and equal power allocation, and \gls{svd}-based deep autoencoder \cite{zhang2021svd}.
{
With a few exceptions, for example \cite{zhou2005mimo}, the above line of work assumes perfect \gls{csi} at the transmitter.}

Recent studies have emphasized the advantages of joint modulation and precoding design strategies compared to their separate counterparts. In \cite{choi2018spatial}, lattice-based symbol layouts have been proposed to enhance spectral efficiency although they face limitations in fully leveraging \gls{mimo} multiplexing gain. The multi-dimensional constellation concept, introduced in \cite{zhang2023multi}, is designed to  fully leverage \gls{mimo} multiplexing gain. This new method designs constellations and precoding by simultaneously optimizing the in-phase and quadrature components for all sub-channels within a \gls{mimo} channel. It exhibits superior performance compared to existing finite-alphabet \gls{mimo} communication techniques, including current \gls{ae}-based constellations, as illustrated in \cite[Fig.~5]{zhang2023multi}. The \gls{ber}  curves obtained by this method can serve as a lower bound for \gls{ae}-based constellation design, indicating potential for further enhancement in \gls{ae}-based end-to-end \gls{mimo} systems. This multi-dimensional constellation approach also holds promise for the development of even more sophisticated \gls{ae}-based end-to-end \gls{mimo}-\gls{noma} systems. Several DAE-based finite-alphabet \gls{mimo} schemes are introduced in \cite{o2017deep,song2020benchmarking,zhang2021svd}.

Lastly, it is crucial to note that a significant portion of {
downlink} \gls{mimo}-\gls{noma} literature attempts to adapt \gls{scsic} decoding of \gls{siso} channel  to \gls{mimo} ones  \cite{ding2015application,ding2017survey,7434594}. However, such strategies are strictly sub-optimal for \gls{mimo}-\gls{noma} with and without secrecy \cite{vaezi2019nomachap,weingartens2006capacity,liu2010multiple,ekrem2012degraded,qi2022signaling,qi2023k}. Despite the common extension of \gls{scsic} (the optimal  strategy of \gls{siso}-\gls{noma})  to \gls{mimo}-\gls{noma}, it is well-established that \gls{scsic} cannot achieve the capacity region in \gls{mimo}-\gls{bc} (\gls{mimo}-\gls{noma}); instead, \gls{dpc} is the optimal choice.   This point has been highlighted in a few recent \gls{mimo}-\gls{noma} works \cite{qi2022signaling,qi2023k,pauls2021secure}. 
In addition,  in these works, various linear precoding and power allocation strategies are developed to achieve the \gls{dpc}-based capacity region of \gls{mimo}-\gls{noma} channels.  More specifically, Table~1 in \cite{qi2022signaling} lists capacity-achieving signaling design for several related problems. {
The fact that linear preceding approaches the \gls{dpc} region is  known from various other works  \cite{fakoorian2013optimality, shi2008rate, park2015weighted, qi2022signaling}. } {
Further, in \cite{chen2016optimal}, the notion of quasi-degraded channels was introduced as a mean to achieve the \gls{dpc} region using linear precoding. This concept has been applied to other settings, such as network \gls{noma}  in \cite{sun2023application}.} Therefore, when designing modulation and coding for \gls{mimo}-\gls{noma}, particularly, when joint design is considered \cite{choi2018spatial,zhang2023multi}, prioritizing the established optimal approach over suboptimal ones is imperative. 

{
The capacity region of uplink MIMO-NOMA, also referred to as MIMO-MAC, is also well-established \cite{goldsmith2003capacity}. This capacity region is the union of pentagons, each corresponding to different transmit covariance matrices, and its boundary is curved, except at the sum-rate point, where it becomes a straight line \cite{goldsmith2003capacity}. Several low-complexity linear methods, such as simultaneous triangularization \cite{krishnamoorthy2021uplink}, zero forcing, and linear MMSE with practical codes \cite{chi2018practical,liu2019capacity} have been developed to approach the boundary of this region. 
}
\section{\gls{rsma}}
\label{sec:RSMA}

\Gls{rsma} is a promising interference management technique in multi-user systems. While \gls{sdma} treats interference as noise and \gls{noma} decodes interference {
of the users with weaker channels}, \gls{rsma} uses \gls{sic} to decode a portion of the interference and treats the remaining interference as noise. This unique feature of \gls{rsma} allows for a gradual transition between decoding interference in \gls{noma} and treating it as noise in \gls{sdma}, enabling a more flexible approach to interference management \cite{mao2018energy}. The idea of rate splitting was first proposed long ago for a two-user \gls{siso} interference channel\cite{Carleial1978interference}. The term \gls{rsma} was first used around 20 years later in \cite{rimoldi1996rate} where rate splitting was proposed for a multiple-access channel. Research on downlink \gls{rsma} \cite{hao2015rate, dai2016rate, clerckx2016rate} and its advantages resulted in the revival of the idea. 

In what follows, we investigate \gls{rsma} under two broad umbrellas of downlink and uplink.
\begin{figure} 
	\centering
	\includegraphics[width=0.49\textwidth, page=1, trim=11.2cm 0.5cm 8cm 2cm, clip=true, scale=0.8]{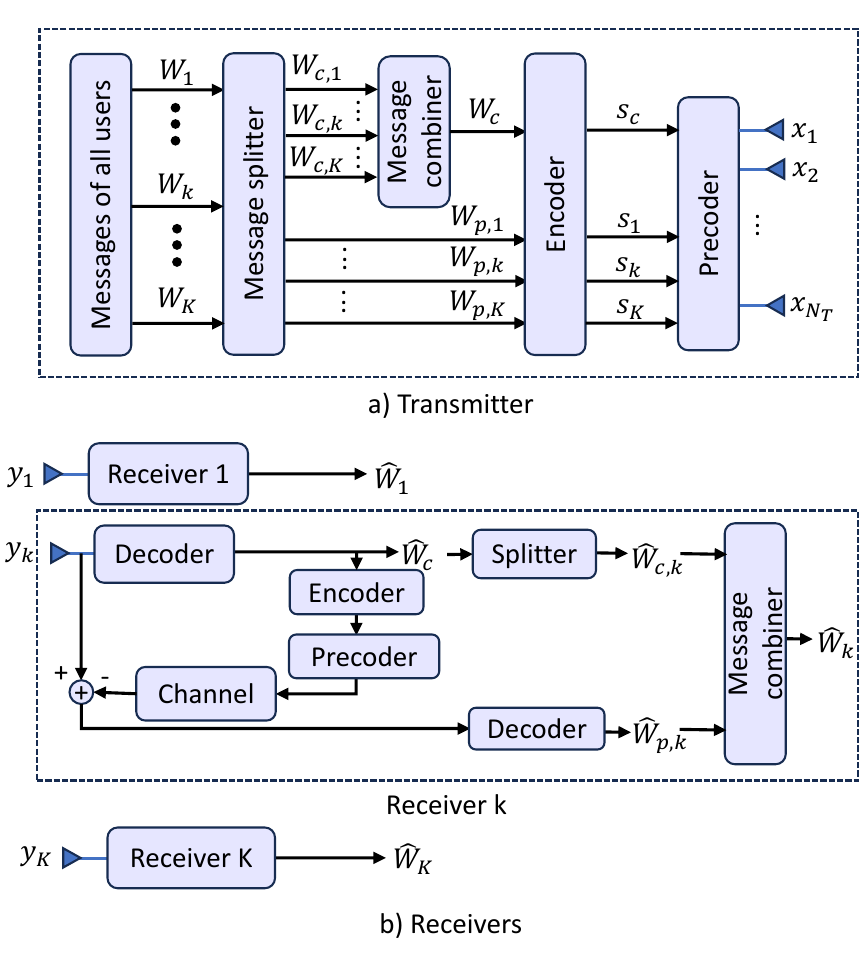}
	\caption{{
 Transmitter and receiver} architecture for a single-layer $K$-user downlink \gls{rsma} system. 
 {{
 a)} Each user's message is divided into private and common parts. The common parts are combined into a single message. The common message and the private messages are coded and transmitted. {
 b)} Each user first decodes the common message and uses \gls{sic} to remove interference, treating the messages from other users as noise.}}
 \label{fig:down_RSMA}
\end{figure}
\subsection{Downlink \gls{rsma}}

To illustrate the framework of downlink \gls{rsma}, let us assume a single-layer scheme \cite{zhou2021rate, matthiesen2021globally}. Consider a system with $K$ single-antenna users where the \gls{bs} is equipped with $M$ antennas. Let $W_1, \dots, W_K$ denote the message of Users $1$ to $K$ and $\mathbf{h}_1^H, \dots, \mathbf{h}_K^H$ be the channels from the transmitter to Users $1$ to $K$, respectively. In a single-layer scheme, the message of the $k$th user, $W_k$, will be split into two parts, namely, common message $W_{c,k}$ and private message $W_{p,k}$. The common parts $\{W_{c,1}, \dots, W_{c,K}\}$ are jointly encoded into the common stream $s_c$ and the private parts $\{W_{p,1}, \dots, W_{p,K}\}$ are encoded in $\{s_1, \dots, s_K\}$. The data stream vector $\mathbf{s} = [s_c, s_1, \dots, s_K]^T\in \mathbb{C}^{K+1}$ is precoded by $\mathbf{P} = [\mathbf{p}_c, \mathbf{p}_1, \dots, \mathbf{p}_K]\in \mathbb{C}^{M \times (K+1)}$ and the transmitted signal can be written as
\begin{equation}
    \label{lin_precoding}
    \mathbf{x} = \mathbf{P}\mathbf{s} = \sum_{k=1}^K s_k \mathbf{p}_k + s_c \mathbf{p}_c.
\end{equation}
The received signal at User $k$ will be
\begin{equation}
    y_k = \mathbf{h}_k^H \mathbf{x} + \eta_k = \sum_{l=1}^K s_l \mathbf{h}_k^H \mathbf{p}_l + s_c \mathbf{h}_k^H \mathbf{p}_c + \eta_k,
\end{equation}
where $\eta_k$ is the complex \gls{awgn} with variance $\sigma^2$. To recover User $k$'s intended message, User $k$ should decode the common message and its own private message. The idea behind downlink \gls{rsma} is to decode the common message at each user and then subtract it from the received signal, to cancel its interference, and then decode the private message by treating the private messages of all other users as noise. This approach can be considered as partial interference cancellation. For   $ k=1,\dots,K$, the rates $ R_{k,c}$ and $ R_{k,p}$ can be calculated as
\begin{align}
    \label{common_rate}
    R_{k,c} &= \log\left(1 + \frac{|\mathbf{h}_k^H \mathbf{p}_c|^2}{\sum_{l=1}^{K}|\mathbf{h}_k^H \mathbf{p}_l|^2  + \sigma^2}\right), \\
    R_{k,p} &= \log\left(1 + \frac{|\mathbf{h}_k^H \mathbf{p}_k|^2}{\sum_{l=1, l\neq k}^{K}|\mathbf{h}_k^H \mathbf{p}_l|^2  + \sigma^2}\right), 
\end{align}
where symbols are assumed to be unit-power, i.e., $\mathbb{E}[|s_k|^2]=1$. To make sure that the common message is decodable by every user, the common rate, $R_c$, should be at most equal to the minimum of rates given in \eqref{common_rate}. That is
\begin{equation}
    R_c \leq \min \{R_{1,c}, \dots, R_{K,c}\}.
\end{equation}
The total rate of User $k$ will be the summation of its private message rate and the corresponding portion of the common message rate. The architecture of transmitter and receiver for this system is shown in Fig.~\ref{fig:down_RSMA}.
More layers can be added to the system by considering common messages that will be decoded by a group of users and treated as noise by the rest \cite{li2020rate}. The transmitted signal in \eqref{lin_precoding} is a linear combination of precoded data streams, but nonlinear precoding is also an option \cite{mao2020beyond}.

\gls{rsma} includes \gls{noma} and \gls{sdma} as special cases. When there is no common message, \gls{rsma} transforms into \gls{sdma}. In a $K$-user system, where the first user has the strongest channel and the {\rm K}th user has the weakest channel, designing an \gls{rsma} system such that the {\rm K}th common message is decoded by Users 1 to K, the ({\rm K-1})th common message is decoded by Users 1 to K-1, $\dots$, the second common message is decoded by Users 1 and 2, and a private message decoded by User 1 turns \gls{rsma} into \gls{noma}. Therefore, performance of \gls{rsma}, as a general framework, is never worse than those of \gls{sdma} and \gls{noma}. 

As a candidate for the \glsentrytext{ngma}, it is insightful to take a look at the advantages and disadvantages of \gls{rsma}. In comparison to other schemes, \gls{rsma} provides robust performance with imperfect \gls{csit} \cite{dai2016rate, medra2018robust}. Having \gls{noma} and \gls{sdma} as special cases, \gls{rsma} results in better spectral efficiency and energy efficiency \cite{mao2018energy, zhou2021rate}. Achieving \gls{urllc} stands out as a key feature in the context of \gls{nr}. A viable solution to enhance latency performance involves the reduction of packet size.  Studies in \cite{dizdar2021rate, xu2022rate} reveal that \gls{rsma} can achieve transmission rates similar to those of  \gls{sdma} and \gls{noma}, yet with shorter block-lengths, resulting in reduced latency. 
Naturally, these enhancements are accompanied by some associated costs. As one example, the decoding complexity of \gls{rsma} is much higher because of \gls{sic} and the need to decode common messages. This extra complexity is also present in \gls{noma}; however, \gls{sdma}, in contrast, does not require \gls{sic}. On the other hand, \gls{sdma} lacks control over interference. Dividing the message at the transmitter introduces the challenge of optimizing message splitting, which becomes an additional task at the transmitter. Moreover, every receiver needs to be aware of the splitting/decoding rule to extract its intended message, adding to the complexity of downlink signaling. It is evident that a larger number of streams resulting from message splitting introduces more challenging optimization problems for beamforming and power allocation in \gls{rsma}. 

{
To illustrate the concept, so far, we have discussed \gls{rsma} for multi-user systems with single-antenna users, i.e., only the transmitter is equipped with multiple antennas. \gls{mimo} has become an essential component of communication systems to significantly enhance their performance. Having multiple antennas at the receiver can further improve the performance of \gls{rsma} systems. While some works, such as \cite{park2022rate}, use the term \gls{mimo} to refer to systems with multiple single-antenna users and a multiple-antenna transmitter, we apply it specifically to systems where both the transmitter and receivers are equipped with multiple antennas. \cite{flores2020linear} introduces practical stream combining methods along with regularized block diagonalization precoding for rate-splitting in \gls{mimo} broadcast channels, focusing on a single common stream and excluding precoding optimization. In \cite{krishnamoorthy2022downlink}, the authors examine a single-layer \gls{mimo}-\gls{rsma} system. To manage the inter-user interference at the receivers, the system employs linear combinations of the null-space basis vectors from the successively augmented MIMO channel matrices of the users as precoding vectors. \cite{mishra2021rate} studies the precoder optimization problem in \gls{mimo}-\gls{rsma} with the goal of maximizing the weighted ergodic sum-rate. Precoder design for MIMO-RSMA has been studied in several other papers such as \cite{diab2022precoding, khamidullina2023rate}. Uplink \gls{mimo}-\gls{rsma} has been studied in \cite{jiang2023rate} where the authors focus on increasing energy efficiency by optimizing the transmit covariance matrices and decoding order using statistical \gls{csi}. In conclusion, these studies demonstrate that while \gls{mimo} systems are highly sensitive to \gls{csi}, \gls{rsma} proves to be much less affected by channel errors.
}

The benefits and promising features of \gls{rsma} have provoked a surge of research in this area. Therefore, effectiveness of \gls{rsma} on many existing problems have been studied. Performance of \gls{rsma} with \gls{ofdm} has been studied in \cite{csahin2023multicarrier} and compared with that of  orthogonal frequency division multiple
access and \gls{ofdm}-\gls{noma}. \gls{ofdm} is a modulation technique that is widely used in modern wireless communication systems, including 4G (\gls{lte}) and 5G networks. It is anticipated that \gls{ofdm} will continue to be a central component in \gls{ngma}. Therefore, ensuring compatibility of new multiple access  techniques with \gls{ofdm} and enhancing overall performance are crucial factors. Application of \gls{rsma} to \gls{isac} has been studied in several works like \cite{chen2023joint, liu2023risac, hu2023joint, xu2021rate}. \gls{isac} is expected to emerge as a pivotal feature in the next generation, seamlessly combining two technologies within a single device. This integration entails the shared utilization of hardware and spectrum for both functionalities. Particularly crucial for smart cities, \gls{isac} enhances continuous environmental monitoring by integrating sensing capabilities into the extensive network of communication base stations. Another rising technology is the \gls{ris}, which enhances communication by establishing new paths between the transmitter and the receiver. Every \gls{ris} consists of a large number of elements, often in the thousands, capable of altering the phase of the incident wave. Through careful control and design of these phase changes, an \gls{ris} can enhance the received power at the receiver. The large number of elements and passive characteristics of these components pose several challenges that need to be addressed. The integration of \gls{ris} in \gls{rsma} systems is an intriguing problem that has been explored in various studies, including \cite{yang2020energy, zhang2023energy, pala2023spectral, niu2023active, dhok2022rate}. \Gls{uav} is another enabling technology for the next generation of wireless communications. These aerial platforms, commonly known as drones, offer a versatile and dynamic solution to address various challenges in wireless communications. Equipped with communication systems, \glspl{uav} can be deployed to enhance network coverage, particularly in distant or disaster-affected regions where conventional infrastructure may be constrained \cite{CTN2022}. Their mobility makes them suitable for duties such as data collection and even acting as relays to extend the network coverage \cite{9930941,10186347,10301558}. As the need for high data rates and low-latency communication increases, incorporating \glspl{uav} into communication networks offers a creative and adaptable approach to fulfill these evolving demands \cite{JGPWHJ20,10594734}. Integration of \gls{rsma} into \gls{uav}-assisted communication has been studied in several works such as \cite{jaafar2020downlink, rahmati2019energy, singh2021outage, xiao2023traffic, liu2023downlink}. 

\subsection{Uplink \gls{rsma}}
\begin{figure}	
	\centering 
	\includegraphics[width=0.4\textwidth, page=1, trim=6.5cm 8cm 16.5cm 2cm, clip=true]{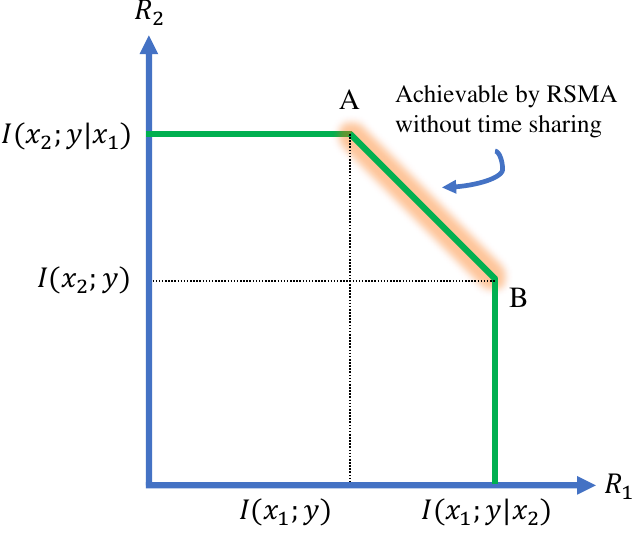}
	\caption{Capacity region of a two-user Gaussian \gls{mac}.}
		\label{fig:time_sharing_RSMA}
\end{figure}
\begin{figure*}[h] 
	\centering
	\includegraphics[width=\textwidth, page=4, trim=1.3cm 3cm 7cm 4cm, clip=true]{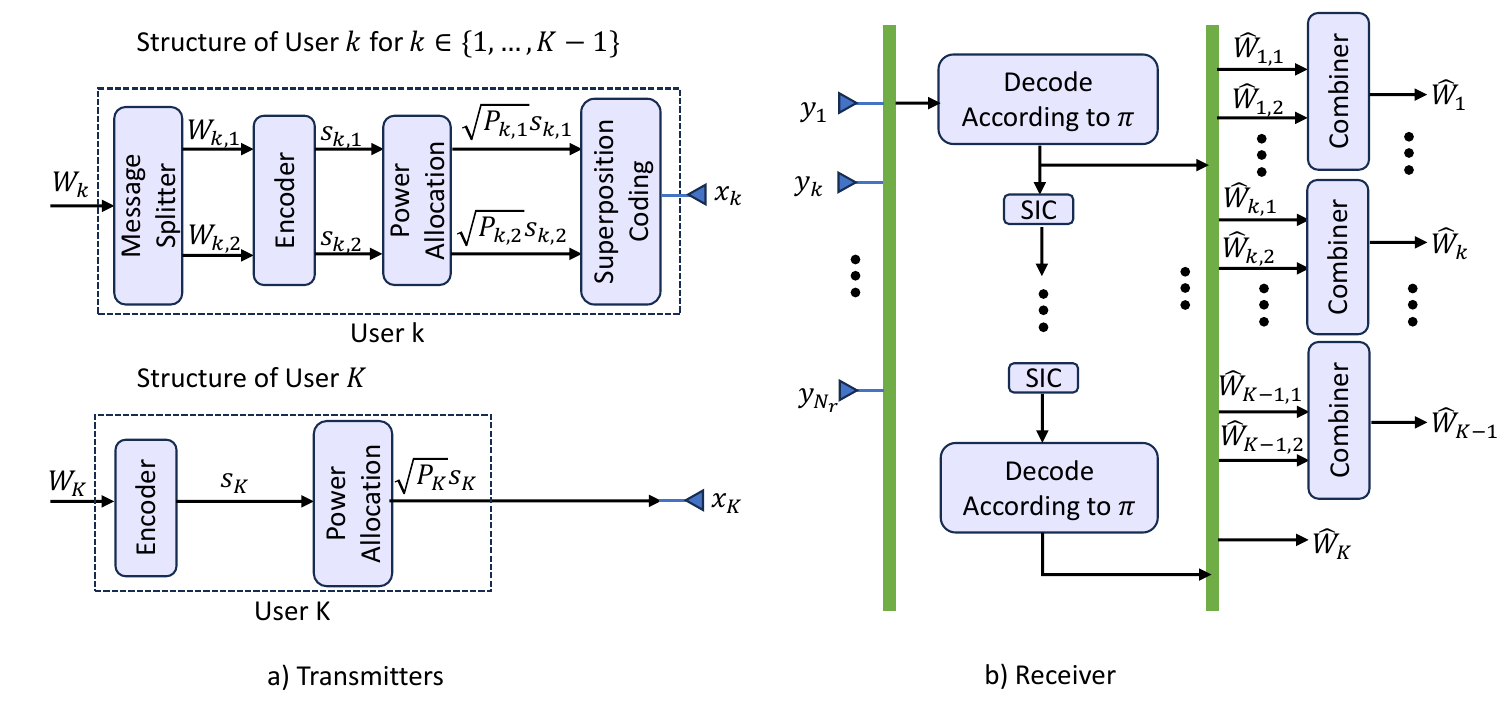}
	\caption{   Uplink \gls{rsma} architecture. 
 {{
 a)} Out of $K$ users, $K-1$ users split their messages into two parts and use superposition coding to transmit them. {
 b)} The receiver determines the optimal decoding order, decodes each user's message, and employs \gls{sic} to remove interference.}}
 \label{fig:Uplink_RSMA}
\end{figure*}
One of the advantages of uplink \gls{rsma} over other multiple access techniques such as \gls{noma} is its ability to achieve the capacity region of the Gaussian \gls{mac} without time sharing. Fig.~\ref{fig:time_sharing_RSMA} shows the rate region of a two-user Gaussian \gls{mac}. Conceptually, this is the same as the black pentagon in Fig.~\ref{rr-sic-uplink}. The horizontal and vertical lines up to points A and B are achievable by \gls{sic} and changing the order of decoding. As shown in Fig.~\ref{fig:time_sharing_RSMA}, \gls{noma} can achieve the line AB by time sharing, while \gls{rsma} can achieve the line AB without time sharing. 
If we continue the plot beyond point A without time sharing (i.e., maintaining the decoding order as \{1,2\}), we obtain a line extending from point A to intersect the horizontal axis. The same occurs for point B and its extension. The resulting pentagon is the convex hull of these plots, which is achieved through time sharing. This involves decoding in the \{1,2\} order for a portion of the time and switching to the \{2,1\} order for the remaining time. Such a switching may not be desirable.

To illustrate the uplink \gls{rsma}'s framework, let us consider an uplink system with K single-antenna users and a receiver with $N_r$ antennas. For K users, it is sufficient to split the messages of K-1 users to avoid time sharing \cite{rimoldi1996rate, mao2022rate}. Without loss of generality, the message of User k for  $k\in \{1, \dots, K-1\}$, denoted by $W_k$, is split into two parts, $W_{k,1}$ and $W_{k,2}$ and the message of User $K$ remains intact. The messages $W_{k,1}$ and $W_{k,2}$ will be encoded into streams $s_{k,1}$ and $s_{k,2}$ with powers $P_{k,1}$ and $P_{k,2}$, respectively. The message of User K will be encoded to the stream $s_K$. We assume unit-power constraint on the symbols, i.e., $\mathbb{E}[|s_{k,1}|^2]=\mathbb{E}[|s_{k,2}|^2]= \mathbb{E}[|s_K|^2]=1$. The transmit signals will be $x_k = \sqrt{P_{k,1}}s_{k,1} + \sqrt{P_{k,2}}s_{k,2}$ for User $k\in \{1, \dots, K-1\}$ and $x_K = \sqrt{P_K}s_K$ for User K. The received signal at the receiver will be
\begin{equation}
    \mathbf{y} = \sum_{k=1}^K \mathbf{h}_k x_k + \mathbf{\eta},
\end{equation}
where $\mathbf{h}_k$ is the channel between User $k$ and the receiver and $\mathbf{\eta}$ is the complex additive white Gaussian noise with the covariance matrix $\sigma^2 \mathbf{I}$.
The receiver uses filters $\mathbf{w}_{k,1}$ and $\mathbf{w}_{k,2}$ to decode the messages of Users 1 to K-1, and uses $\mathbf{w}_K$ to decode the message of User K. Single-layer downlink \gls{rsma}, discussed in the previous section, does not require ordering for decoding, because each user simply decodes the common message first and then its own private message. However, ordering is required for downlink \gls{rsma} with more than one layer \cite{li2020rate}. Similarly, uplink \gls{rsma} requires the receiver to decide which message should be decoded first to achieve the optimal performance. This problem has been studied in \cite{yang2020sum}.  Assuming a decoding order $\pi$ such that $\pi_{k,i} < \pi_{k^{'}, i^{'}}$ indicates stream $s_{k,i}$ is decoded before stream $s_{k^{'},i^{'}}$ (for User K we only have $\pi_K$), the achievable rates will be
\begin{align}
    R_{k,i} =& \log_2\left(1 + \frac{P_{k,i}|\mathbf{w}_{k,i}^H \mathbf{h}_k|^2}{\sum_{\pi_{k,i} < \pi_{k^{'}, i^{'}}}P_{k^{'},i^{'}}|\mathbf{w}_{k^{'},i^{'}}^H \mathbf{h}_k|^2  + \sigma^2}\right), \\
    R_{K} =& \log_2\left(1 + \frac{P_{K}|\mathbf{w}_{K}^H \mathbf{h}_k|^2}{\sum_{\pi_{K} < \pi_{k^{'}, i^{'}}}P_{k^{'},i^{'}}|\mathbf{w}_{k^{'},i^{'}}^H \mathbf{h}_k|^2  + \sigma^2}\right)\\
    &i, i^{'}=1,2, \quad k, k^{'} = 1,\dots,K-1. \nonumber
\end{align}
Fig.~\ref{fig:Uplink_RSMA} shows the architecture of the uplink \gls{rsma} system.
While \gls{rsma} for uplink communication was proposed prior to the downlink, the body of literature in this area is very small \cite{abbasi2022transmission, katwe2022rate, lu2023outage, katwe2023improved, zhang2023fairness, sun2023joint}. This may be viewed as a drawback of uplink \gls{rsma}; however, it also represents a relatively unexplored research area with many open problems. In particular, existing works focus on uplink systems with a single antenna. The next generation is set to deploy multiple antennas for both uplink and downlink. Consequently, tackling multi-antenna uplink \gls{rsma} problems can establish a significant and valuable research domain.

\section{Open Problems \& Future Directions}
\label{sec:open}

In this section, we briefly discuss some open problems and future research directions in designing coding and modulation methods for \glsentrytext{ngma}.
{
While \gls{ngma} systems offer numerous advantages, they may introduce new security and privacy challenges, compared to existing communication systems, because of the shared symbols and models.
}

\subsection{Non-Uniform Modulation for \gls{noma}}
\label{sec:nonunifNOMA}
It is known that  superposition of two uniform constellations generally leads to a non-uniform super-constellations. As this is inherent in \gls{noma}, non-uniform constellations with \gls{bicm} are transformative shifts for addressing the rate losses observed when implemented with uniform constellations \cite{qi2021over}, rendering \gls{noma} a practically viable scheme. 
One  possible approach is to have non-uniform constellations for each user and superimpose them to create a non-uniform super-constellation, but this approach may be overly complicated, as the resulting super-constellations would be distinct for each values of the power allocation coefficient $\alpha$.
On the other hand, the direct design of non-uniform constellations for different ranges of $\alpha$ appears to be more promising. In addition, the integration of non-uniform constellations with \gls{bicm} represents an intriguing and promising avenue for enhancing the practical performance of \gls{noma}. {
\gls{bicm} implementation in low-resource \gls{iot} devices is challenging due to increased receiver complexity. Therefore, efficient algorithms and hardware are essential to make \gls{bicm} feasible for \gls{iot}. Exploring deep learning-based \gls{bicm} presents a promising new research avenue \cite{rovella2024scalable}.
 }

\subsection{Modulation for \gls{mimo}-\gls{noma}}
\label{sec:MIMO-NOMA}
As discussed earlier, modulation schemes for \gls{mimo} channels often borrow concepts from \gls{siso} channels, employing methods like \gls{svd} decomposition to treat each channel individually \cite{zhou2005mimo,xiao2011globally}. These techniques  treat   modulation and precoding separately using various bit allocation strategies. Recent studies advocate for joint modulation and precoding design \cite{choi2018spatial,zhang2023multi}. The joint multi-dimensional constellation and precoding design proposed in \cite{zhang2023multi} optimizes constellations and precoding simultaneously for all \gls{mimo} sub-channels, demonstrating superior performance compared to existing techniques. This approach is based on two key factors: 1) the collaborative design of modulation and precoding, and 2) leveraging the Mahalanobis distance \cite{de2000mahalanobis} to effectively utilize sub-channels with varying gains. By treating the \gls{mimo} system as a unified multi-dimensional space, rather than separate 2D spaces for in-phase and quadrature-phase, this strategy enhances overall performance.  The potential application of these techniques to \gls{mimo}-\gls{noma} systems raises intriguing questions with substantial possibilities. Specifically, the design of \gls{mimo}-\gls{noma} constellations can draw inspiration in the aforementioned multi-dimensional constellation concept presented in \cite{zhang2023multi}. 

{
Additionally, designing modulation schemes based on imperfect or quantized \gls{csi} is another promising research direction \cite{zhou2005mimo}. Also, like regular \gls{mimo} systems, highly correlated channels may result in rank-deficient matrices that require their own studies in \gls{mimo}-\gls{noma} systems.}

\subsection{\gls{ae}-Based \gls{noma}} There are several research questions to explore in \gls{ae}-based constellation design for  \gls{noma}. 
Particularly, future research  in this domain should address key challenges, including refining network construction, training, and loss function design.  Additionally, a more effective \gls{ber} may be achieved by exploring bit-interleaved coded modulation with iterative decoding, allowing for overlapping super-symbols. The multi-dimensional constellation design offers a promising avenue for any \gls{mimo} problem including \gls{mimo}-\gls{noma}. Besides, addressing the complex challenges of inter-cell interference in multi-cell environments is another crucial future research direction for finite-alphabet \gls{noma}. In addition, evaluating the above questions with {
imperfect \gls{csi} and channels with severe multipath and Doppler effects} is crucial for understanding the potential improvement in \glspl{ber} achievable in real-world \gls{noma} transmission.
{
Lastly, non-Gaussian noise channels, such as impulsive noise, can lead to diverse outcomes and offer promising directions for future research. Researchers have investigated \gls{noma} in the presence of impulsive noise \cite{selim2020noma}. Extending these studies to \gls{ae}-based \gls{noma} would be useful.  
}

	 \subsection{Limited Feedback/\gls{csi} }


Acquiring and distributing \gls{csi} to adapt the precoding/beamforming and modulation have been used in different wireless communication systems. 
For example, \gls{3gpp} Release 15 explains the role of \gls{csi} in its description of “5G Phase 1 Specifications” in Section 5.2.2 of TS 138.214 \cite{3GPPrelease15}. The \gls{ue} reports \gls{csi} parameters to the base station using limited
feedback, i.e., a few bits. 
In most of the discussions so far, it has been assumed that the \gls{csi} is known perfectly. 
Obviously, the number of available feedback bits affects the performance of the \gls{ngma} systems, similar to the case of single-user systems \cite{STC2005}. 
Channel values can be estimated at the receiver by transmitting pilots. The accuracy of the \gls{csi} estimation is limited and its error is usually modeled by a Guassian distribution. I
n \gls{fdd} scenarios, to use \gls{csi} at the transmitter, the receiver should quantize and send back the estimated \gls{csi}. Since the channel values are real numbers and only a limited number of feedback bits are available, there will be a quantization error in addition to the estimation error. In \gls{noma} systems that use \gls{sic}, the benefits of \gls{noma} heavily depend on the knowledge of channel orders for decoding. As such, it is very important to design the limited/quantized feedback systems to maintain the channel orders.
Recently, there have been some studies on the effects of limited/quantized feedback on \gls{noma} systems~\cite{XLHJ17,XZMGHJwcl20,7434594, 7506136,9094017,8999638,MAHJ23}. In addition to designing appropriate uniform quantizers to maximize the minimum rate in \gls{siso}-\gls{noma} systems, \cite{XLHJ17} analyzes the performance and demonstrates the catastrophic effects of an incorrect channel order estimation. A more general optimal scalar quantizer that works for both \gls{noma} and A-\gls{noma} is designed in \cite{XZMGHJwcl20}. 
Designing \gls{noma} systems with very small number of feedback bits that can achieve outage probabilities close to those of full-\gls{csi} \gls{noma} systems is possible if the bits are mainly utilized to preserve the user ordering~\cite{7434594, 7506136, 9094017,8999638}. An \gls{ris}-aided \gls{noma} system with limited feedback is designed in \cite{MAHJ23} and the rate loss due to quantization is analyzed. Also, there has been some limited research on \gls{rsma} systems with limited feedback \cite{wu2023deep, lu2017mmse, hao2015rate}.

Nevertheless, there are still many open problems in this area. For example, the {
capacity region of \gls{noma} for a given number of feedback bits is not known.} Future research directions include the design of the feedback link with optimal quantizers, studying the trade-off between the number of feedback bits and the performance, and analyzing the rate loss due to the limited feedback. One important challenge is to keep the transmitted throughput less than the capacity of the channel. Since the transmitter has only access to a quantized version of the channel gains, its estimate of the possible throughput rates may exceed the real capacity of the channel. This is more sever for downlink \gls{noma} where channels are estimated at different nodes without knowing other channels. 
{If the transmitter of downlink \gls{noma} does not know the correct ordering, because of feedback error or quantization noise, a receiver may try to decode at a rate which is higher than its capacity, resulting in catastrophic outcomes and error propagation.}
Another challenge is the robust design of \gls{mimo}-\glsentrytext{ngma} systems for estimated and quantized \gls{csi}, especially taking into account the impact of precoding/beamforming vectors in designing user grouping methods and decoding order algorithms for each group.
{
When \gls{ris} components exist in \gls{noma} systems, estimating the corresponding cascaded channels is a major challenge, especially since \gls{ris} elements are usually passive and cannot transmit pilots \cite{10384715}. Designing appropriate limited feedback mechanisms for such  \gls{ris}-assisted \gls{noma} systems is another interesting open problem.}

{
\subsection{Effects of \gls{mmwave} and \Gls{thz} Channels}

By enabling ultra-high data rates up to terabits per second, and massive connectivity,  \gls{mmwave} and \gls{thz} frequencies are critical for next-generation wireless systems.  However, they face challenges like limited coverage, hardware limitations, and  susceptibility to blockages \cite{akyildiz2022terahertz}.
These channels are sparse, with few significant propagation paths from transmitter to receiver. Their smaller wavelengths allow highly directional antennas and beamforming, focusing energy in specific directions. Thus, communication in these bands is more directional, with signals traveling in defined paths rather than spreading widely. 

Both \gls{noma} and \gls{rsma}  are used with these bands \cite{zhu2019millimeter,wei2018multi,MALJHJHM21, cai2022experimental,cho2023coverage}. The potential for highly directional beamforming in these bands makes \gls{noma} user grouping different from sub-6 GHz channels. Additionally, beamforming will shift from digital to analog or hybrid to reduce power consumption  when the number of antennas is high \cite{almasi2019impact,LJHJ19,LJHJ20}. 
Similarly, challenges related to modulation and coding in \gls{mmwave} and \gls{thz} bands are mostly hardware-related. High-speed \glspl{adc} and \glspl{dac} are power-hungry and costly \cite{etemadi2018analog,wang2019analysis}. As an alternative, \gls{qam} demodulation in the analog domain has been proposed \cite{etemadi2018analog}. Power dissipation of \glspl{adc} and \glspl{dac} reduces by lowering their resolution bits. While low-resolution \glspl{adc} and \glspl{dac}, such as one-bit converters, can simplify hardware design and reduce power consumption, they also  increase quantization noise, introducing challenges in signal quality, beamforming precision, and overall  performance \cite{mo2015capacity}.
\gls{noma} and \gls{rsma} are more sensitive to channel imperfections, compared to \gls{oma}, since they require \gls{csi} for decoding. It is crucial to study the impact of low-resolution converters on both \gls{noma} and \gls{rsma} decoders, particularly on \gls{sic}. Designing signal processing and machine learning techniques to mitigate these effects is essential for achieving acceptable performance levels.
}

\balance


\begin{thebibliography}{100}
	
	\bibitem{vaezi2022cellular}
	M.~Vaezi, A.~Azari, S.~R. Khosravirad, M.~Shirvanimoghaddam, M.~M. Azari,
	D.~Chasaki, and P.~Popovski, ``{Cellular, wide-area, and non-terrestrial IoT:
		A survey on 5G advances and the road toward 6G},'' {\em IEEE Commun. Surv.
		Tut.}, vol.~24, no.~2, pp.~1117--1174, 2022.
	
	\bibitem{viswanathan2020communications}
	H.~Viswanathan and P.~E. Mogensen, ``{Communications in the 6G era},'' {\em
		IEEE Access}, vol.~8, pp.~57063--57074, 2020.
	
	\bibitem{wang20236g}
	C.-X. Wang, X.~You, X.~Gao, X.~Zhu, Z.~Li, C.~Zhang, H.~Wang, Y.~Huang,
	Y.~Chen, H.~Haas, J.~S. Thompson, E.~G. Larsson, M.~D. Renzo, W.~Tong,
	P.~Zhu, X.~Shen, H.~V. Poor, and L.~Hanzo, ``{On the road to {6G: V}isions,
		requirements, key technologies, and testbeds},'' {\em IEEE Commun. Surv.
		Tut.}, vol.~25, no.~2, pp.~905--974, 2023.
	
	\bibitem{saito2013non}
	Y.~Saito, Y.~Kishiyama, A.~Benjebbour, T.~Nakamura, A.~Li, and K.~Higuchi,
	``{Non-orthogonal multiple access (NOMA) for cellular future radio access},''
	in {\em Proc. IEEE Veh. Technol. Conf. (VTC Spring)}, pp.~1--5, Jun 2013.
	
	\bibitem{vaezi2018book}
	M.~Vaezi, Z.~Ding, and H.~V. Poor, {\em {Multiple Access Techniques for 5G
			Wireless Networks and Beyond}}.
	\newblock {Cham, Switzerland: Springer}, 2019.
	
	\bibitem{dai2015non}
	L.~Dai, B.~Wang, Y.~Yuan, S.~Han, C.-L. I, and Z.~Wang, ``{Non-orthogonal
		multiple access for 5G: Solutions, challenges, opportunities, and future
		research trends},'' {\em {IEEE} Commun. Mag.}, vol.~53, no.~9, pp.~74--81,
	2015.
	
	\bibitem{liu2017non}
	Y.~Liu, Z.~Qin, M.~Elkashlan, Z.~Ding, A.~Nallanathan, and L.~Hanzo,
	``{Non-orthogonal multiple access for 5G and beyond},'' {\em Proc. IEEE},
	vol.~105, no.~12, pp.~2347--2381, 2017.
	
	\bibitem{ding2017survey}
	Z.~Ding, X.~Lei, G.~K. Karagiannidis, R.~Schober, J.~Yuan, and V.~K. Bhargava,
	``{A survey on non-orthogonal multiple access for 5G networks: Research
		challenges and future trends},'' {\em {IEEE} J. Sel. Areas Commun.}, vol.~35,
	no.~10, pp.~2181--2195, 2017.
	
	\bibitem{vaezi2019interplay}
	M.~Vaezi, G.~A.~A. Baduge, Y.~Liu, A.~Arafa, F.~Fang, and Z.~Ding, ``{Interplay
		between NOMA and other emerging technologies: A survey},'' {\em IEEE Trans.
		Cogn. Commun. Netw.}, vol.~5, no.~4, pp.~900--919, 2019.
	
	\bibitem{NOMA3GPP}
	{3GPP TD RP-150496}, ``{Study on downlink multiuser superposition
		transmission},'' March 2015.
	\newblock [Online]. Available:
	\url{https://portal.3gpp.org/desktopmodules/Specifications/SpecificationDetails.aspx?specificationId=2912}.
	
	\bibitem{yuan20205g}
	Y.~Yuan, Z.~Yuan, and L.~Tian, ``{5G non-orthogonal multiple access study in
		3GPP},'' {\em {IEEE} Commun. Mag.}, vol.~58, no.~7, pp.~90--96, 2020.
	
	\bibitem{saito2013system}
	Y.~Saito, A.~Benjebbour, Y.~Kishiyama, and T.~Nakamura, ``{System-level
		performance evaluation of downlink non-orthogonal multiple access (NOMA)},''
	in {\em Proc. Int. Symp. Pers. Indoor Mob. Radio Commun. (PIMRC)},
	pp.~611--615, Sep 2013.
	
	\bibitem{benjebbour2015noma}
	A.~Benjebbour, A.~Li, K.~Saito, Y.~Saito, Y.~Kishiyama, and T.~Nakamura,
	``{NOMA: From concept to standardization},'' in {\em {Proc. IEEE Conf. Stand.
			Commun. Netw. (CSCN)}}, pp.~18--23, 2015.
	
	\bibitem{qi2021over}
	Y.~Qi, X.~Zhang, and M.~Vaezi, ``{Over-the-air implementation of NOMA: new
		experiments and future directions},'' {\em {IEEE Access}}, vol.~9,
	pp.~135828--135844, 2021.
	
	\bibitem{cover1972broadcast}
	T.~Cover, ``{Broadcast channels},'' {\em {IEEE} Trans. Inf. Theory}, vol.~18,
	pp.~2--14, Jan 1972.
	
	\bibitem{foschini1974optimization}
	G.~Foschini, R.~Gitlin, and S.~Weinstein, ``{Optimization of two-dimensional
		signal constellations in the presence of Gaussian noise},'' {\em {IEEE}
		Trans. Commun.}, vol.~22, no.~1, pp.~28--38, 1974.
	
	\bibitem{forney1984efficient}
	G.~Forney, R.~Gallager, G.~Lang, F.~Longstaff, and S.~Qureshi, ``{Efficient
		modulation for band-limited channels},'' {\em {IEEE} J. Sel. Areas Commun.},
	vol.~2, no.~5, pp.~632--647, 1984.
	
	\bibitem{goldsmith1997variable}
	A.~J. Goldsmith and S.-G. Chua, ``{Variable-rate variable-power MQAM for fading
		channels},'' {\em {IEEE} Trans. Commun.}, vol.~45, no.~10, pp.~1218--1230,
	1997.
	
	\bibitem{barsoum2007constellation}
	M.~F. Barsoum, C.~Jones, and M.~Fitz, ``{Constellation design via capacity
		maximization},'' in {\em Proc. IEEE Int. Symp. Inf. Theory (ISIT)},
	pp.~1821--1825, 2007.
	
	\bibitem{andrews2005interference}
	J.~G. Andrews, ``{Interference cancellation for cellular systems: A
		contemporary overview},'' {\em {IEEE} Wireless Commun.}, vol.~12, no.~2,
	pp.~19--29, 2005.
	
	\bibitem{forney1998modulation}
	G.~D. Forney and G.~Ungerboeck, ``{Modulation and coding for linear {Gaussian}
		channels},'' {\em {IEEE} Trans. Inf. Theory}, vol.~44, no.~6, pp.~2384--2415,
	1998.
	
	\bibitem{o2017deep}
	T.~J. O'Shea, T.~Erpek, and T.~C. Clancy, ``{Deep learning based MIMO
		communications},'' {\em arXiv preprint arXiv:1707.07980}, 2017.
	
	\bibitem{song2020benchmarking}
	J.~Song, C.~H{\"a}ger, J.~Schr{\"o}der, T.~O'Shea, and H.~Wymeersch,
	``{Benchmarking end-to-end eearning of {MIMO} physical-layer
		communication},'' in {\em Proc. IEEE Global Commun. Conf. (GLOBECOM)},
	pp.~1--6, 2020.
	
	\bibitem{o2017introduction}
	T.~O’Shea and J.~Hoydis, ``{An introduction to deep learning for the physical
		layer},'' {\em IEEE Trans. Cogn. Commun. Netw.}, vol.~3, no.~4, pp.~563--575,
	2017.
	
	\bibitem{zhang2021svd}
	X.~Zhang, M.~Vaezi, and T.~J. O'Shea, ``{SVD-embedded deep autoencoder for MIMO
		communications},'' in {\em Proc. IEEE Int. Commun. Conf. (ICC)}, pp.~1--5,
	2021.
	
	\bibitem{nikopour2013sparse}
	H.~Nikopour and H.~Baligh, ``{Sparse code multiple access},'' in {\em Proc.
		Int. Symp. Pers. Indoor Mob. Radio Commun. (PIMRC)}, pp.~332--336, Sep 2013.
	
	\bibitem{chen2016pattern}
	S.~Chen, B.~Ren, Q.~Gao, S.~Kang, S.~Sun, and K.~Niu, ``{Pattern division
		multiple access: a novel nonorthogonal multiple access for fifth-generation
		radio networks},'' {\em {IEEE} Trans. Veh. Technol.}, vol.~66,
	pp.~3185--3196, Jul 2016.
	
	\bibitem{mohammed2012performance}
	A.-I. Mohammed, M.~A. Imran, R.~Tafazolli, and D.~Chen, ``{Performance
		evaluation of low density spreading multiple access},'' in {\em {Proc. IEEE
			Int. Wireless Commun. Mob. Comput. Conf. (IWCMC)}}, pp.~383--388, Aug 2012.
	
	\bibitem{mao2022rate}
	Y.~Mao, O.~Dizdar, B.~Clerckx, R.~Schober, P.~Popovski, and H.~V. Poor,
	``{Rate-splitting multiple access: Fundamentals, survey, and future research
		trends},'' {\em IEEE Commun. Surv. Tut.}, vol.~24, no.~4, pp.~2073--2126,
	2022.
	
	\bibitem{cai2017modulation}
	Y.~Cai, Z.~Qin, F.~Cui, G.~Y. Li, and J.~A. McCann, ``{Modulation and multiple
		access for 5G networks},'' {\em IEEE Commun. Surv. Tut.}, vol.~20, no.~1,
	pp.~629--646, 2017.
	
	\bibitem{yahya2023error}
	H.~Yahya, A.~Ahmed, E.~Alsusa, A.~Al-Dweik, and Z.~Ding, ``{Error rate analysis
		of NOMA: Principles, survey and future directions},'' {\em IEEE Open J.
		Commun. Soc.}, vol.~4, no.~7, pp.~1682--1727, 2023.
	
	\bibitem{shahab2020grant}
	M.~B. Shahab, R.~Abbas, M.~Shirvanimoghaddam, and S.~J. Johnson, ``{Grant-free
		non-orthogonal multiple access for IoT: A survey},'' {\em IEEE Commun. Surv.
		Tut.}, vol.~22, no.~3, pp.~1805--1838, 2020.
	
	\bibitem{vamos}
	3GPP, ``{Modulation (Version 9.1.0}),'' {\em TS 45.004}, 2010.
	\newblock
	\url{https://telecomfiles.com/servidor/views/img/extractos/vamos.pdf}.
	
	\bibitem{vamosEricsson}
	Ericsson, ``{Evolution of GSM voice},''
	\newblock
	\url{https://www.etsi.org/deliver/etsi_ts/145000_145099/145004/09.01.00_60/ts_145004v090100p.pdf}.
	
	\bibitem{3gppTR36859}
	{3rd generation partnership project (3GPP)}, ``{Study on downlink multiuser
		superposition transmission (MUST) for LTE (Release 13)},'' {Technical report}
	36.859, November 2015.
	
	\bibitem{3gppTR38812}
	{3GPP}, ``{Study on non-orthogonal multiple access (NOMA) for NR (Release
		16)},'' {Technical report} 38.812, December 2018.
	
	\bibitem{elbayoumi2020noma}
	M.~Elbayoumi, M.~Kamel, W.~Hamouda, and A.~Youssef, ``{NOMA-assisted
		machine-type communications in UDN: State-of-the-art and challenges},'' {\em
		IEEE Commun. Surv. Tut.}, vol.~22, no.~2, pp.~1276--1304, 2020.
	
	\bibitem{maraqa2020survey}
	O.~Maraqa, A.~S. Rajasekaran, S.~Al-Ahmadi, H.~Yanikomeroglu, and S.~M. Sait,
	``{A survey of rate-optimal power domain NOMA with enabling technologies of
		future wireless networks},'' {\em IEEE Commun. Surv. Tut.}, vol.~22, no.~4,
	pp.~2192--2235, 2020.
	
	\bibitem{hussain2020machine}
	F.~Hussain, R.~Hussain, S.~A. Hassan, and E.~Hossain, ``{Machine learning in
		IoT security: Current solutions and future challenges},'' {\em IEEE Commun.
		Surv. Tut.}, vol.~22, no.~3, pp.~1686--1721, 2020.
	
	\bibitem{cover2012elements}
	T.~M. Cover and J.~A. Thomas, {\em Elements of Information Theory}.
	\newblock John Wiley \& Sons, 2012.
	
	\bibitem{yang2016general}
	Z.~Yang, Z.~Ding, P.~Fan, and N.~Al-Dhahir, ``A general power allocation scheme
	to guarantee quality of service in downlink and uplink {NOMA} systems,'' {\em
		IEEE Trans. Wireless Commun.}, vol.~15, pp.~7244--7257, Aug 2016.
	
	\bibitem{wang2016power}
	C.-L. Wang, J.-Y. Chen, and Y.-J. Chen, ``{Power allocation for a downlink
		non-orthogonal multiple access system},'' {\em {IEEE Wireless Commun.
			Lett.}}, vol.~5, pp.~532--535, Aug 2016.
	
	\bibitem{zhu2017optimal}
	J.~Zhu, J.~Wang, Y.~Huang, S.~He, X.~You, and L.~Yang, ``{On optimal power
		allocation for downlink non-orthogonal multiple access systems},'' {\em
		{IEEE} J. Sel. Areas Commun.}, vol.~35, pp.~2744--2757, Dec 2017.
	
	\bibitem{doan2019power}
	K.~N. Doan, M.~Vaezi, W.~Shin, H.~V. Poor, H.~Shin, and T.~Q. Quek, ``{Power
		allocation in cache-aided NOMA systems: Optimization and deep reinforcement
		learning approaches},'' {\em {IEEE} Trans. Commun.}, vol.~68, no.~1,
	pp.~630--644, 2020.
	
	\bibitem{ali2016dynamic}
	M.~S. Ali, H.~Tabassum, and E.~Hossain, ``{Dynamic user clustering and power
		allocation for uplink and downlink non-orthogonal multiple access (NOMA)
		systems},'' {\em IEEE Access}, vol.~4, pp.~6325--6343, 2016.
	
	\bibitem{MGHJ16}
	M.~Ganji and H.~Jafarkhani, ``{Interference mitigation using asynchronous
		transmission and sampling diversity},'' in {\em Proc. IEEE Global Commun.
		Conf. (GLOBECOM)}, Dec 2016.
	
	\bibitem{SPHJ15}
	S.~Poorkasmaei and H.~Jafarkhani, ``{Asynchronous orthogonal differential
		decoding for multiple access channels},'' {\em {IEEE} Trans. Wireless
		Commun.}, vol.~14, pp.~481--493, Jan. 2014.
	
	\bibitem{MAHJ15}
	M.~Avendi and H.~Jafarkhani, ``{Differential distributed space-time coding with
		imperfect synchronization in frequency-selective channels},'' {\em {IEEE}
		Trans. Wireless Commun.}, vol.~14, pp.~1811--1822, April 2014.
	
	\bibitem{cui2017asynchronous}
	J.~Cui, G.~Dong, S.~Zhang, H.~Li, and G.~Feng, ``{Asynchronous {NOMA} for
		downlink transmissions},'' {\em {IEEE} Commun. Lett.}, vol.~21, no.~2,
	pp.~402--405, 2017.
	
	\bibitem{XZBHHJ19}
	X.~Zou, B.~He, and H.~Jafarkhani, ``{An analysis of two-user uplink
		asynchronous non-orthogonal multiple access systems},'' {\em {IEEE} Trans.
		Wireless Commun.}, vol.~18, pp.~1404--1418, Feb. 2019.
	
	\bibitem{MGHJwcnc19}
	M.~Ganji and H.~Jafarkhani, ``{Time asynchronous {NOMA} for downlink
		transmission},'' in {\em Proc. IEEE Wireless Commun. Net. Conf. (WCNC)}, Apr.
	2019.
	
	\bibitem{MGXZHJ21}
	M.~Ganji, X.~Zou, and H.~Jafarkhani, ``{Asynchronous transmission for multiple
		access channels: Rate-region analysis and system design for uplink {NOMA}},''
	{\em {IEEE} Trans. Wireless Commun.}, vol.~20, pp.~4364--4378, July 2021.
	
	\bibitem{STC2005}
	H.~Jafarkhani, {\em Space-Time Coding: Theory and Practice}.
	\newblock Cambridge University Press, 2005.
	
	\bibitem{MGXZHJ20}
	M.~Ganji, X.~Zou, and H.~Jafarkhani, ``{Exploiting time asynchrony in
		multi-user transmit beamforming},'' {\em {IEEE} Trans. Wireless Commun.},
	vol.~19, pp.~3156--3169, May 2020.
	
	\bibitem{NOMA3GPPNR}
	{3GPP TR 38.812}, ``{Study on non-orthogonal multiple access (NOMA) for NR},''
	\newblock [Online]. Available:
	\url{https://portal.3gpp.org/desktopmodules/Specifications/SpecificationDetails.aspx?specificationId=3236}.
	
	\bibitem{YuanIndustry}
	Y.~Yuan, ``{Industry Perspective: 5G non-orthogonal multiple access study},''
	{\em {IEEE} Commun. Mag.}, vol.~25, pp.~4--6, October 2018.
	
	\bibitem{hoshyar2008novel}
	R.~Hoshyar, F.~P. Wathan, and R.~Tafazolli, ``{Novel low-density signature for
		synchronous CDMA systems over AWGN channel},'' {\em {IEEE} Trans. Signal
		Process.}, vol.~56, no.~4, pp.~1616--1626, 2008.
	
	\bibitem{chen2018toward}
	Y.~Chen, A.~Bayesteh, Y.~Wu, B.~Ren, S.~Kang, S.~Sun, Q.~Xiong, C.~Qian, B.~Yu,
	Z.~Ding, {\em et~al.}, ``{Toward the standardization of non-orthogonal
		multiple access for next generation wireless networks},'' {\em {IEEE} Commun.
		Mag.}, vol.~56, no.~3, pp.~19--27, 2018.
	
	\bibitem{zaidel2018sparse}
	B.~M. Zaidel, O.~Shental, and S.~S. Shitz, ``{Sparse NOMA: A closed-form
		characterization},'' in {\em Proc. IEEE Int. Symp. Inf. Theory (ISIT)},
	pp.~1106--1110, 2018.
	
	\bibitem{hu2018nonorthogonal}
	S.~Hu, B.~Yu, C.~Qian, Y.~Xiao, Q.~Xiong, C.~Sun, and Y.~Gao, ``{Nonorthogonal
		interleave-grid multiple access scheme for industrial Internet of things in
		5G network},'' {\em {IEEE Trans. Industr. Inform.}}, vol.~14, no.~12,
	pp.~5436--5446, 2018.
	
	\bibitem{bayesteh2015low}
	A.~Bayesteh, H.~Nikopour, M.~Taherzadeh, H.~Baligh, and J.~Ma, ``{Low
		complexity techniques for SCMA detection},'' in {\em Proc. IEEE Globecom
		Workshops (GC Wkshps)}, pp.~1--6, 2015.
	
	\bibitem{nikopour2014scma}
	H.~Nikopour, E.~Yi, A.~Bayesteh, K.~Au, M.~Hawryluck, H.~Baligh, and J.~Ma,
	``{SCMA for downlink multiple access of 5G wireless networks},'' in {\em
		Proc. IEEE Global Commun. Conf. (GLOBECOM)}, pp.~3940--3945, 2014.
	
	\bibitem{NOCA}
	{Nokia, Alcatel-Lucent}, ``{Non-orthogonal multiple access for New Radio, {3GPP
			TSG-RAN WG1 85,R1-165019}},'' May 2016.
	\newblock [Online]. Available:
	\url{https://www.3gpp.org/DynaReport/TDocExMtg--R1-85--31662.htm}.
	
	\bibitem{yuan2016multi}
	Z.~Yuan, G.~Yu, W.~Li, Y.~Yuan, X.~Wang, and J.~Xu, ``{Multi-user shared access
		for Internet of things},'' in {\em Proc. IEEE Veh. Technol. Conf. (VTC
		Spring)}, pp.~1--5, 2016.
	
	\bibitem{PDMA}
	S.~Chen, B.~Ren, Q.~Gao, S.~Kang, S.~Sun, and K.~Niu, ``Pattern division
	multiple access—a novel nonorthogonal multiple access for fifth-generation
	radio networks,'' {\em {IEEE} Trans. Veh. Technol.}, vol.~66, no.~4,
	pp.~3185--3196, 2016.
	
	\bibitem{hoshyar2010lds}
	R.~Hoshyar, R.~Razavi, and M.~Al-Imari, ``{LDS-OFDM an efficient multiple
		access technique},'' in {\em {Proc. IEEE Vehi Technol. Conf.}}, pp.~1--5,
	2010.
	
	\bibitem{RSMA}
	{Qualcomm}, ``{Resource spread multiple access (RSMA), {3GPP TSG RAN WG1
			85,R1‑164688}},'' May 2016.
	\newblock [Online]. Available:
	\url{https://www.3gpp.org/DynaReport/TDocExMtg--R1-85--31662.htm}.
	
	\bibitem{RDMA}
	{https://www.3gpp.org/DynaReport/TDocExMtg--R1-86--31663.htm}, ``{New uplink
		non-orthogonal multiple access schemes for NR, {3GPP TSG RAN WG1 86,
			167535}},'' August 2016.
	\newblock [Online]. Available:
	\url{https://www.3gpp.org/DynaReport/TDocExMtg--R1-86--31663.htm}.
	
	\bibitem{ping2006interleave}
	L.~Ping, L.~Liu, K.~Wu, and W.~K. Leung, ``{Interleave division
		multiple-access},'' {\em {IEEE Trans. Wireless Commun.}}, vol.~5,
	pp.~938--947, Apr 2006.
	
	\bibitem{LCRS}
	{Intel}, ``{Multiple access schemes for new radio interface, {3GPP TSG-RAN WG1
			84b,R1-162385}},'' April 2016.
	\newblock [Online]. Available:
	\url{https://www.3gpp.org/DynaReport/TDocExMtg--R1-84b--31661.htm}.
	
	\bibitem{fang2016lattice}
	D.~Fang, Y.-C. Huang, Z.~Ding, G.~Geraci, S.-L. Shieh, and H.~Claussen,
	``{Lattice partition multiple access: A new method of downlink non-orthogonal
		multiuser transmissions},'' in {\em Proc. IEEE Global Commun. Conf.
		(GLOBECOM)}, pp.~1--6, 2016.
	
	\bibitem{taherzadeh2014scma}
	M.~Taherzadeh, H.~Nikopour, A.~Bayesteh, and H.~Baligh, ``{SCMA codebook
		design},'' in {\em Proc. IEEE Veh. Technol. Conf. (VTC Fall)}, pp.~1--5,
	2014.
	
	\bibitem{sun2019ldpc}
	W.-C. Sun, Y.-C. Su, Y.-L. Ueng, and C.-H. Yang, ``{An LDPC-coded SCMA receiver
		with multi-user iterative detection and decoding},'' {\em IEEE Trans.
		Circuits Syst. I, Reg. Papers.}, vol.~66, no.~9, pp.~3571--3584, 2019.
	
	\bibitem{tang2016low}
	S.~Tang, L.~Hao, and Z.~Ma, ``{Low complexity joint MPA detection for downlink
		MIMO-SCMA},'' in {\em Proc. IEEE Global Commun. Conf. (GLOBECOM)}, pp.~1--4,
	2016.
	
	\bibitem{Ma2019}
	Z.~Ma and J.~Bao, {\em {Sparse code multiple access (SCMA)}}, pp.~369--416.
	\newblock Cham: Springer International Publishing, 2019.
	
	\bibitem{roy1997spatial}
	R.~H. Roy, ``{Spatial division multiple access technology and its application
		to wireless communication systems},'' in {\em Proc. IEEE Veh. Technol. Conf.
		(VTC Spring)}, vol.~2, pp.~730--734, 1997.
	
	\bibitem{mao2018rate}
	Y.~Mao, B.~Clerckx, and V.~O. Li, ``{Rate-splitting multiple access for
		downlink communication systems: Bridging, generalizing, and outperforming
		{SDMA} and {NOMA}},'' {\em EURASIP J. Wireless Commun. Netw.}, vol.~2018,
	no.~1, pp.~1--54, 2018.
	
	\bibitem{tusha2020hybrid}
	A.~Tusha, S.~Do{\u{g}}an, and H.~Arslan, ``{A hybrid downlink NOMA with OFDM
		and OFDM-IM for beyond 5G wireless networks},'' {\em {IEEE} Signal Process.
		Lett.}, vol.~27, pp.~491--495, 2020.
	
	\bibitem{parida2014power}
	P.~Parida and S.~S. Das, ``{Power allocation in OFDM based NOMA systems: A DC
		programming approach},'' in {\em Proc. IEEE Globecom Workshops (GC Wkshps)},
	pp.~1026--1031, 2014.
	
	\bibitem{xie2021deep}
	Y.~Xie, K.~C. Teh, and A.~C. Kot, ``{Deep learning-based joint detection for
		OFDM-NOMA scheme},'' {\em {IEEE} Commun. Lett.}, vol.~25, no.~8,
	pp.~2609--2613, 2021.
	
	\bibitem{liu2021sparse}
	Z.~Liu and L.-L. Yang, ``{Sparse or dense: A comparative study of code-domain
		NOMA system}s,'' {\em {IEEE} Trans. Wireless Commun.}, vol.~20, no.~8,
	pp.~4768--4780, 2021.
	
	\bibitem{rajasekaran2019resource}
	A.~S. Rajasekaran, M.~Vameghestahbanati, M.~Farsi, H.~Yanikomeroglu, and
	H.~Saeedi, ``{Resource allocation-based PAPR analysis in uplink SCMA-OFDM
		systems},'' {\em IEEE Access}, vol.~7, pp.~162803--162817, 2019.
	
	\bibitem{yang2018clipping}
	L.~Yang, X.~Lin, X.~Ma, and K.~Song, ``{Clipping noise-aided message passing
		algorithm for SCMA-OFDM system},'' {\em {IEEE} Commun. Lett.}, vol.~22,
	no.~10, pp.~2156--2159, 2018.
	
	\bibitem{essiambre2010capacity}
	R.-J. Essiambre, G.~Kramer, P.~J. Winzer, G.~J. Foschini, and B.~Goebel,
	``{Capacity limits of optical fiber networks},'' {\em {J. Lightwave
			Technol.}}, vol.~28, no.~4, pp.~662--701, 2010.
	
	\bibitem{davarian1989multiple}
	F.~Davarian and J.~T. Sumida, ``{A multiple digital modulator},'' {\em {IEEE}
		Commun. Mag.}, vol.~27, no.~2, pp.~36--45, 1989.
	
	\bibitem{massey1974coding}
	J.~L. Massey, ``{Coding and modulation in digital communication},'' in {\em
		{Zurich Sem. Digital Commun., 1974}}, vol.~2, 1974.
	
	\bibitem{ungerboeck1982channel}
	G.~Ungerboeck, ``{Channel coding with multilevel/phase signals},'' {\em {IEEE}
		Trans. Inf. Theory}, vol.~28, no.~1, pp.~55--67, 1982.
	
	\bibitem{viterbi1989pragmatic}
	A.~J. Viterbi, J.~K. Wolf, E.~Zehavi, and R.~Padovani, ``{A pragmatic approach
		to trellis-coded modulation},'' {\em {IEEE} Commun. Mag.}, vol.~27, no.~7,
	pp.~11--19, 1989.
	
	\bibitem{costello2007channel}
	D.~J. Costello and G.~D. Forney, ``{Channel coding: The road to channel
		capacity},'' {\em Proc. IEEE}, vol.~95, no.~6, pp.~1150--1177, 2007.
	
	\bibitem{zehavi19928}
	E.~Zehavi, ``{8-PSK trellis codes for a Rayleigh channel},'' {\em {IEEE} Trans.
		Commun.}, vol.~40, no.~5, pp.~873--884, 1992.
	
	\bibitem{caire1998bit}
	G.~Caire, G.~Taricco, and E.~Biglieri, ``{Bit-interleaved coded modulation},''
	{\em {IEEE} Trans. Inf. Theory}, vol.~44, no.~3, pp.~927--946, 1998.
	
	\bibitem{li1999bit}
	X.~Li and J.~A. Ritcey, ``{Bit-interleaved coded modulation with iterative
		decoding},'' in {\em Proc. IEEE Int. Commun. Conf. (ICC)}, vol.~2,
	pp.~858--863, 1999.
	
	\bibitem{i2008bit}
	A.~G. i~Fabregas, A.~Martinez, G.~Caire, {\em et~al.}, ``{Bit-interleaved coded
		modulation},'' {\em {Foundations and Trends{\textregistered} in
			Communications and Information Theory}}, vol.~5, no.~1--2, pp.~1--153, 2008.
	
	\bibitem{ElGamal2011network}
	A.~El~Gamal and Y.~H. Kim, {\em {Network Information Theory}}.
	\newblock Cambridge University Press, 2011.
	
	\bibitem{vaezi2019nomachap}
	M.~Vaezi and H.~V. Poor, ``{NOMA: An information-theoretic perspective},'' in
	{\em {Multiple Access Techniques for 5G Wireless Networks and Beyond}},
	pp.~167--193, Cham, Switzerland: Springer, 2019.
	
	\bibitem{he2019closed}
	Q.~He, Y.~Hu, and A.~Schmeink, ``{Closed-form symbol error rate expressions for
		non-orthogonal multiple access systems},'' {\em {IEEE Trans. Veh. Technol.}},
	vol.~68, no.~7, pp.~6775--6789, 2019.
	
	\bibitem{cejudo2017power}
	E.~C. Cejudo, H.~Zhu, and O.~Alluhaibi, ``{On the power allocation and
		constellation selection in downlink NOMA},'' in {\em Proc. IEEE Veh. Technol.
		Conf. (VTC Fall)}, pp.~1--5, 2017.
	
	\bibitem{assaf2020exact}
	T.~Assaf, A.~J. Al-Dweik, M.~S. El~Moursi, H.~Zeineldin, and M.~Al-Jarrah,
	``{Exact bit error-rate analysis of two-user NOMA using QAM with arbitrary
		modulation orders},'' {\em {IEEE} Commun. Lett.}, vol.~24, no.~12,
	pp.~2705--2709, 2020.
	
	\bibitem{aldababsa2020bit}
	M.~Aldababsa, C.~G{\"o}ztepe, G.~K. Kurt, and O.~Kucur, ``{Bit error rate for
		NOMA network},'' {\em {IEEE} Commun. Lett.}, vol.~24, no.~6, pp.~1188--1191,
	2020.
	
	\bibitem{han2021study}
	W.~Han, X.~Ma, D.~Tang, and N.~Zhao, ``{Study of SER and BER in NOMA
		systems},'' {\em {IEEE} Trans. Veh. Technol.}, vol.~70, no.~4,
	pp.~3325--3340, 2021.
	
	\bibitem{kara2018ber}
	F.~Kara and H.~Kaya, ``{BER performances of downlink and uplink NOMA in the
		presence of SIC errors over fading channels},'' {\em {IET Commun.}}, vol.~12,
	no.~15, pp.~1834--1844, 2018.
	
	\bibitem{vaezi2019non}
	M.~Vaezi, R.~Schober, Z.~Ding, and H.~V. Poor, ``{Non-orthogonal multiple
		access: Common myths and critical questions},'' {\em {IEEE} Wireless
		Commun.}, 2019.
	
	\bibitem{huppert2007achievable}
	C.~Huppert and M.~Bossert, ``{On achievable rates in the two user {AWGN}
		broadcast channel with finite input alphabets},'' in {\em Proc. IEEE Int.
		Symp. Inf. Theory (ISIT)}, pp.~2581--2585, 2007.
	
	\bibitem{almohamad2021novel}
	A.~Almohamad, M.~O. Hasna, S.~Althunibat, and K.~Qaraqe, ``{A novel downlink
		{IM-NOMA} scheme},'' {\em {IEEE Open J. Commun. Soc.}}, vol.~2, pp.~235--244,
	2021.
	
	\bibitem{qi2024ciss}
	Y.~Qi and M.~Vaezi, ``{NOMA decoding: Successive interference cancellation or
		maximum likelihood detection},'' in {\em Proc. Annual Conf. Inf. Sci. Sys.
		(CISS)}, pp.~1--6, 2024.
	
	\bibitem{hoeher2011superposition}
	P.~A. Hoeher and T.~Wo, ``{Superposition modulation: myths and facts},'' {\em
		{IEEE} Commun. Mag.}, vol.~49, no.~12, pp.~110--116, 2011.
	
	\bibitem{wo2011superposition}
	T.~Wo, {\em {Superposition mapping \& related coding techniques}}.
	\newblock PhD thesis, 2011.
	\newblock PhD Dissertation.
	
	\bibitem{ma2004coded}
	X.~Ma and L.~Ping, ``{Coded modulation using superimposed binary codes},'' {\em
		{IEEE} Trans. Inf. Theory}, vol.~50, no.~12, pp.~3331--3343, 2004.
	
	\bibitem{fuentes2015low}
	M.~Fuentes, D.~Vargas, and D.~Gomez-Barquero, ``{Low-complexity demapping
		algorithm for two-dimensional non-uniform constellations},'' {\em IEEE Trans.
		Mob. Comput.}, vol.~62, no.~2, pp.~375--383, 2015.
	
	\bibitem{mesleh2008spatial}
	R.~Y. Mesleh, H.~Haas, S.~Sinanovic, C.~W. Ahn, and S.~Yun, ``Spatial
	modulation,'' {\em {IEEE} Trans. Veh. Technol.}, vol.~57, no.~4,
	pp.~2228--2241, 2008.
	
	\bibitem{guo2017generalized}
	S.~Guo, H.~Zhang, P.~Zhang, D.~Wu, and D.~Yuan, ``{Generalized 3-D
		constellation design for spatial modulation},'' {\em {IEEE} Trans. Commun.},
	vol.~65, no.~8, pp.~3316--3327, 2017.
	
	\bibitem{choi2018spatial}
	J.~Choi, Y.~Nam, and N.~Lee, ``{Spatial lattice modulation for {MIMO}
		systems},'' {\em {IEEE} Trans. Signal Process.}, vol.~66, no.~12,
	pp.~3185--3198, 2018.
	
	\bibitem{zhang2023multi}
	X.~Zhang and M.~Vaezi, ``{Multi-dimensional joint constellation and precoding
		design for {MIMO} systems},'' {\em {IEEE} Wireless Commun. Lett.}, vol.~12,
	no.~4, pp.~713--717, 2023.
	
	\bibitem{kang2011construction}
	S.~G. Kang, Z.~Chen, J.~Y. Kim, J.~S. Bae, and J.-S. Lim, ``{Construction of
		higher-level 3-D signal constellations and their accurate symbol error
		probabilities in AWGN},'' {\em {IEEE} Trans. Signal Process.}, vol.~59,
	no.~12, pp.~6267--6272, 2011.
	
	\bibitem{4432260}
	F.~Fazel and H.~Jafarkhani, ``Quasi-orthogonal space-frequency and
	space-time-frequency block codes for {MIMO OFDM} channels,'' vol.~7, no.~1,
	pp.~184--192, 2008.
	
	\bibitem{5351720}
	F.~Fazel, A.~Grau, H.~Jafarkhani, and F.~D. Flaviis, ``{Space-time-state block
		coded {MIMO} communication systems using reconfigurable antennas},'' {\em
		{IEEE} Trans. Wireless Commun.}, vol.~8, no.~12, pp.~6019--6029, 2009.
	
	\bibitem{loghin2016non}
	N.~S. Loghin, J.~Z{\"o}llner, B.~Mouhouche, D.~Ansorregui, J.~Kim, and S.-I.
	Park, ``{Non-uniform constellations for ATSC 3.0},'' {\em IEEE Trans. Mob.
		Comput.}, vol.~62, no.~1, pp.~197--203, 2016.
	
	\bibitem{fragouli2001turbo}
	C.~Fragouli, R.~D. Wesel, D.~Sommer, and G.~P. Fettweis, ``{Turbo codes with
		non-uniform constellations},'' in {\em Proc. IEEE Int. Commun. Conf. (ICC)},
	vol.~1, pp.~70--73, 2001.
	
	\bibitem{gomez2016mimo}
	D.~G{\'o}mez-Barquero, D.~Vargas, M.~Fuentes, P.~Klenner, S.~Moon, J.-Y. Choi,
	D.~Schneider, and K.~Murayama, ``{MIMO for ATSC 3.0},'' {\em IEEE Trans. Mob.
		Comput.}, vol.~62, no.~1, pp.~298--305, 2016.
	
	\bibitem{ozaydin2022grand}
	B.~Ozaydin, M.~M{\'e}dard, and K.~R. Duffy, ``{GRAND-assisted optimal
		modulation},'' in {\em Proc. IEEE Global Commun. Conf. (GLOBECOM)},
	pp.~813--818, 2022.
	
	\bibitem{abbas2018multi}
	R.~Abbas, M.~Shirvanimoghaddam, Y.~Li, and B.~Vucetic, ``{A multi-layer
		grant-free NOMA scheme for short packet transmissions},'' in {\em Proc. IEEE
		Global Commun. Conf. (GLOBECOM)}, pp.~1--6, 2018.
	
	\bibitem{fuentes2018non}
	M.~Fuentes, L.~Christodoulou, and B.~Mouhouche, ``{Non-uniform constellations
		for broadcast and multicast in {5G} new radio},'' in {\em {Proc. IEEE
			International Symposium on Broadband Multimedia Systems and Broadcasting
			(BMSB)}}, pp.~1--5, 2018.
	
	\bibitem{barrueco2021constellation}
	J.~Barrueco, J.~Montalban, E.~Iradier, and P.~Angueira, ``{Constellation design
		for future communication systems: A comprehensive survey},'' {\em {IEEE
			Access}}, vol.~9, pp.~89778--89797, 2021.
	
	\bibitem{vitthaladevuni2003recursive}
	M.-S. Alouini, ``{A recursive algorithm for the exact BER computation of
		generalized hierarchical QAM constellations},'' {\em {IEEE} Trans. Inf.
		Theory}, vol.~49, no.~1, pp.~297--307, 2003.
	
	\bibitem{xiao2018joint}
	B.~Xiao, K.~Xiao, Z.~Chen, B.~Xia, and H.~Liu, ``{Joint design for modulation
		and constellation labels in multiuser superposition transmission},'' {\em
		IEEE Trans. Mob. Comput.}, vol.~65, no.~2, pp.~245--259, 2018.
	
	\bibitem{yang2020spatially}
	Z.~Yang, Y.~Fang, G.~Han, and K.~M.~S. Huq, ``{Spatially coupled protograph
		LDPC-coded hierarchical modulated BICM-ID systems: A promising transmission
		technique for 6G-enabled Internet of Things},'' {\em IEEE Internet Things
		J.}, vol.~8, no.~7, pp.~5149--5163, 2020.
	
	\bibitem{zhang2016downlink}
	J.~Zhang, X.~Wang, T.~Hasegawa, and T.~Kubo, ``{Downlink non-orthogonal
		multiple access (NOMA) constellation rotation},'' in {\em Proc. IEEE Veh.
		Technol. Conf. (VTC Fall)}, pp.~1--5, 2016.
	
	\bibitem{ye2017constellation}
	N.~Ye, A.~Wang, X.~Li, W.~Liu, X.~Hou, and H.~Yu, ``{On constellation rotation
		of NOMA with SIC receiver},'' {\em {IEEE} Commun. Lett.}, vol.~22, no.~3,
	pp.~514--517, 2017.
	
	\bibitem{khorov2022phase}
	E.~Khorov, A.~Kureev, I.~Levitsky, and I.~F. Akyildiz, ``{A phase noise
		resistant constellation rotation method and its experimental validation for
		NOMA Wi-Fi},'' {\em {IEEE} J. Sel. Areas Commun.}, vol.~40, no.~4,
	pp.~1346--1354, 2022.
	
	\bibitem{Baker}
	{P. A. Baker}, ``{Phase-modulation data sets for serial transmission at 2000
		and 2400 bits per second},'' {\em {Trans. Am. Inst. Electr. Eng., Part I:
			Commun. Electron.}}, vol.~81, pp.~166--171, July 1962.
	
	\bibitem{1193802}
	H.~Jafarkhani and N.~Seshadri, ``Super-orthogonal space-time trellis codes,''
	vol.~49, no.~4, pp.~937--950, 2003.
	
	\bibitem{1381439}
	H.~Jafarkhani and N.~Hassanpour, ``Super-quasi-orthogonal space-time trellis
	codes for four transmit antennas,'' vol.~4, no.~1, pp.~215--227, 2005.
	
	\bibitem{1545875}
	Y.~Zhu and H.~Jafarkhani, ``Differential modulation based on quasi-orthogonal
	codes,'' vol.~4, no.~6, pp.~3005--3017, 2005.
	
	\bibitem{stierstorfer2010optimizing}
	C.~Stierstorfer, R.~F. Fischer, and J.~B. Huber, ``{Optimizing BICM with
		convolutional codes for transmission over the AWGN channel},'' in {\em Proc.
		Int. Zurich Sem. Commun.}, 2010.
	
	\bibitem{yang2021protograph}
	Z.~Yang, Y.~Fang, Y.~Cheng, P.~Chen, and D.~J. Almakhles, ``{Protograph
		LDPC-coded BICM-ID with irregular mapping: An emerging transmission technique
		for massive Internet of Things},'' {\em IEEE Trans. Green Commun. Netw.},
	vol.~5, no.~3, pp.~1051--1065, 2021.
	
	\bibitem{rowshan2024channel}
	M.~Rowshan, M.~Qiu, Y.~Xie, X.~Gu, and J.~Yuan, ``Channel coding towards {6G:
		T}echnical overview and outlook,'' {\em IEEE Open J. Commun. Soc.}, vol.~5,
	no.~4, pp.~2585-- 2685, 2024.
	
	\bibitem{blahut2003algebraic}
	R.~E. Blahut, {\em {Algebraic Codes for Data Transmission}}.
	\newblock New York: Cambridge Univ. Press, 2003.
	
	\bibitem{viterbi1971convolutional}
	A.~Viterbi, ``Convolutional codes and their performance in communication
	systems,'' {\em IEEE Trans. Commun. Technol.}, vol.~19, no.~5, pp.~751--772,
	1971.
	
	\bibitem{berrou1993near}
	C.~Berrou, A.~Glavieux, and P.~Thitimajshima, ``{Near Shannon limit
		error-correcting coding and decoding: Turbo-codes. 1},'' in {\em Proc. IEEE
		Int. Commun. Conf. (ICC)}, vol.~2, pp.~1064--1070, 1993.
	
	\bibitem{gallager1962low}
	R.~Gallager, ``Low-density parity-check codes,'' {\em IRE Trans. Inf. Theory.},
	vol.~8, no.~1, pp.~21--28, 1962.
	
	\bibitem{shokrollahi2004ldpc}
	A.~Shokrollahi, ``{LDPC codes: An introduction},'' in {\em Coding, Cryptography
		and Combinatorics}, pp.~85--110, Springer, 2004.
	
	\bibitem{arikan2009channel}
	E.~Arikan, ``{Channel polarization: A method for constructing
		capacity-achieving codes for symmetric binary-input memoryless channels},''
	{\em {IEEE} Trans. Inf. Theory}, vol.~55, no.~7, pp.~3051--3073, 2009.
	
	\bibitem{lin20215g}
	X.~Lin and N.~Lee, ``{5G and Beyond},'' {\em Cham, Switzerland: Springer Nature
		Switzerland AG}, 2021.
	
	\bibitem{3GPPNR}
	``{NR; Physical layer procedures for data (Version 18.2.0)},'' {\em 3GPP TS
		38.214}, 2024.
	\newblock
	\url{https://portal.3gpp.org/desktopmodules/Specifications/SpecificationDetails.aspx?specificationId=3216}.
	
	\bibitem{popovski2022perspective}
	P.~Popovski, F.~Chiariotti, K.~Huang, A.~E. Kal{\o}r, M.~Kountouris, N.~Pappas,
	and B.~Soret, ``{A perspective on time toward wireless 6G},'' {\em Proc.
		IEEE}, vol.~110, no.~8, pp.~1116--1146, 2022.
	
	\bibitem{fantacci2021end}
	R.~Fantacci and B.~Picano, ``{End-to-end delay bound for wireless uVR services
		over 6G terahertz communications},'' {\em IEEE Internet Things J.}, vol.~8,
	no.~23, pp.~17090--17099, 2021.
	
	\bibitem{arikan2019sequential}
	E.~Ar{\i}kan, ``From sequential decoding to channel polarization and back
	again,'' {\em arXiv preprint arXiv:1908.09594}, 2019.
	
	\bibitem{moradi2020performance}
	M.~Moradi, A.~Mozammel, K.~Qin, and E.~Arikan, ``{Performance and complexity of
		sequential decoding of PAC codes},'' {\em arXiv preprint arXiv:2012.04990},
	2020.
	
	\bibitem{choi2024deep}
	G.~Choi and N.~Lee, ``Deep polar codes,'' {\em {IEEE} Trans. Commun.},
	vol.~72,, 2024.
	
	\bibitem{shirvanimoghaddam2018short}
	M.~Shirvanimoghaddam, M.~S. Mohammadi, R.~Abbas, A.~Minja, C.~Yue, B.~Matuz,
	G.~Han, Z.~Lin, W.~Liu, Y.~Li, {\em et~al.}, ``Short block-length codes for
	ultra-reliable low latency communications,'' {\em {IEEE} Commun. Mag.},
	vol.~57, no.~2, pp.~130--137, 2018.
	
	\bibitem{persson2011joint}
	D.~Persson, J.~Kron, M.~Skoglund, and E.~G. Larsson, ``{Joint source-channel
		coding for the MIMO broadcast channel},'' {\em {IEEE} Trans. Signal
		Process.}, vol.~60, no.~4, pp.~2085--2090, 2011.
	
	\bibitem{vaezi2014distributed}
	M.~Vaezi and F.~Labeau, ``{Distributed source-channel coding based on
		real-field BCH codes},'' {\em {IEEE} Trans. Signal Process.}, vol.~62, no.~5,
	pp.~1171--1184, 2014.
	
	\bibitem{yilmaz2023distributed}
	S.~F. Yilmaz, C.~Karamanl{\i}, and D.~G{\"u}nd{\"u}z, ``Distributed deep joint
	source-channel coding over a multiple access channel,'' in {\em Proc. IEEE
		Int. Commun. Conf. (ICC)}, pp.~1400--1405, 2023.
	
	\bibitem{ungerboeck1987trellis}
	G.~Ungerboeck, ``{Trellis-coded modulation with redundant signal sets Part {I}:
		Introduction},'' {\em {IEEE} Trans. Commun. Technol.}, vol.~25, pp.~5--11,
	Feb. 1987.
	
	\bibitem{765488}
	T.~Aulin and R.~Espineira, ``{Trellis coded multiple access (TCMA)},'' in {\em
		Proc. IEEE Int. Commun. Conf. (ICC)}, vol.~2, pp.~1177--1181, 1999.
	
	\bibitem{XZMGHJwcl}
	X.~Zou, M.~Ganji, and H.~Jafarkhani, ``{Trellis-coded non-orthogonal multiple
		access},'' vol.~9, pp.~538--542, Apr. 2020.
	
	\bibitem{jafarkhani1999multiple}
	H.~Jafarkhani and V.~Tarokh, ``{Multiple description trellis-coded
		quantization},'' {\em {IEEE} Trans. Commun.}, vol.~47, pp.~799 -- 803, June
	1999.
	
	\bibitem{jafarkhani1999design}
	H.~Jafarkhani and V.~Tarokh, ``{Design of successively refinable trellis-coded
		quantizers},'' {\em {IEEE} Trans. Inf. Theory}, vol.~45, pp.~1490--1497, July
	1999.
	
	\bibitem{goodfellow2016deep}
	I.~Goodfellow, Y.~Bengio, A.~Courville, and Y.~Bengio, {\em Deep learning},
	vol.~1.
	\newblock MIT press Cambridge, 2016.
	
	\bibitem{zhang2021multi}
	X.~Zhang and M.~Vaezi, ``{Multi-objective DNN-based precoder for MIMO
		communications},'' {\em {IEEE} Trans. Commun.}, vol.~69, no.~7,
	pp.~4476--4488, 2021.
	
	\bibitem{lin2022overview}
	X.~Lin, ``{An overview of 5G advanced evolution in 3GPP release 18},'' {\em
		{IEEE Commun. Stand. Mag.}}, vol.~6, no.~3, pp.~77--83, 2022.
	
	\bibitem{o2018over}
	T.~J. O’Shea, T.~Roy, and T.~C. Clancy, ``{Over-the-air deep learning based
		radio signal classification},'' {\em IEEE J. Sel. Topic Signal Process.},
	vol.~12, no.~1, pp.~168--179, 2018.
	
	\bibitem{peng2021survey}
	S.~Peng, S.~Sun, and Y.-D. Yao, ``{A survey of modulation classification using
		deep learning: Signal representation and data preprocessing},'' {\em {IEEE
			Trans. Neural Netw. Learn. Syst.}}, vol.~33, no.~12, pp.~7020--7038, 2021.
	
	\bibitem{tekbiyik2020robust}
	K.~Tekb{\i}y{\i}k, A.~R. Ekti, A.~G{\"o}r{\c{c}}in, G.~K. Kurt, and
	C.~Ke{\c{c}}eci, ``{Robust and fast automatic modulation classification with
		CNN under multipath fading channels},'' in {\em Proc. IEEE Veh. Technol.
		Conf. (VTC Spring)}, pp.~1--6, 2020.
	
	\bibitem{wang2023strategies}
	R.~Wang, Y.~Qi, M.~Vaezi, X.~Jiao, and M.~Amin, ``{Strategies for enhanced
		signal modulation classifications under unknown symbol rates and noise
		conditions},'' in {\em Proc. IEEE Int. Acoust. Speech Signal Process.
		(ICASSP)}, pp.~1--5, 2023.
	
	\bibitem{sohrabi2018one}
	F.~Sohrabi, Y.-F. Liu, and W.~Yu, ``{One-bit precoding and constellation range
		design for massive MIMO with QAM signaling},'' {\em IEEE J. Sel. Topic Signal
		Process.}, vol.~12, no.~3, pp.~557--570, 2018.
	
	\bibitem{han2021deep}
	M.~Han, H.~Seo, A.~T. Abebe, and C.~G. Kang, ``{Deep learning-based codebook
		design for code-domain non-orthogonal multiple access: Approaching
		single-user bit-error rate performance},'' {\em IEEE Trans. Cogn. Commun.
		Netw.}, vol.~8, no.~2, pp.~1159--1173, 2021.
	
	\bibitem{lopez2020survey}
	M.~J. L{\'o}pez-Morales, K.~Chen-Hu, and A.~G. Armada, ``{A survey about deep
		learning for constellation design in communications},'' in {\em {Proc. IEEE
			Int. Symp. Communication Syst. Netw. Digital Signal Process. (CSNDSP)}},
	pp.~1--5, 2020.
	
	\bibitem{ma2021joint}
	Z.~Ma, W.~Wu, M.~Jian, F.~Gao, and X.~Shen, ``{Joint constellation design and
		multiuser detection for grant-free NOMA},'' {\em {IEEE} Trans. Wireless
		Commun.}, vol.~21, no.~3, pp.~1973--1988, 2021.
	
	\bibitem{madadi2023ai}
	P.~Madadi, J.~Cho, C.~J. Zhang, and D.~Burghal, ``{AI/ML optimized high-order
		modulations},'' in {\em Proc. IEEE Int. Commun. Conf. (ICC)}, pp.~6373--6378,
	2023.
	
	\bibitem{zhang2023deep}
	X.~Zhang and M.~Vaezi, ``{Deep autoencoder-based Z-interference channels with
		perfect and imperfect CSI},'' {\em {IEEE} Trans. Commun.}, vol.~72, no.~2,
	pp.~861--873, 2024.
	
	\bibitem{nartasilpa2018communications}
	N.~Nartasilpa, A.~Salim, D.~Tuninetti, and N.~Devroye, ``Communications system
	performance and design in the presence of radar interference,'' {\em {IEEE}
		Trans. Commun.}, vol.~66, no.~9, pp.~4170--4185, 2018.
	
	\bibitem{deshpande2009constellation}
	N.~Deshpande and B.~S. Rajan, ``{Constellation constrained capacity of two-user
		broadcast channels},'' in {\em Proc. IEEE Global Commun. Conf. (GLOBECOM)},
	pp.~1--6, 2009.
	
	\bibitem{islam2016power}
	S.~R. Islam, N.~Avazov, O.~A. Dobre, and K.-S. Kwak, ``{Power-domain
		non-orthogonal multiple access (NOMA) in 5G systems: Potentials and
		challenges},'' {\em IEEE Commun. Surv. Tut.}, vol.~19, no.~2, pp.~721--742,
	2016.
	
	\bibitem{qiu2019downlink}
	M.~Qiu, Y.-C. Huang, and J.~Yuan, ``{Downlink non-orthogonal multiple access
		without SIC for block fading channels: An algebraic rotation approach},''
	{\em {IEEE Trans. on Wireless Commun.}}, vol.~18, no.~8, pp.~3903--3918,
	2019.
	
	\bibitem{baldi2012autoencoders}
	P.~Baldi, ``{Autoencoders, unsupervised learning, and deep architectures},'' in
	{\em {Proc. ICML Workshop on Unsupervised and Transfer Learning}},
	pp.~37--49, 2012.
	
	\bibitem{alberge2018constellation}
	F.~Alberge, ``Constellation design with deep learning for downlink
	non-orthogonal multiple access,'' in {\em Proc. Int. Symp. Pers. Indoor Mob.
		Radio Commun. (PIMRC)}, pp.~1--5, 2018.
	
	\bibitem{ninkovic2023weighted}
	V.~Ninkovic, D.~Vukobratovic, A.~Pastore, and C.~Anton-Haro, ``A weighted
	autoencoder-based approach to downlink {NOMA} constellation design,'' in {\em
		Proc. IEEE Int. Works. Signal Process. Adv. Wireless Commun. (SPAWC)},
	pp.~1--6, 2023.
	
	\bibitem{Aboutaleb2014NOMA}
	A.~Aboutaleb, M.~Torabi, B.~Belzer, and K.~Sivakuamar, ``Deep learning-based
	auto-encoder for time-offset sub-faster-than-{Nyquist downlink NOMA} with
	timing errors and imperfect {CSI},'' {\em IEEE J. Sel. Topic Signal
		Process.}, 2024.
	
	\bibitem{haykin2008communication}
	S.~Haykin, {\em Communication Systems}.
	\newblock John Wiley \& Sons, 2008.
	
	\bibitem{yahya2021exact}
	H.~Yahya, E.~Alsusa, and A.~Al-Dweik, ``{Exact BER analysis of NOMA with
		arbitrary number of users and modulation orders},'' {\em {IEEE} Trans.
		Commun.}, vol.~69, no.~9, pp.~6330--6344, 2021.
	
	\bibitem{pradhan2003distributed}
	S.~S. Pradhan and K.~Ramchandran, ``{Distributed source coding using syndromes
		(DISCUS): Design and construction},'' {\em {IEEE} Trans. Inf. Theory},
	vol.~49, no.~3, pp.~626--643, 2003.
	
	\bibitem{el2013practical}
	O.~El~Ayach, S.~W. Peters, and R.~W. Heath, ``{The practical challenges of
		interference alignment},'' {\em {IEEE} Wireless Commun.}, vol.~20, no.~1,
	pp.~35--42, 2013.
	
	\bibitem{sun2013interference}
	S.~Sun, Q.~Gao, Y.~Peng, Y.~Wang, and L.~Song, ``{Interference management
		through CoMP in 3GPP LTE-advanced networks},'' {\em {IEEE} Wireless Commun.},
	vol.~20, no.~1, pp.~59--66, 2013.
	
	\bibitem{vaezi2023drl}
	M.~Vaezi, X.~Lin, H.~Zhang, W.~Saad, and H.~V. Poor, ``{Deep reinforcement
		learning for interference management in {UAV}-based {3D} networks:
		{Potentials} and challenges},'' {\em {IEEE} Commun. Mag.}, vol.~62, no.~2,
	pp.~134--140, 2024.
	
	\bibitem{shin2016coordinated}
	W.~Shin, M.~Vaezi, B.~Lee, D.~Love, J.~Lee, and H.~Poor, ``Coordinated
	beamforming for multi-cell {MIMO-NOMA},'' {\em {IEEE} Commun. Lett.},
	vol.~21, no.~1, pp.~84--87, 2016.
	
	\bibitem{shin2017non}
	W.~Shin, M.~Vaezi, B.~Lee, D.~Love, J.~Lee, and H.~Poor, ``{Non-orthogonal
		multiple access in multi-cell networks: Theory, performance, and practical
		challenges},'' {\em {{IEEE} Commun. Mag.}}, vol.~55, no.~10, pp.~176--183,
	2017.
	
	\bibitem{guo2021qos}
	F.~Guo, H.~Lu, X.~Jiang, M.~Zhang, J.~Wu, and C.~W. Chen, ``{QoS-aware user
		grouping strategy for downlink multi-cell NOMA systems},'' {\em {IEEE} Trans.
		Wireless Commun.}, vol.~20, no.~12, pp.~7871--7887, 2021.
	
	\bibitem{rezvani2021optimal}
	S.~Rezvani, E.~A. Jorswieck, N.~M. Yamchi, and M.~R. Javan, ``{Optimal SIC
		ordering and power allocation in downlink multi-cell NOMA systems},'' {\em
		{IEEE} Trans. Wireless Commun.}, vol.~21, no.~6, pp.~3553--3569, 2021.
	
	\bibitem{sung2021distributed}
	C.~W. Sung, Y.~Chen, and Y.~Gu, ``{Distributed dual optimization for the uplink
		of multi-cell NOMA},'' {\em {IEEE} Trans. Commun.}, vol.~69, no.~5,
	pp.~3135--3146, 2021.
	
	\bibitem{erpek2018learning}
	T.~Erpek, T.~J. O'Shea, and T.~C. Clancy, ``{Learning a physical layer scheme
		for the MIMO interference channel},'' in {\em Proc. IEEE Int. Commun. Conf.
		(ICC)}, pp.~1--5, 2018.
	
	\bibitem{vishwakarma2018mitigation}
	S.~Vishwakarma, V.~Ummalaneni, M.~S. Iqbal, A.~Majumdar, and S.~S. Ram,
	``{Mitigation of through-wall interference in radar images using denoising
		autoencoders},'' in {\em {Proc. IEEE Radar Conf.}}, pp.~1543--1548, 2018.
	
	\bibitem{wu2020deep}
	D.~Wu, M.~Nekovee, and Y.~Wang, ``{Deep learning-based autoencoder for m-User
		wireless interference channel physical layer design},'' {\em {IEEE Access}},
	vol.~8, pp.~174679--174691, 2020.
	
	\bibitem{zhang2022daezic}
	X.~Zhang and M.~Vaezi, ``Deep autoencoder-based {Z}-interference channels,'' in
	{\em Proc. IEEE Wireless Commun. Net. Conf. (WCNC)}, pp.~1--6, 2023.
	
	\bibitem{zhang2023interference}
	X.~Zhang, M.~Vaezi, and L.~Zheng, ``{Interference-aware constellation design
		for Z-interference channels with imperfect CSI},'' in {\em Proc. IEEE Int.
		Commun. Conf. (ICC)}, pp.~1--6, 2023.
	
	\bibitem{vaezi2016simplified}
	M.~Vaezi and H.~V. Poor, ``{Simplified Han-Kobayashi region for one-sided and
		mixed Gaussian interference channels},'' in {\em Proc. IEEE Int. Commun.
		Conf. (ICC)}, pp.~1--6, 2016.
	
	\bibitem{alamouti1998simple}
	S.~M. Alamouti, ``A simple transmit diversity technique for wireless
	communications,'' {\em {IEEE} J. Sel. Areas Commun.}, vol.~16, no.~8,
	pp.~1451--1458, 1998.
	
	\bibitem{VTHJRC1999}
	V.~Tarokh, H.~Jafarkhani, and A.~Calderbank, ``Space-time block coding for
	wireless communications: performance results,'' {\em {IEEE} J. Sel. Areas
		Commun.}, vol.~17, no.~3, pp.~451--460, 1999.
	
	\bibitem{VTHJRC99}
	V.~Tarokh, H.~Jafarkhani, and A.~Calderbank, ``Space-time block codes from
	orthogonal designs,'' vol.~45, no.~5, pp.~1456--1467, 1999.
	
	\bibitem{cox2014introduction}
	C.~Cox, {\em {An Introduction to LTE: LTE, LTE-Advanced, SAE, VoLTE and 4G
			Mobile Communications}}.
	\newblock John Wiley \& Sons, 2014.
	
	\bibitem{857917}
	V.~Tarokh and H.~Jafarkhani, ``A differential detection scheme for transmit
	diversity,'' {\em {IEEE} J. Sel. Areas Commun.}, vol.~18, no.~7,
	pp.~1169--1174, 2000.
	
	\bibitem{945280}
	H.~Jafarkhani and V.~Tarokh, ``Multiple transmit antenna differential detection
	from generalized orthogonal designs,'' vol.~47, no.~6, pp.~2626--2631, 2001.
	
	\bibitem{9363693}
	``{IEEE standard for information technology--Telecommunications and information
		exchange between systems - Local and metropolitan area networks--Specific
		requirements - Part 11: Wireless LAN medium access control (MAC) and physical
		layer (PHY) specifications},'' {\em {IEEE Std 802.11-2020 (Revision of IEEE
			Std 802.11-2016)}}, pp.~1--4379, 2021.
	
	\bibitem{3GPPrelease15}
	3GPP, ``{5G NR Physical layer procedures for data (Version 15.3.0 Release
		15}),'' {\em TS 138 214}, 2018.
	\newblock
	\url{https://www.etsi.org/deliver/etsi_ts/138200_138299/138214/15.03.00_60/ts_138214v150300p.pdf}.
	
	\bibitem{xiao2011globally}
	C.~Xiao, Y.~R. Zheng, and Z.~Ding, ``{Globally optimal linear precoders for
		finite alphabet signals over complex vector {Gaussian} channels},'' {\em
		{IEEE} Trans. Signal Process.}, vol.~59, no.~7, pp.~3301--3314, 2011.
	
	\bibitem{zhou2005mimo}
	Z.~Zhou, B.~Vucetic, M.~Dohler, and Y.~Li, ``{{MIMO} systems with adaptive
		modulation},'' {\em {IEEE} Trans. Veh. Technol.}, vol.~54, no.~5,
	pp.~1828--1842, 2005.
	
	\bibitem{zhou2010adaptive}
	Z.~Zhou and B.~Vucetic, ``{Adaptive coded {MIMO} systems with near full
		multiplexing gain using outdated {CSI}},'' {\em {IEEE} Trans. Wireless
		Commun.}, vol.~10, no.~1, pp.~294--302, 2010.
	
	\bibitem{xia2020note}
	J.~Xia, D.~Deng, and D.~Fan, ``{A note on implementation methodologies of deep
		learning-based signal detection for conventional {MIMO} transmitters},'' {\em
		IEEE Trans. Mob. Comput.}, vol.~66, no.~3, pp.~744--745, 2020.
	
	\bibitem{prabhu2010energy}
	R.~S. Prabhu and B.~Daneshrad, ``{Energy-efficient power loading for a
		{MIMO-SVD} system and its performance in flat fading},'' in {\em Proc. IEEE
		Global Commun. Conf. (GLOBECOM)}, pp.~1--5, 2010.
	
	\bibitem{ding2015application}
	Z.~Ding, F.~Adachi, and H.~V. Poor, ``{The application of MIMO to
		non-orthogonal multiple access},'' {\em {IEEE Trans. Wireless Commun.}},
	vol.~15, no.~1, pp.~537--552, 2016.
	
	\bibitem{7434594}
	Z.~Ding and H.~V. Poor, ``{Design of massive-MIMO-NOMA With limited
		feedback},'' {\em {IEEE} Signal Process. Lett.}, vol.~23, no.~5,
	pp.~629--633, 2016.
	
	\bibitem{weingartens2006capacity}
	H.~Weingarten, Y.~Steinberg, and S.~S. Shamai, ``The capacity region of the
	{Gaussian} multiple-input multiple-output broadcast channel,'' {\em {IEEE}
		Trans. Inf. Theory}, vol.~52, no.~9, pp.~3936--3964, 2006.
	
	\bibitem{liu2010multiple}
	R.~Liu, T.~Liu, H.~V. Poor, and S.~Shamai, ``{Multiple-input multiple-output
		Gaussian broadcast channels with confidential messages},'' {\em {IEEE} Trans.
		Inf. Theory}, vol.~56, no.~9, pp.~4215--4227, 2010.
	
	\bibitem{ekrem2012degraded}
	E.~Ekrem and S.~Ulukus, ``{Degraded compound multi-receiver wiretap
		channels},'' {\em {IEEE} Trans. Inf. Theory}, vol.~58, no.~9, pp.~5681--5698,
	2012.
	
	\bibitem{qi2022signaling}
	Y.~Qi and M.~Vaezi, ``{Signaling design for MIMO-NOMA with different security
		requirements},'' {\em {IEEE} Trans. Signal Process.}, vol.~70, no.~3,
	pp.~1389--1401, 2022.
	
	\bibitem{qi2023k}
	Y.~Qi, M.~Vaezi, and H.~V. Poor, ``{K-receiver wiretap channel: Optimal
		encoding order and signaling design},'' {\em {IEEE} Trans. Wireless Commun.},
	vol.~22, no.~12, pp.~8575--8586, 2023.
	
	\bibitem{pauls2021secure}
	J.~Pauls and M.~Vaezi, ``{Secure precoding in MIMO-NOMA: A deep learning
		approach},'' {\em {IEEE} Wireless Commun. Lett.}, vol.~11, no.~1, pp.~77--80,
	2021.
	
	\bibitem{fakoorian2013optimality}
	S.~A.~A. Fakoorian and A.~L. Swindlehurst, ``{On the optimality of linear
		precoding for secrecy in the MIMO broadcast channel},'' {\em {IEEE} J. Sel.
		Areas Commun.}, vol.~31, no.~9, pp.~1701--1713, 2013.
	
	\bibitem{shi2008rate}
	S.~Shi, M.~Schubert, and H.~Boche, ``{Rate optimization for multiuser MIMO
		systems with linear processing},'' {\em {IEEE} Trans. Signal Process.},
	vol.~56, no.~8, pp.~4020--4030, 2008.
	
	\bibitem{park2015weighted}
	D.~Park, ``Weighted sum rate maximization of {MIMO} broadcast and interference
	channels with confidential messages,'' {\em {IEEE} Trans. Wireless Commun.},
	vol.~15, no.~3, pp.~1742--1753, 2015.
	
	\bibitem{chen2016optimal}
	Z.~Chen, Z.~Ding, P.~Xu, and X.~Dai, ``{Optimal precoding for a QoS
		optimization problem in two-user MISO-NOMA downlink},'' {\em {IEEE} Commun.
		Lett.}, vol.~20, no.~6, pp.~1263--1266, 2016.
	
	\bibitem{sun2023application}
	Y.~Sun, Z.~Ding, X.~Dai, M.~Zhou, and Z.~Ding, ``{On the application of
		quasi-degradation to network NOMA in downlink CoMP systems},'' {\em {IEEE}
		Trans. Wireless Commun.}, vol.~23, no.~2, pp.~978--993, 2024.
	
	\bibitem{goldsmith2003capacity}
	A.~Goldsmith, S.~A. Jafar, N.~Jindal, and S.~Vishwanath, ``Capacity limits of
	{MIMO} channels,'' {\em {IEEE} J. Sel. Areas Commun.}, vol.~21, no.~5,
	pp.~684--702, 2003.
	
	\bibitem{krishnamoorthy2021uplink}
	A.~Krishnamoorthy and R.~Schober, ``{Uplink and downlink MIMO-NOMA with
		simultaneous triangularization},'' {\em {IEEE} Trans. Wireless Commun.},
	vol.~20, no.~6, pp.~3381--3396, 2021.
	
	\bibitem{chi2018practical}
	Y.~Chi, L.~Liu, G.~Song, C.~Yuen, Y.~L. Guan, and Y.~Li, ``{Practical
		MIMO-NOMA: Low complexity and capacity-approaching solution},'' {\em {IEEE}
		Trans. Wireless Commun.}, vol.~17, no.~9, pp.~6251--6264, 2018.
	
	\bibitem{liu2019capacity}
	L.~Liu, Y.~Chi, C.~Yuen, Y.~L. Guan, and Y.~Li, ``{Capacity-achieving
		MIMO-NOMA: iterative LMMSE detection},'' {\em {IEEE} Trans. Signal Process.},
	vol.~67, no.~7, pp.~1758--1773, 2019.
	
	\bibitem{mao2018energy}
	Y.~Mao, B.~Clerckx, and V.~O. Li in {\em {Proc. IEEE Int. Symp. Wireless
			Commun. Syst. (ISWCS)}, title={{Energy efficiency of rate-splitting multiple
				access, and performance benefits over SDMA and NOMA}}, year={2018},
		pages={1-5}}.
	
	\bibitem{Carleial1978interference}
	A.~Carleial, ``{Interference channels},'' {\em {IEEE} Trans. Inf. Theory},
	vol.~24, no.~1, pp.~60--70, 1978.
	
	\bibitem{rimoldi1996rate}
	B.~Rimoldi and R.~Urbanke, ``{A rate-splitting approach to the Gaussian
		multiple-access channel},'' {\em {IEEE} Trans. Inf. Theory}, vol.~42, no.~2,
	pp.~364--375, 1996.
	
	\bibitem{hao2015rate}
	C.~Hao, Y.~Wu, and B.~Clerckx, ``{Rate analysis of two-receiver MISO broadcast
		channel with finite rate feedback: A rate-splitting approach},'' {\em {IEEE}
		Trans. Commun.}, vol.~63, no.~9, pp.~3232--3246, 2015.
	
	\bibitem{dai2016rate}
	M.~Dai, B.~Clerckx, D.~Gesbert, and G.~Caire, ``{A rate splitting strategy for
		massive MIMO with imperfect CSIT},'' {\em {IEEE} Trans. Wireless Commun.},
	vol.~15, no.~7, pp.~4611--4624, 2016.
	
	\bibitem{clerckx2016rate}
	B.~Clerckx, H.~Joudeh, C.~Hao, M.~Dai, and B.~Rassouli, ``{Rate splitting for
		MIMO wireless networks: A promising PHY-layer strategy for LTE evolution},''
	{\em {IEEE} Commun. Mag.}, vol.~54, no.~5, pp.~98--105, 2016.
	
	\bibitem{zhou2021rate}
	G.~Zhou, Y.~Mao, and B.~Clerckx, ``{Rate-splitting multiple access for
		multi-antenna downlink communication systems: Spectral and energy efficiency
		tradeoff},'' vol.~21, no.~7, pp.~4816--4828, 2022.
	
	\bibitem{matthiesen2021globally}
	B.~Matthiesen, Y.~Mao, P.~Popovski, and B.~Clerckx, ``{Globally optimal
		beamforming for rate splitting multiple access},'' in {\em Proc. IEEE Int.
		Acoust. Speech Signal Process. (ICASSP)}, pp.~4775--4779, 2021.
	
	\bibitem{li2020rate}
	Z.~Li, C.~Ye, Y.~Cui, S.~Yang, and S.~Shamai {\em {IEEE} J. Sel. Areas Commun.}
	
	\bibitem{mao2020beyond}
	Y.~Mao and B.~Clerckx, ``{Beyond dirty paper coding for multi-antenna broadcast
		channel with partial CSIT: A rate-splitting approach},'' {\em {IEEE} Trans.
		Commun.}, vol.~68, no.~11, pp.~6775--6791, 2020.
	
	\bibitem{medra2018robust}
	M.~Medra and T.~N. Davidson, ``{Robust downlink transmission: An offset-based
		single-rate-splitting approach},'' in {\em Proc. IEEE Int. Works. Signal
		Process. Adv. Wireless Commun. (SPAWC)}, pp.~1--5, 2018.
	
	\bibitem{dizdar2021rate}
	O.~Dizdar, Y.~Mao, Y.~Xu, P.~Zhu, and B.~Clerckx, ``{Rate-splitting multiple
		access for enhanced URLLC and eMBB in 6G},'' in {\em {Proc. IEEE Int. Symp.
			Wireless Commun. Syst. (ISWCS)}}, pp.~1--6, 2021.
	
	\bibitem{xu2022rate}
	Y.~Xu, Y.~Mao, O.~Dizdar, and B.~Clerckx, ``{Rate-splitting multiple access
		with finite blocklength for short-packet and low-latency downlink
		communications},'' {\em {IEEE} Trans. Veh. Technol.}, vol.~71, no.~11,
	pp.~12333--12337, 2022.
	
	\bibitem{park2022rate}
	J.~Park, J.~Choi, N.~Lee, W.~Shin, and H.~V. Poor, ``{Rate-splitting multiple
		access for downlink MIMO: A generalized power iteration approach},'' {\em
		{IEEE} Trans. Wireless Commun.}, vol.~22, no.~3, pp.~1588--1603, 2022.
	
	\bibitem{flores2020linear}
	A.~R. Flores, R.~C. de~Lamare, and B.~Clerckx, ``{Linear precoding and stream
		combining for rate splitting in multiuser MIMO systems},'' {\em {IEEE}
		Commun. Lett.}, vol.~24, no.~4, pp.~890--894, 2020.
	
	\bibitem{krishnamoorthy2022downlink}
	A.~Krishnamoorthy and R.~Schober, ``{Downlink MIMO-RSMA with successive
		null-space precoding},'' {\em {IEEE} Trans. Wireless Commun.}, vol.~21,
	no.~11, pp.~9170--9185, 2022.
	
	\bibitem{mishra2021rate}
	A.~Mishra, Y.~Mao, O.~Dizdar, and B.~Clerckx, ``{Rate-splitting multiple access
		for downlink multiuser MIMO: Precoder optimization and PHY-layer design},''
	{\em {IEEE} Trans. Commun.}, vol.~70, no.~2, pp.~874--890, 2021.
	
	\bibitem{diab2022precoding}
	R.~Diab, A.~Krishnamoorthy, and R.~Schober, ``{Precoding and decoding schemes
		for downlink MIMO-RSMA with simultaneous diagonalization and user
		exclusion},'' in {\em Proc. IEEE Int. Commun. Conf. Workshops (ICC Wkshps)},
	pp.~586--591, 2022.
	
	\bibitem{khamidullina2023rate}
	L.~Khamidullina, A.~L. de~Almeida, and M.~Haardt, ``{Rate splitting and
		precoding strategies for multi-user MIMO broadcast channels with common and
		private streams},'' in {\em Proc. IEEE Int. Acoust. Speech Signal Process.
		(ICASSP)}, pp.~1--5, 2023.
	
	\bibitem{jiang2023rate}
	H.~Jiang, L.~You, A.~Elzanaty, J.~Wang, W.~Wang, X.~Gao, and M.-S. Alouini,
	``{Rate-splitting multiple access for uplink massive MIMO with
		electromagnetic exposure constraints},'' {\em {IEEE} J. Sel. Areas Commun.}
	
	\bibitem{csahin2023multicarrier}
	M.~M. Şahin, O.~Dizdar, B.~Clerckx, and H.~Arslan, ``{Multicarrier
		rate-splitting multiple access: Superiority of OFDM-RSMA over OFDMA and
		OFDM-NOMA},'' {\em {IEEE} Commun. Lett.}, vol.~27, no.~11, pp.~3088--3092,
	2023.
	
	\bibitem{chen2023joint}
	Z.~Chen, J.~Wang, Z.~Tian, M.~Wang, Y.~Jia, and T.~Q.~S. Quek, ``{Joint rate
		splitting and beamforming design for RSMA-RIS-assisted ISAC system},'' {\em
		{IEEE} Wireless Commun. Lett.}, pp.~1--1, 2023.
	
	\bibitem{liu2023risac}
	Z.~Liu, Y.~Jint, B.~Cao, and R.~Lu, ``{RISAC: Rate-splitting multiple access
		enabled integrated sensing and communication systems},'' in {\em Proc. IEEE
		Int. Commun. Conf. (ICC)}, pp.~6449--6454, 2023.
	
	\bibitem{hu2023joint}
	C.~Hu, Y.~Fang, and L.~Qiu, ``{Joint transmit and receive beamforming design
		for uplink RSMA enabled integrated sensing and communication systems},'' in
	{\em Proc. IEEE Wireless Commun. Net. Conf. (WCNC)}, pp.~1--6, 2023.
	
	\bibitem{xu2021rate}
	C.~Xu, B.~Clerckx, S.~Chen, Y.~Mao, and J.~Zhang, ``{Rate-splitting multiple
		access for multi-antenna joint radar and communications},'' {\em IEEE J. Sel.
		Topic Signal Process.}, vol.~15, no.~6, pp.~1332--1347, 2021.
	
	\bibitem{yang2020energy}
	Z.~Yang, J.~Shi, Z.~Li, M.~Chen, W.~Xu, and M.~Shikh-Bahaei, ``{Energy
		efficient rate splitting multiple access (RSMA) with reconfigurable
		intelligent surface},'' in {\em Proc. IEEE Int. Commun. Conf. (ICC)},
	pp.~1--6, 2020.
	
	\bibitem{zhang2023energy}
	R.~Zhang, K.~Xiong, Y.~Lu, P.~Fan, D.~W.~K. Ng, and K.~B. Letaief, ``{Energy
		efficiency maximization in RIS-assisted SWIPT networks with RSMA: A PPO-based
		approach},'' {\em {IEEE} J. Sel. Areas Commun.}, vol.~41, no.~5,
	pp.~1413--1430, 2023.
	
	\bibitem{pala2023spectral}
	S.~Pala, M.~Katwe, K.~Singh, B.~Clerckx, and C.-P. Li, ``{Spectral-efficient
		RIS-aided RSMA URLLC: Toward mobile broadband reliable low latency
		communication (mBRLLC) system},'' {\em {IEEE} Trans. Wireless Commun.},
	pp.~1--1, 2023.
	
	\bibitem{niu2023active}
	H.~Niu, Z.~Lin, K.~An, J.~Wang, G.~Zheng, N.~Al-Dhahir, and K.-K. Wong {\em
		{IEEE} J. Sel. Areas Commun.}
	
	\bibitem{dhok2022rate}
	S.~Dhok and P.~K. Sharma, ``{Rate-splitting multiple access with STAR RIS over
		spatially-correlated channels},'' {\em {IEEE} Trans. Commun.}, vol.~70,
	no.~10, pp.~6410--6424, 2022.
	
	\bibitem{CTN2022}
	H.~Jafarkhani, ``Taking to the air to help on the ground: How {UAVs} can help
	fight wildfires,'' in {\em IEEE ComSoc Technology News}, Oct. 2022.
	
	\bibitem{9930941}
	C.~Diaz-Vilor, A.~Lozano, and H.~Jafarkhani, ``Cell-free {UAV} networks:
	Asymptotic analysis and deployment optimization,'' vol.~22, no.~5,
	pp.~3055--3070, 2023.
	
	\bibitem{10186347}
	C.~Diaz-Vilor, A.~Lozano, and H.~Jafarkhani, ``Cell-free {UAV} networks with
	wireless fronthaul: Analysis and optimization,'' vol.~23, no.~3,
	pp.~2054--2069, 2024.
	
	\bibitem{10301558}
	C.~Diaz-Vilor, M.~A. Almasi, A.~M. Abdelhady, A.~Celik, A.~M. Eltawil, and
	H.~Jafarkhani, ``Sensing and communication in {UAV} cellular networks: Design
	and optimization,'' vol.~23, no.~6, pp.~5456--5472, 2024.
	
	\bibitem{JGPWHJ20}
	J.~Guo, P.~Walk, and H.~Jafarkhani, ``Optimal deployments of {UAVs} with
	directional antennas for a power-efficient coverage,'' vol.~68, no.~8,
	pp.~5159--5174, 2020.
	
	\bibitem{10594734}
	S.~Karimi-Bidhendi, G.~Geraci, and H.~Jafarkhani, ``{Optimizing cellular
		networks for {UAV} corridors via quantization theory},'' {\em {IEEE} Trans.
		Wireless Commun.}, 2024.
	
	\bibitem{jaafar2020downlink}
	W.~Jaafar, S.~Naser, S.~Muhaidat, P.~C. Sofotasios, and H.~Yanikomeroglu, ``{On
		the downlink performance of RSMA-based UAV communications},'' {\em {IEEE}
		Trans. Veh. Technol.}, vol.~69, no.~12, pp.~16258--16263, 2020.
	
	\bibitem{rahmati2019energy}
	A.~Rahmati, Y.~Yapici, N.~Rupasinghe, I.~Guvenc, H.~Dai, and A.~Bhuyan,
	``{Energy efficiency of RSMA and NOMA in cellular-connected mmWave UAV
		networks},'' in {\em Proc. IEEE Int. Commun. Conf. Workshops (ICC Wkshps)},
	pp.~1--6, 2019.
	
	\bibitem{singh2021outage}
	S.~K. Singh, K.~Agrawal, K.~Singh, and C.-P. Li, ``{Outage probability and
		throughput analysis of UAV-assisted rate-splitting multiple access},'' {\em
		{IEEE} Wireless Commun. Lett.}, vol.~10, no.~11, pp.~2528--2532, 2021.
	
	\bibitem{xiao2023traffic}
	M.~Xiao, H.~Cui, D.~Huang, Z.~Zhao, X.~Cao, and D.~O. Wu, ``{Traffic-aware
		energy-efficient resource allocation for RSMA based UAV communications},''
	{\em {IEEE Trans. Netw. Sci. Eng.}}, pp.~1--12, 2023.
	
	\bibitem{liu2023downlink}
	X.~Liu, J.~Feng, F.~Li, and V.~C.~M. Leung, ``{Downlink energy efficiency
		maximization for RSMA-UAV assisted communications},'' {\em {IEEE} Wireless
		Commun. Lett.}, pp.~1--1, 2023.
	
	\bibitem{yang2020sum}
	Z.~Yang, M.~Chen, W.~Saad, W.~Xu, and M.~Shikh-Bahaei, ``{Sum-rate maximization
		of uplink rate splitting multiple access (RSMA) communication},'' {\em IEEE
		Trans. Mob. Comput.}, vol.~21, no.~7, pp.~2596--2609, 2022.
	
	\bibitem{abbasi2022transmission}
	O.~Abbasi and H.~Yanikomeroglu, ``{Transmission scheme, detection and power
		allocation for uplink user cooperation with {NOMA} and {RSMA}},'' {\em {IEEE}
		Trans. Wireless Commun.}, vol.~22, no.~1, pp.~471--485, 2022.
	
	\bibitem{katwe2022rate}
	M.~Katwe, K.~Singh, B.~Clerckx, and C.-P. Li, ``{Rate-splitting multiple access
		and dynamic user clustering for sum-rate maximization in multiple RISs-aided
		uplink mmWave system},'' {\em {IEEE} Trans. Commun.}, vol.~70, no.~11,
	pp.~7365--7383, 2022.
	
	\bibitem{lu2023outage}
	H.~Lu, X.~Xie, Z.~Shi, H.~Lei, N.~Zhao, and J.~Cai, ``{Outage performance of
		uplink rate splitting multiple access with randomly deployed users},'' {\em
		{IEEE} Trans. Wireless Commun.}, pp.~1--1, 2023.
	
	\bibitem{katwe2023improved}
	M.~Katwe, K.~Singh, B.~Clerckx, and C.-P. Li, ``{Improved spectral efficiency
		in STAR-RIS aided uplink communication using rate splitting multiple
		access},'' {\em {IEEE} Trans. Wireless Commun.}, vol.~22, no.~8,
	pp.~5365--5382, 2023.
	
	\bibitem{zhang2023fairness}
	S.~Zhang and W.~Chen, ``{Fairness optimization of RSMA for uplink communication
		based on intelligent reflecting surface},'' {\em arXiv preprint
		arXiv:2309.02264}, 2023.
	
	\bibitem{sun2023joint}
	Q.~Sun, H.~Liu, S.~Yan, T.~A. Tsiftsis, and J.~Yuan, ``{Joint receive and
		passive beamforming optimization for RIS-assisted uplink RSMA systems},''
	{\em {IEEE} Wireless Commun. Lett.}, vol.~12, no.~7, pp.~1204--1208, 2023.
	
	\bibitem{rovella2024scalable}
	G.~D.~B. Rovella, M.~Benammar, T.~Benaddi, and H.~Meric, ``Scalable
	syndrome-based neural decoders for bit-interleaved coded modulations,'' {\em
		arXiv preprint 2403.02850}, 2024.
	
	\bibitem{de2000mahalanobis}
	R.~De~Maesschalck, D.~Jouan-Rimbaud, and D.~L. Massart, ``{The {Mahalanobis}
		distance},'' {\em {Chemometrics and intelligent laboratory systems}},
	vol.~50, no.~1, pp.~1--18, 2000.
	
	\bibitem{selim2020noma}
	B.~Selim, M.~S. Alam, J.~V. Evangelista, G.~Kaddoum, and B.~L. Agba,
	``{NOMA-based IoT networks: Impulsive noise effects and mitigation},'' {\em
		{IEEE} Commun. Mag.}, vol.~58, no.~11, pp.~69--75, 2020.
	
	\bibitem{XLHJ17}
	X.~Liu and H.~Jafarkhani, ``{Downlink non-orthogonal multiple access with
		limited feedback},'' {\em {IEEE} Trans. Wireless Commun.}, vol.~16,
	pp.~6151--6164, Sept. 2017.
	
	\bibitem{XZMGHJwcl20}
	X.~Zou, M.~Ganji, and H.~Jafarkhani, ``{Downlink asynchronous non-orthogonal
		multiple access with quantizer optimization},'' vol.~9, pp.~1606--1610, Oct.
	2020.
	
	\bibitem{7506136}
	P.~Xu, Y.~Yuan, Z.~Ding, X.~Dai, and R.~Schober, ``On the outage performance of
	non-orthogonal multiple access with 1-bit feedback,'' {\em {IEEE} Trans.
		Wireless Commun.}, vol.~15, no.~10, pp.~6716--6730, 2016.
	
	\bibitem{9094017}
	Y.~Yapıcı, I.~Güvenç, and H.~Dai, ``{Low-resolution limited-feedback {NOMA}
		for mm{W}ave communications},'' {\em {IEEE} Trans. Wireless Commun.},
	vol.~19, no.~8, pp.~5433--5446, 2020.
	
	\bibitem{8999638}
	P.~Swami, M.~K. Mishra, V.~Bhatia, and T.~Ratnarajah, ``{Performance analysis
		of {NOMA} enabled hybrid network with limited feedback},'' {\em {IEEE} Trans.
		Veh. Technol.}, vol.~69, no.~4, pp.~4516--4521, 2020.
	
	\bibitem{MAHJ23}
	M.~Almasi and H.~Jafarkhani, ``{Reconfigurable intelligent surface-aided {NOMA}
		with limited feedback},'' in {\em Proc. IEEE Int. Commun. Conf. (ICC)}, 2023.
	
	\bibitem{wu2023deep}
	M.~Wu, Z.~Gao, Y.~Huang, Z.~Xiao, D.~W.~K. Ng, and Z.~Zhang, ``{Deep
		learning-based rate-splitting multiple access for reconfigurable intelligent
		surface-aided tera-hertz massive MIMO},'' {\em {IEEE} J. Sel. Areas Commun.},
	vol.~41, no.~5, pp.~1431--1451, 2023.
	
	\bibitem{lu2017mmse}
	G.~Lu, L.~Li, H.~Tian, and F.~Qian, ``{MMSE-based precoding for rate splitting
		systems with finite feedback},'' {\em {IEEE} Commun. Lett.}, vol.~22, no.~3,
	pp.~642--645, 2017.
	
	\bibitem{10384715}
	M.~A. Almasi and H.~Jafarkhani, ``{Rate loss analysis of reconfigurable
		intelligent surface-aided {NOMA} With limited feedback},'' {\em IEEE Open J.
		Commun. Soc.}, vol.~5, pp.~856--871, 2024.
	
	\bibitem{akyildiz2022terahertz}
	I.~F. Akyildiz, C.~Han, Z.~Hu, S.~Nie, and J.~M. Jornet, ``{Terahertz band
		communication: An old problem revisited and research directions for the next
		decade},'' {\em {IEEE} Trans. Commun.}, vol.~70, no.~6, pp.~4250--4285, 2022.
	
	\bibitem{zhu2019millimeter}
	L.~Zhu, J.~Zhang, Z.~Xiao, X.~Cao, D.~O. Wu, and X.-G. Xia, ``{Millimeter-wave
		NOMA with user grouping, power allocation and hybrid beamforming},'' {\em
		{IEEE} Trans. Wireless Commun.}, vol.~18, no.~11, pp.~5065--5079, 2019.
	
	\bibitem{wei2018multi}
	Z.~Wei, L.~Zhao, J.~Guo, D.~W.~K. Ng, and J.~Yuan, ``{Multi-beam NOMA for
		hybrid mmWave systems},'' {\em {IEEE} Trans. Commun.}, vol.~67, no.~2,
	pp.~1705--1719, 2018.
	
	\bibitem{MALJHJHM21}
	M.~A. Almasi, L.~Jiang, { H. Jafarkhani}, and H.~Mehrpouyan, ``{Joint beamwidth
		and power optimization in mmWave hybrid beamforming-{NOMA} systems},'' {\em
		{IEEE} Trans. Wireless Commun.}, vol.~20, pp.~1934--1947, April 2021.
	
	\bibitem{cai2022experimental}
	Y.~Cai, M.~Chen, A.~Deng, D.~Wang, L.~Wang, X.~Gao, J.~Zhou, Y.~Liu, and
	C.~Xiang, ``{Experimental demonstration of 16QAM/QPSK OFDM-NOMA VLC with LDPC
		codes and analog pre-equalization},'' {\em Applied Optics}, vol.~61, no.~19,
	pp.~5585--5591, 2022.
	
	\bibitem{cho2023coverage}
	H.~Cho, B.~Ko, B.~Clerckx, and J.~Choi, ``{Coverage increase at THz
		frequencies: A cooperative rate-splitting approach},'' {\em {IEEE} Trans.
		Wireless Commun.}, vol.~22, no.~12, pp.~9821--9834, 2023.
	
	\bibitem{almasi2019impact}
	M.~A. Almasi, M.~Vaezi, and H.~Mehrpouyan, ``{Impact of beam misalignment on
		hybrid beamforming NOMA for mmWave communications},'' {\em {IEEE} Trans.
		Commun.}, vol.~67, no.~6, pp.~4505--4518, 2019.
	
	\bibitem{LJHJ19}
	L.~Jiang and H.~Jafarkhani, ``Multi-user analog beamforming in {mmWave} {MIMO}
	systems based on path angle information,'' {\em {IEEE} Trans. Wireless
		Commun.}, vol.~18, pp.~608--619, Jan. 2019.
	
	\bibitem{LJHJ20}
	L.~Jiang and H.~Jafarkhani, ``{MmWave} amplify-and-forward {MIMO} relay
	networks with hybrid precoding/combining design,'' {\em {IEEE} Trans.
		Wireless Commun.}, vol.~19, pp.~1333--1346, Feb. 2020.
	
	\bibitem{etemadi2018analog}
	F.~Etemadi, P.~Heydari, and H.~Jafarkhani, ``{On analog QAM demodulation for
		millimeter-wave communications},'' {\em IEEE Trans. Circuits Syst. II, Exp.
		Briefs.}, vol.~66, no.~3, pp.~402--406, 2018.
	
	\bibitem{wang2019analysis}
	H.~Wang, H.~Mohammadnezhad, and P.~Heydari, ``{Analysis and design of
		high-order QAM direct-modulation transmitter for high-speed point-to-point
		mm-wave wireless links},'' {\em IEEE J. Solid-State Circuits}, vol.~54,
	no.~11, pp.~3161--3179, 2019.
	
	\bibitem{mo2015capacity}
	J.~Mo and R.~W. Heath, ``{Capacity analysis of one-bit quantized MIMO systems
		with transmitter channel state information},'' {\em {IEEE} Trans. Signal
		Process.}, vol.~63, no.~20, pp.~5498--5512, 2015.
	
\end{thebibliography}
\end{document}